\documentclass[%
 reprint,
 amsmath,amssymb,
 aps,
]{revtex4-2}

\usepackage{graphicx}
\usepackage{dcolumn}
\usepackage{bm}

\usepackage{slashed}
\usepackage{setspace} 
\usepackage{ulem}
\usepackage{braket}
\usepackage{color}
\usepackage{subfigure}
\usepackage{float}

\usepackage{amsfonts}
\usepackage[bookmarks=true,bookmarksopen=true,
bookmarksnumbered=true,
colorlinks=true,linkcolor=black,
citecolor=red]{hyperref}

\usepackage{caption}

\begin{document}

\preprint{APS/123-QED}

\title{Hunting a charged Higgs boson pair\\in proton-proton collisions}

\author{M. A. Arroyo-Ure\~na}
\email{marco.arroyo@fcfm.buap.mx}
\affiliation{Facultad de Ciencias F\'isico-Matem\'aticas, Benem\'erita Universidad Aut\'onoma de Puebla, C.P. 72570, Puebla, M\'exico,}
\affiliation{Centro Interdisciplinario de Investigaci\'on y Ense\~nanza de la Ciencia (CIIEC), Benem\'erita Universidad Aut\'onoma de Puebla, C.P. 72570, Puebla, M\'exico.}

\author{E. A. Herrera-Chac\'on}
\email{edwin.a.herrera-chacon@durham.ac.uk}
\affiliation{Institute for Particle Physics Phenomenology, Durham University, South Road, Durham, DH1 3LE}

\author{S. Rosado-Navarro}
\email{sebastian.rosado@protonmail.com}
\affiliation{Centro Interdisciplinario de Investigaci\'on y Ense\~nanza de la Ciencia (CIIEC), Benem\'erita Universidad Aut\'onoma de Puebla, C.P. 72570, Puebla, M\'exico.}

\author{Humberto Salazar}
\email{hsalazar@fcfm.buap.mx}
\affiliation{Facultad de Ciencias F\'isico-Matem\'aticas, Benem\'erita Universidad Aut\'onoma de Puebla, C.P. 72570, Puebla, M\'exico,}
\affiliation{Centro Interdisciplinario de Investigaci\'on y Ense\~nanza de la Ciencia (CIIEC), Benem\'erita Universidad Aut\'onoma de Puebla, C.P. 72570, Puebla, M\'exico. }


\begin{abstract}

We explore the production and possible detection of a charged scalar Higgs pair $H^-H^+$ decaying into the final state $\mu\nu_{\mu}cb$ in proton-proton collisions at the LHC and its next step, the High Luminosity LHC (HL-LHC). The charged scalars are predicted within the theoretical framework of the Two-Higgs Doublet Model of type III (2HDM-III). As a test and validation of the model, we identify regions of the 2HDM-III parameter space that accommodate the current excess of events at $3\sigma$ in the process $\mathcal{BR}(t\to H^{\pm}b)\times \mathcal{BR}(H^{\pm}\to cb)$ for $M_{H^{\pm}}=130$ GeV, as reported by the ATLAS collaboration. Theoretical and additional experimental constraints are also included. Based on this, we propose realistic scenarios that could be brought under experimental scrutiny at the HL-LHC. Assuming the most favorable scenario, we predict a signal significance at the level of $5\sigma$ for a charged scalar boson mass $M_{H^{\pm}}$ in the $100-350$ GeV range.
\end{abstract}

\maketitle


\section{\label{sec:Introduction}Introduction}

Unveiling the nature of charged Higgs bosons, a cornerstone of theories beyond the Standard Model (SM), is a key objective in high-energy particle physics. This study explores a production channel of two charged Higgs bosons at the Large Hadron Collider (LHC) and the future High Luminosity Large Hadron Collider (HL-LHC) via $pp \rightarrow H^- H^+ $, which is a fascinating phenomenon in the field of particle physics. In this study, we specifically focus on the decay channels $H^{\pm} \rightarrow \mu \nu_\mu$ and $H^{\pm} \rightarrow cb$ in the final states. By analyzing this production process within the framework of the Two-Higgs Doublet Model Type III (2HDM-III), we aim to deepen our understanding of the properties and behavior of charged Higgs bosons.
	
	Building upon recent searches for a charged Higgs boson carried by the ATLAS \cite{ ATLAS:2023ofo,ATLAS:2012nhc} and CMS \cite{CMS:2012fgz} collaborations at CERN. Yet, the production channel we choose in our work may produce two charged Higgs bosons.  By focusing on their specific decay channels, we explore their behavior within the framework of the 2HDM-III. Our investigation pushes the boundaries of current knowledge, not only by quantifying the LHC's sensitivity to these elusive particles, but also by offering valuable insights into the parameter space of the 2HDM-III. This opens doors for future experimental pursuits and theoretical explorations, potentially leading to the long-awaited discovery of charged Higgs bosons.
	
	Recently, the ATLAS collaboration searched for a charged Higgs boson, $H^{\pm}$, via the $t\to H^{\pm}b (H^{\pm}\to cb)$ decay \cite{ATLAS:2023bzb}. The study was focused on a selection enriched in top-quark pair production, where one of them decays into a leptonically decaying $W$ boson and a bottom quark, while the other top quark decays into $H^{\pm}b$. The search used a dataset of proton-proton collisions at a center-of-mass energy $\sqrt{s}=13$ TeV and an integrated luminosity of 139 fb$^{-1}$. A light excess in data with a significance of about 3$\sigma$ for $m_{H^{\pm}}$ was reported by the ATLAS collaboration. 
	
	The 2HDM-III provides a rich theoretical framework that allows different Yukawa couplings, which can significantly impact the production and decay processes of charged Higgs bosons \cite{HernandezSanchez:2012eg,Felix-Beltran:2013tra,Hernandez-Sanchez:2020vax}. Previous works have addressed the flavor-violating $H^{\pm} \rightarrow cb$ decay with interesting phenomenology, so we decided to add the $H^{\pm} \rightarrow \ell \nu_{\ell}$ decay in order to have cleaner events. To assess the production rates and kinematic properties of the $H^- H^+$ system, we employ advanced Monte Carlo simulations that incorporate detector effects and account for relevant background processes. This enables us to determine the potential of the LHC to observe the signals of these charged Higgs bosons and discern them from the dominant irreducible background, providing crucial insights into their existence and behavior in the collider environment.
	
	The implications of our study extend beyond the immediate experimental prospects. By quantifying the sensitivity of the LHC experiments to these $H^\pm$ signals, we contribute to the ongoing efforts to explore the 2HDM-III parameter space. The insights gained from our analysis serve as valuable guidance for future experimental searches, aiding in the design and optimization of search strategies to maximize the chances of discovering charged Higgs bosons.
	
	In summary, our comprehensive analysis sheds light on the rich phenomenology associated with charged Higgs bosons at the LHC. Through the investigation of their production and decay modes, we deepen our understanding of the underlying physics and pave the way for further theoretical explorations and experimental discoveries in the realm of particle physics.
	
	This work is structured as follows. In Sec. \ref{SecII}, we conduct a comprehensive review of the 2HDM-III with a particular emphasis on the theoretical implications of employing a four-zero texture in the Yukawa Lagrangian.  Section \ref{SecIII} is devoted to meticulously examining the constraints imposed on the model's parameter space by both established theoretical frameworks and current experimental data. Section \ref{SecIV} is focused on leveraging the insights gained from the previous steps, and then we perform a computational analysis of charged Higgs boson production and decay processes at the LHC and the HL-LHC. Finally, the conclusions are presented in Sec. \ref{SecV}.

\section{\label{SecII}The model}

For our research, we consider the 2HDM-III using a four-zero texture. This extension introduces the charged Higgs phenomenology. The theoretical framework is presented in this section, analyzing the Yukawa Lagrangian of the 2HDM-III and obtaining the Feynman rules for our calculations. Through this in-depth exploration, we establish a solid foundation for our subsequent analysis of charged Higgs boson production and decay at the Large Hadron Collider (LHC).

\subsection{Scalar potential}

The 2HDM-III introduces an additional Higgs doublet beyond the SM. This doublet can be denoted as $\Phi_a^T=( \phi_{a}^{+},\phi_{a}^{0})$ for $a=1, 2$, and carries a hypercharge of $+1$. During the process of Spontaneous Symmetry Breaking (SSB), these Higgs doublets acquire non-zero vacuum expectation values (VEV), given as
	
	\begin{center}
		$\braket{\Phi_a}=\frac{1}{\sqrt{2}}\left(\begin{array}{c}
			0\\
			\upsilon_a
		\end{array}\right)$.
	\end{center}
	
	
	In the 2HDM-III, the presence of additional Higgs doublets introduces new interactions between these doublets and all types of fermions. However, these interactions can lead to the emergence of Flavor-Changing Neutral Currents (FCNCs) at the tree level. Since experimental observations place stringent constraints on FCNCs, mechanisms are needed to suppress them within the model.
	
	An effective approach to achieve this suppression involves implementing a specific four-zero texture in the Yukawa sector of the 2HDM-III. This texture essentially acts as a simplified theory that governs the flavor interactions of fermions with the Higgs bosons. By employing a four-zero texture in our works \cite{Hernandez-Sanchez:2020vax, Arroyo-Urena:2020mgg, Arroyo-Urena:2019qhl, Arroyo-Urena:2013cyf}, we can effectively control the strength of FCNCs and ensure consistency with the experimental data.
	
	The most general $SU(2)_L \times U(1)_Y$ invariant scalar potential reads:  

 \begin{eqnarray}
V(\Phi_1,\Phi_2) &=& \mu_{1}^{2}(\Phi_{1}^{\dag}\Phi_{1}^{}) + \mu_{2}^{2}(\Phi_{2}^{\dag}\Phi_{2}^{}) - \mu_{12}^{2}(\Phi_{1}^{\dag}\Phi_{2}^{} + h.c.) \nonumber \\ 
&+& \frac{1}{2} \lambda_{1}(\Phi_{1}^{\dag}\Phi_{1}^{})^2 + \frac{1}{2} \lambda_{2}(\Phi_{2}^{\dag}\Phi_{2}^{})^2 \nonumber \\ 
&+& \lambda_{3}(\Phi_{1}^{\dag}\Phi_{1}^{})(\Phi_{2}^{\dag}\Phi_{2}^{}) + \lambda_{4}(\Phi_{1}^{\dag}\Phi_{2}^{})(\Phi_{2}^{\dag}\Phi_{1}^{}) \nonumber \\
&+& \frac{1}{2} \lambda_{5}(\Phi_{1}^{\dag}\Phi_{2}^{})^2  + \lambda_{6}(\Phi_{1}^{\dag}\Phi_{1}^{})(\Phi_{1}^{\dag}\Phi_{2}^{}) \nonumber \\
&+& \lambda_{7}(\Phi_{2}^{\dag}\Phi_{2}^{})(\Phi_{1}^{\dag}\Phi_{2}^{}) + h.c.
\label{potential}
\end{eqnarray}

	In the present work, we assume the potential parameters to be real, even the VEV's of the Higgs doublets. Thus, the CP symmetry is conserved.

	The model has additional independent parameters, namely the mixing angle $\alpha$, which is related to the mass matrix corresponding to the CP-even sector, and it defines the transition from gauge to mass eigenstates:
	
	\begin{eqnarray}
		H&=&\rm Re(\phi_1^0)\cos\alpha+\rm Re(\phi_2^0)\sin\alpha,\nonumber\\
		h&=&\rm Re(\phi_1^0)\sin\alpha+\rm Re (\phi_2^0)\cos\alpha,
		\label{eq:Hh_definition}
	\end{eqnarray}
	with 
	\begin{equation*}
		\tan2\alpha=\frac{2m_{12}}{m_{11}-m_{22}},
	\end{equation*}
	where $m_{11,\,12,\,22}$ are elements of the real part of the mass matrix $\bold{M}$,
	\begin{center}
		$\rm Re(\bold{M})=\left(\begin{array}{cc}
			m_{11} & m_{12}\\
			m_{12} & m_{22}
		\end{array}\right)$,
		\par\end{center}
	where
	\begin{eqnarray}
		m_{11}&=&m_A^2 \sin^2\beta+v^2(\lambda_1\cos^2\beta+\lambda_{5}\sin^2\beta)\nonumber
		\\
		m_{12}&=&-m_A^2 \cos\beta\sin\beta+v^2(\lambda_{3}+\lambda_{4})\cos\beta\sin\beta\nonumber 
		\\
		m_{22}&=&m_A^2\cos^2\beta+v^2(\lambda_{2}\sin^2\beta+\lambda_{5}\cos^2\beta).
	\end{eqnarray}
	The Eqs. \eqref{eq:Hh_definition} define a physical charged scalar boson, a pseudo-Goldstone boson associated with the $W$ gauge fields, a CP-odd state, and the pseudo-Goldstone boson related to the $Z$ gauge boson.

	There is another angle, $\beta$, which mixes the charged components of $\Phi_a$ and the imaginary part of the neutral components of $\Phi_a$ as follows,
	\begin{eqnarray}
		G_W^{\pm}&=&\phi_1^{\pm}\cos\beta+\phi_2^{\pm}\sin\beta,\nonumber\\
		H^{\pm}&=&-\phi_1^{\pm}\sin\beta+\phi_2^{\pm}\cos\beta,\nonumber\\	
		G_Z&=&\rm Im(\phi_1^0)\cos\beta+\rm Im(\phi_2^0)\sin\beta,\nonumber\\
		A^0&=&-\rm Im(\phi_1^0)\sin\beta+\rm Im(\phi_2^0)\cos\beta.
	\end{eqnarray}
	
	The mixing angle $\beta$ parameterizes the VEV ratio as $\tan\beta=\upsilon_2/\upsilon_1$.

	In this model, we will work with the mixing angles $\alpha$ and $\beta$ and the physical masses $M_H^{\pm}$, $M_h$, $M_H$ and $M_A$, which are given by
	
	\begin{eqnarray}
		M_{H^{\pm}}^2&=&\frac{\mu_{12}}{\sin\beta\cos\beta}\\&-&\frac{1}{2}v^2\Bigg(\lambda_4+\lambda_{5}+\cot\beta\lambda_{6}+\tan\beta\lambda_7\Bigg), \nonumber
		\\
		M_{h,H}^2&=&\frac{1}{2}\Bigg(m_{11}+m_{22}\mp\sqrt{(m_{11}-m_{22})^2+4m_{12}^2}\Bigg),  
		\\
		M_A^2&=&M_H^{\pm}+\frac{1}{2}v^2(\lambda_4-\lambda_{5}).
	\end{eqnarray}

\subsection{Yukawa Lagrangian}
	The equation for the Yukawa Lagrangian in the 2HDM-III is given by \cite{HernandezSanchez:2012eg}:
	\begin{align}
\label{YukawaLagrangian} 
\mathcal{L}_Y = & -\left( Y_{1}^{u} \bar{Q}_{L} \tilde{\Phi}_{1} u_{R} + Y_{2}^{u} \bar{Q}_{L} \tilde{\Phi}_{2} u_{R} \right. \nonumber \\
& \left. + Y_{1}^{d} \bar{Q}_{L} \Phi_{1} d_{R} + Y_{2}^{d} \bar{Q}_{L} \Phi_{2} d_{R} \right. \nonumber \\
& \left. + Y_{1}^{l} \bar{L}_{L} \tilde{\Phi}_{1} l_{R} + Y_{2}^{l} \bar{L}_{L} \tilde{\Phi}_{2} l_{R} \right),
\end{align}

	where  $\tilde{\Phi}_{a} = i\sigma_2 \Phi_{a}^{*}$ $(a=1, 2)$.  After the Electroweak Symmetry Breaking (EWSB), the fermion mass matrices have the form:
	
	\begin{equation}
		M_f = \frac{1}{\sqrt{2}} \left( v_1 Y_{1}^{f} + v_2 Y_{2}^{f} \right),\; f=u,d,\ell.	
	\end{equation}
	
	In this step, we assume that both Yukawa matrices have the mentioned four-zero texture form and that they are Hermitian.
	
	After the diagonalization process, we have
	\begin{eqnarray}
		\bar{M}_f = V_{fL}^{\dag} M_f V_{fR}&=& \tfrac{1}{\sqrt{2}} \left( v_1 \tilde{Y}_{1}^{f} + v_2 \tilde{Y}_{2}^{f} \right),\\
		\tilde{Y}_{a}^{f} &=& V_{fL}^{\dag} Y_{a}^{f} V_{fR}
	\end{eqnarray}
	from where we can deduce the following,
	\begin{equation}\label{RotateYukawas}
		\left[ \tilde{Y}_a^f \right]_{ij} = \frac{\sqrt{2}}{v_a}\delta_{ij}\bar{M}_{ij}^f-\frac{v_b}{v_a}\left[ \tilde{Y}_b^f \right]_{ij}
	\end{equation}
	where $f$ stands for massive fermions and the parameters $\left[\tilde{\chi}_a^f \right]_{ij}$'s are unknown dimensionless quantities of the present model. It is important to mention that from Eq. \ref{RotateYukawas} we can obtain different kinds of interactions. By choosing specific structures \cite{HernandezSanchez:2012eg}, the following models are defined as
	\begin{itemize}
		\item 2HDM-I-like
		\begin{eqnarray}\label{I}
			\left[ \tilde{Y}_2^d \right]_{ij} &=& \frac{\sqrt{2}}{v\sin\beta}\delta_{ij}\bar{M}_{ij}^d-\cot\beta\left[ \tilde{Y}_1^d \right]_{ij}\nonumber\\
			\left[ \tilde{Y}_2^u \right]_{ij} &=& \frac{\sqrt{2}}{v\sin\beta}\delta_{ij}\bar{M}_{ij}^u-\cot\beta\left[ \tilde{Y}_1^u \right]_{ij}\nonumber\\
			\left[ \tilde{Y}_2^\ell \right]_{ij} &=& \left[ \tilde{Y}_1^d \right]_{ij}(d\to\ell).
		\end{eqnarray}  
		\item 2HDM-II-like
		\begin{eqnarray}\label{II}
			\left[ \tilde{Y}_1^d \right]_{ij} &=& \frac{\sqrt{2}}{v\cos\beta}\delta_{ij}\bar{M}_{ij}^d-\tan\beta\left[ \tilde{Y}_2^d \right]_{ij}\nonumber\\
			\left[ \tilde{Y}_2^u \right]_{ij} &=& \frac{\sqrt{2}}{v\sin\beta}\delta_{ij}\bar{M}_{ij}^u-\cot\beta\left[ \tilde{Y}_1^u \right]_{ij}\nonumber\\
			\left[ \tilde{Y}_1^\ell \right]_{ij} &=& \left[ \tilde{Y}_1^d \right]_{ij}(d\to\ell).
		\end{eqnarray}  
		\item 2HDM-Lepton Specific-like
		\begin{eqnarray}\label{LS}
			\left[ \tilde{Y}_2^d \right]_{ij} &=& \frac{\sqrt{2}}{v\sin\beta}\delta_{ij}\bar{M}_{ij}^d-\cot\beta\left[ \tilde{Y}_1^d \right]_{ij}\nonumber\\
			\left[ \tilde{Y}_2^u \right]_{ij} &=& \frac{\sqrt{2}}{v\sin\beta}\delta_{ij}\bar{M}_{ij}^u-\cot\beta\left[ \tilde{Y}_1^u \right]_{ij}\nonumber\\
			\left[ \tilde{Y}_1^\ell \right]_{ij} &=& \left[ \tilde{Y}_1^d \right]_{ij}(d\to\ell).
		\end{eqnarray}  
		\item 2HDM-Flipped-like
		\begin{eqnarray}\label{F}
			\left[ \tilde{Y}_1^d \right]_{ij} &=& \frac{\sqrt{2}}{v\cos\beta}\delta_{ij}\bar{M}_{ij}^d-\tan\beta\left[ \tilde{Y}_2^d \right]_{ij}\nonumber\\
			\left[ \tilde{Y}_2^u \right]_{ij} &=& \frac{\sqrt{2}}{v\sin\beta}\delta_{ij}\bar{M}_{ij}^u-\cot\beta\left[ \tilde{Y}_1^u \right]_{ij}\nonumber\\
			\left[ \tilde{Y}_2^\ell \right]_{ij} &=& \left[ \tilde{Y}_2^d \right]_{ij}(d\to\ell).
		\end{eqnarray}  
		
	\end{itemize}
	From Eqs. \eqref{YukawaLagrangian}-\eqref{RotateYukawas}, we obtain 
	\begin{equation}
		\mathcal{L}_Y^\phi=\phi\bar{f}_i(S_{ij}^{\phi}+iP_{ij}^\phi\gamma^5)f_j,
	\end{equation}
	where $\phi=h,\,H,\,A$.
	The CP-conserving and CP-violating factors $S_{ij}^{\phi}$ and $P_{ij}^\phi$, respectively, include the flavor physics. They are written as:
	\begin{eqnarray}
		S_{ij}^{\phi}&=&\frac{gm_f}{2M_W}c_f^\phi\delta_{ij}+d^\phi_f\left[ \tilde{Y}_a^f \right]_{ij},\nonumber\\
		P_{ij}^\phi&=&\frac{gm_f}{2M_W}e_f^\phi\delta_{ij}+g^\phi_f\left[ \tilde{Y}_a^f \right]_{ij}.\label{SijPij}
	\end{eqnarray}
	The coefficients $c_f^\phi,\,d_f^\phi,\,e_f^\phi,\,g_f^\phi$ depend on the new physics in the Higgs sector. Within the Standard Model $c_f^{\phi=h}=1$ and $d_f^{\phi=h}=e_f^{\phi=h}=g_f^{\phi=h}=0$, while in the 2HDM-III these coefficients are shown in Table \ref{couplings}. In Eqs. \eqref{SijPij} we can select $a$ for the Yukawa matrices according \eqref{I}-\eqref{F}. We assume a 2HDM-II like of the 2HDM-III. For simplicity, we will refer to the theoretical framework only as 2HDM-III.  
	\begin{center}
		\begin{table}
			\begin{tabular}{|c|c|c|c|c|c|c|}
				\hline 
				Coefficient & $c_{f}^{h}$ & $c_{f}^{A}$ & $c_{f}^{H}$ & $d_{f}^{h}$ & $d_{f}^{A}$ & $d_{f}^{H}$\tabularnewline
				\hline 
				\hline 
				$d$-type & $-\frac{\sin\alpha}{\cos\beta}$ & $-\tan\beta$ & $\frac{\cos\alpha}{\sin\beta}$ & $\frac{\cos(\alpha-\beta)}{\cos\beta}$ & $\csc\beta$ & $\frac{\sin(\alpha-\beta)}{\cos\beta}$\tabularnewline
				\hline 
				$u$-type & $\frac{\cos\alpha}{\sin\beta}$ & $-\cot\beta$ & $\frac{\sin\alpha}{\sin\beta}$ & $-\frac{\cos(\alpha-\beta)}{\sin\beta}$ & $\sec\beta$ & $\frac{\sin(\alpha-\beta)}{\sin\beta}$\tabularnewline
				\hline 
				leptons $\ell$ & $-\frac{\sin\alpha}{\cos\beta}$ & $-\tan\beta$ & $\frac{\cos\alpha}{\sin\beta}$ & $\frac{\cos(\alpha-\beta)}{\cos\beta}$ & $\csc\beta$ & $\frac{\sin(\alpha-\beta)}{\cos\beta}$\tabularnewline
				\hline 
			\end{tabular}
			\caption{Coefficients for $\phi$-Fermion couplings in 2HDM-III with $\mathcal{CP}$-conserving Higgs potential.}	\label{couplings}
			
		\end{table}
	\end{center}
	
	As far the couplings of the charged scalar boson with quarks are given as follows,
	\begin{widetext}
	\begin{align}\label{YukChargedQuarks}
		\mathcal{L}_Y^{H^{\pm}q_i q_j} 
		& =\frac{\sqrt{2}}{v}\Bigg\{\left[\bar{d}_i\left(m_u\cot \beta   - \frac{v}{\sqrt{2}} g(\beta)\Big[\tilde{Y}_2^u\Big]_{ij} \right) u_j H^{-}V_{\rm CKM}^{ij*} -\bar{u}_i\left(m_d\tan \beta  -\frac{v}{\sqrt{2}} f(\beta) \Big[\tilde{Y}^d_1\Big]_{ij} \right) d_j H^{+}V_{\rm CKM}^{ij}\right] P_R \\ \notag
		& -\left[\bar{d}_i\left(m_d\tan \beta  -\frac{v}{\sqrt{2}} f(\beta) \Big[\tilde{Y}^d_1\Big]_{ij} \right) u_j H^{-}V_{\rm CKM}^{ij*} +\bar{u}_i\left(m_u\cot\beta  -\frac{v}{\sqrt{2}} g(\beta)\Big[\tilde{Y}_2^u\Big]_{ij}\right) d_j H^{+}V_{\rm CKM}^{ij}\right] P_L \notag \Bigg\}
	\end{align}
\end{widetext}
where
	\begin{eqnarray}		\Big[\tilde{Y}_a^f\Big]_{ij}&=&\frac{\sqrt{m_{f_i}m_{f_j}}}{v}\chi_{ij}, (a=1,2),\\
		f(\beta)&=&\sqrt{1+\tan^2\beta},\\
		g(\beta)&=&\sqrt{1+\cot^2\beta},
	\end{eqnarray}
	while the couplings of the charged scalar boson with leptons read:
	\begin{align}
\mathcal{L}_Y^{{H^{\pm}}\ell \nu_{{\ell}}} &= \frac{\sqrt{2}m_{\ell_i}}{v}\bar{\nu}_{L}\Bigg(\tan\beta\frac{m_{\ell_i}}{m_{\ell_j}}\delta_{ij}\notag \\&-\frac{f(\beta)}{\sqrt{2}}\sqrt{\frac{m_{\ell_i}}{m_{\ell_j}}}\chi_{ij}^{\ell}\Bigg)\ell^-_{R}H^+  
+ H.c.
\end{align}

	
	\section{Constraints on the 2HDM-III parameter space}
	\label{SecIII}
	
	In order to achieve realistic predictions, we conduct a comprehensive analysis of various experimental constraints, namely
	\begin{itemize}
		\item Process induced for the neutral scalar (pseudo-scalar) bosons $h,\,H,\,A$
		\begin{itemize}
			\item LHC Higgs boson data \cite{CMS:2017con, ATLAS:2019pmk},
			\item Neutral meson physics $B_{s}^0\to\mu^+\mu^-$ \cite{CMS:2022mgd}, $B_{d}^0\to\mu^+\mu^-$ \cite{CMS:2022mgd},
			\item $\ell_i\to \ell_j\ell_k\bar{\ell}_k$ and $\ell_i\to \ell_j\gamma$ \cite{Workman:2022ynf},
			\item muon anomalous magnetic moment $a_\mu$\footnote{$a_\mu$ also receives contributions coming from a charged scalar boson, however its contribution is subdominant.} \cite{Muong-2:2021ojo}
			\item $\sigma(pp\to H \to hh\to b\bar{b}\gamma\gamma)$ \cite{ATLAS:2021ifb}
			\item $\sigma(gb\to \phi\to \tau\tau)$ ($\phi=H,\,A$) \cite{ATLAS-CONF, Sirunyan:2018zut} 
		\end{itemize}
		\item Process induced for the charged scalar boson $H^{\pm}$
		\begin{itemize}
			\item Limits on $\sigma(pp\to tbH^{\pm})\times \rm BR(H^{\pm}\to\tau^{\pm}\nu)$ \cite{ATLAS:2018gfm},
			\item Limits on $\mathcal{BR}(t\to H^{\pm}b)\times \mathcal{BR}(H^{\pm}\to cb)$ \cite{ATLAS:2023bzb},
			\item Decay $b\to s\gamma$ \cite{Ciuchini:1997xe, Misiak:2017bgg}.
		\end{itemize} 
	\end{itemize}
	
	The free model parameters that have a direct impact on our predictions are as follows,
	\begin{enumerate}
		\item Cosine of the difference of mixing angles: $\cos(\alpha-\beta)$,
		\item Ratio of the VEV's: $\tan\beta$,
		\item Boson masses: $M_H$, $M_A$, $M_{H^{\pm}}$,
		\item Matrix elements $\chi_{cb}$, $\chi_{tb}$, $\chi_{\mu\mu}$, $\chi_{tt}$, $\chi_{bb}$. 
	\end{enumerate}
	There are several parameters $\chi_{ij}$ necessary for the calculation of the $\mathcal{BR}(H^+\to bc)$ and $\mathcal{BR}(t\to H^+b)$. We set all the $\chi_{ij}=1$, except where otherwise indicated. Indices $i$ and $j$ denote fermions, in general $i\neq j$.
	
	\subsection{Constraint on $\tan\beta$ and $\cos(\alpha-\beta)$}
	
	\subsection*{LHC Higgs boson data: Signal strength modifiers}\label{HiggsConstraints}
	For a decay $S\to X$ or a production process $\sigma(pp\to S)$, the signal strength is parameterized as
	\begin{equation}\label{muX}
		\mathcal{\mu}_{X}=\frac{\sigma(pp\to h)\cdot\mathcal{BR}(h\to X)}{\sigma(pp\to h^{\text{SM}})\cdot\mathcal{BR}(h^{\text{SM}}\to X)},
	\end{equation}
	where $\sigma(pp\to S)$ is the production cross-section of $S$, with $S=h,\,h^{\text{SM}}$; here $h$ is the SM-like Higgs boson coming from an extension of the SM and $h^{\text{SM}}$ is the SM Higgs boson; $\mathcal{BR}(S\to X)$ is the branching ratio of the decay $S\to X$, with $X=b\bar{b},\;\tau^-\tau^+,\;\mu^-\mu^+,\;WW^*,\;ZZ^*,\;\gamma\gamma$.  
	\subsection*{Neutral meson physics}\label{NeutralMesonsConstraints}
	In order to complement constraints coming from LHC Higgs boson data, we consider Lepton Flavor Violation (LFV) processes, which are mediated by neutral scalar bosons ($H,\,h$) and a pseudoscalar ($A$)\footnote{ In general, all of them can induce FCNC at tree-level.} and a charged scalar boson $H^{\pm}$. These observables are 1) the decay $B_s^0\to\mu^+\mu^-$, 2) muon anomalous magnetic moment $a_{\mu}$, 3) radiative processes $\ell_i\to \ell_j \gamma$ ($\ell_i=\tau,\,\mu$, $\ell_j=\mu,\,e$, $i\neq j$), 4)  $\ell_i\to\ell_j\ell_k\bar{\ell}_k$ ($k=\mu,\,e$) and the decay of the Higgs boson changing flavor $h\to \ell_i\ell_j$. 
	\subsubsection{Decays $B_{s}^0\to\mu^+\mu^-$ and $B_{d}^0\to\mu^+\mu^-$}
\begin{figure}[!ht]
		\centering
		\includegraphics[width=6cm]{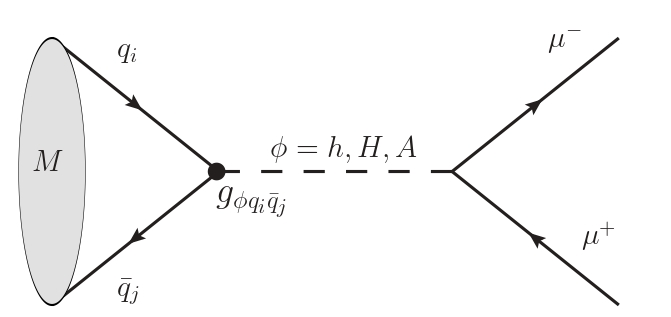}
		\caption{Generic Feynman diagram for the decays $M\to\mu^+\mu^-$. The black circle denotes a flavor-changing in the quark sector.}\label{FDMmumu} 
	\end{figure}

	Decays of neutral mesons into muons are strongly suppressed in the theoretical framework of the SM. This suppression arises due to three key factors: (i) they occur through loop diagrams, which are inherently weaker than tree-level processes; (ii) helicity suppression reduces the probability of the interaction; and (iii) the small values of certain CKM matrix elements further diminish the decay rate. Consequently, the branching ratios for these decays are extremely low. While other channels such as electron and $\tau$ decays might be theoretically possible, they are strongly suppressed and difficult to reconstruct, respectively. A representative Feynman diagram illustrating this process is shown in Fig. \ref{FDMmumu}
	
	$B_{s, d}^0$ meson decay into $\mu^+\mu^-$ pair, is both, interesting and stringent due to its sensitivity to constrain BSM theories. Within the theoretical framework of the SM the branching ratios read $\mathcal{BR}(B_s^0\to \mu^+\mu^-)=(3.66\pm 0.14)\times 10^{-9}$ and $\mathcal{BR}(B_d^0\to \mu^+\mu^-)=(1.03\pm 0.05)\times 10^{-10}$ \cite{Beneke:2019slt}, while the current experimental value reported by CMS collaboration is $\mathcal{BR}(B_s^0\to \mu^+\mu^-)=(4.02^{+0.40}_{-0.38}(\rm stat)^{+0.28}_{-0.23}(\rm syst)^{+0.18}_{-0.15}(B))\times 10^{-9}$, meanwhile $\mathcal{BR}(B_d^0\to \mu^+\mu^-)<1.9\times 10^{-10}$ at $95\%$ C.L. \cite{CMS:2022mgd}. In the context of the 2HDM-III, the decays $B_{s,d}^0\to\mu^+\mu^-$ are mediated by the SM-like Higgs boson, the heavy scalar $H$ and the pseudoscalar $A$ and it can arise at tree level. Feynman diagram for these decays is depicted in Fig. \ref{FDMmumu}. For the decay $B_s^0\to \mu^+\mu^-$ ($B_d^0\to \mu^+\mu^-$) corresponds to $q_i=s,\,\bar{q}_j=\bar{b}$ ($q_i=d,\,\bar{q}_j=\bar{b}$)
	
	The effective Hamiltonian governing the transition $B_s^0\to \mu^+\mu^-$ is
	\begin{align}
\mathcal{H}_{\rm eff}^{B_{s}^0\to \mu^+\mu^-}=-\frac{G_F^2m_W^2}{\pi^2}\Big(C_A^{bs}\mathcal{O}_A^{bs}+C_S^{bs}\mathcal{O}_S^{bs}+C_P^{bs}\mathcal{O}_P^{bs} \notag \\
+C_A^{\prime bs}\mathcal{O}_A^{\prime bs}+C_S^{\prime bs}\mathcal{O}_S^{\prime bs}+C_P^{\prime bs}\mathcal{O}_P^{\prime bs}\Big)+h.c.,
\end{align}

	where the Wilson operators are given as follows
	\begin{eqnarray}\label{WilsonOperators}
		\mathcal{O}_A^{bs}&=&\left(\bar{b}\gamma_\mu P_Ls\right)\left(\mu^+\gamma_\mu\gamma_5\mu^-\right),\nonumber \\
		\mathcal{O}_S^{bs}&=&\left(\bar{b} P_Ls\right)\left(\mu^+\mu^-\right),\\
		\mathcal{O}_P^{bs}&=&\left(\bar{b} P_Ls\right)\left(\mu^+\gamma_5\mu^-\right)\nonumber.
	\end{eqnarray}
	The primed operators are obtained replacing $P_L\leftrightarrows P_R$.
	The branching ratio for this decay is given by 
	\begin{widetext}
	\begin{align}\label{BRMlili}
		\begin{array}{ccl}
			\mathcal{BR}\left(M\rightarrow\ell^{+}\ell^{-}\right) & = & \frac{G_{F}^{4}m_{W}^{4}}{8\pi^{5}}\sqrt{1-4\frac{m_{\ell}^{2}}{m_{M}^{2}}}m_{M}f_{M}^{2}m_{\ell}^{2}\tau_{M}\\
			& \times & \left[\left|\frac{m_{M}^{2}\left(C_{P}^{ij}-C_{P}^{\prime ij}\right)}{2\left(m_{i}+m_{j}\right)m_{\ell}}-C_{A}^{\rm SM}\right|^{2}+\left|\frac{m_{M}^{2}\left(C_{S}^{ij}-C_{S}^{\prime ij}\right)}{2\left(m_{i}+m_{j}\right)m_{\ell}}\right|^{2}\left(1-4\frac{m_{\ell}^{2}}{m_{M}^{2}}\right)\right]
		\end{array}
	\end{align}
	\end{widetext}

	where $\ell^{+(-)}=\mu^{+(-)}$, $i(j)=s(\bar{b})$, $m_M=m_{B_s^0}=$5.36692 GeV is the $B_s^0$ meson mass, $f_M=f_{B_s^0}=0.2303$ is the $B_s^0$ meson decay constant, $\tau_M=\tau_{B_s^0}=2.311\times 10^{12}$ GeV is the lifetime of the $B_s^0$ meson, $G_F$ the Fermi constant and the SM contribution at one loop $C_A^{\rm SM}$ is given by
	\begin{equation}
		C_A^{\rm SM}=-V_{tb}^* V_{ts} Y\Bigg(\frac{m_t^2}{m_W^2}\Bigg)-V_{cb}^* V_{cs} Y\Bigg(\frac{m_c^2}{m_W^2}\Bigg),
	\end{equation}
	where the function $Y$ is defined as $Y=\eta_Y Y_0$ such that the NLO QCD effects are included in $\eta_Y=1.0113$ \cite{Buras:2012ru} and the loop Inami-Lim function \cite{Inami:1980fz} reads
	\begin{equation}
		Y_0(x)=\frac{x}{8}\Big[\frac{4-x}{1-x}+\frac{3x}{(1-x)^2}\ln(x)\Big].	
	\end{equation}
Finally, the form factors are given by 
	\begin{eqnarray}\label{FormFactors}
	C_{S}^{ij}&=&\frac{\pi^{2}}{2G_{F}^{2}m_{W}^{2}}\underset{\phi=h,H}{\sum}\frac{2g_{\phi\ell^+\ell^-}g_{\phi ij}}{M_{\phi}^{2}},\nonumber\\
	C_{P}^{ij}&=&\frac{\pi^{2}}{2G_{F}^{2}m_{W}^{2}}\frac{2g_{\phi\ell^+\ell^-}g_{\phi ij}}{M_{A}^{2}},\\
	C_{S}^{\prime ij}&=&C_{S}^{ij}\left(g_{\phi ij}\leftrightarrows g_{\phi ji}\right),\nonumber\\
	C_{P}^{\prime ij}&=&C_{S}^{ij}\left(g_{\phi ij}\leftrightarrows g_{\phi ji}\right)\nonumber.
\end{eqnarray}

	To obtain the corresponding $\mathcal{BR}(B_d^0\to \mu^+\mu^-)$ we carry out in Eq. \eqref{BRMlili} the replacements $m_{B_s^0}\to m_{B_d^0}=5.27966$ GeV, $f_{B_s^0}\to f_{B_d^0}=0.190$, $\tau_{B_s^0}\to\tau_{B_d^0}=2.312\times 10^{12}$ GeV; In Eqs. \eqref{FormFactors}  $g_{\phi \bar{b}s}\to g_{\phi \bar{b}d}$ and in Eqs. \eqref{WilsonOperators} $s\to d$.
	
	\subsection*{Lepton flavor violating processes}\label{LFVconstraints}
	\subsubsection{Muon anomalous magnetic moment $a_\mu$}
\begin{figure}[!htb]
		\centering
		\includegraphics[width=8.5cm]{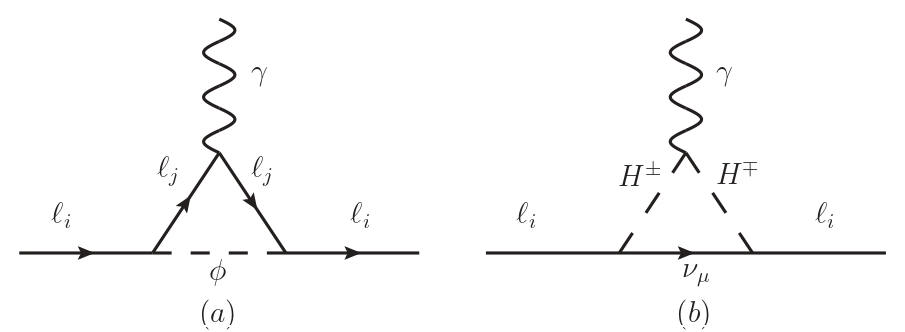}
         \includegraphics[width=4.5cm]{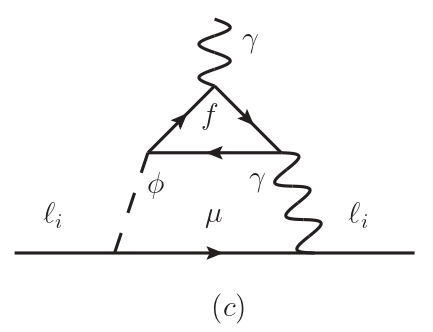}
		\caption{Feynman diagrams that contribute to $a_\mu$. Here $\phi$ represents a CP-even scalar, CP-odd scalar and the SM-like Higgs boson. $H^{\pm}$ stand for charged scalar bosons. Here $\ell_i=\mu$.}\label{FDmuonAMDM}
	\end{figure}

	Given (still) the current discrepancy between the experimental measurement and the SM theoretical prediction of $a_\mu$, we use it to constrain the free model parameters that have an impact in that observable. The Feynman diagrams that contribute to $a_\mu$ are shown in Fig. \ref{FDmuonAMDM}.
	The contributions at one-loop level reads
	\begin{equation}
		\delta a_\mu^\phi=\frac{G_F m_\mu^2}{4\pi^2 \sqrt{2}}\sum_{\ell_i} g_{\mu\ell_i}^2 R_{\phi}F_{\phi}(R_{\phi}),	
	\end{equation}
	with
	\begin{eqnarray}
		F_{h,\,H}&=&\int_{0}^{1}dx\frac{x^2(2-x)}{R_{h,\,H}x^2-x+1},\\
		F_A&=&\int_{0}^{1}dx\frac{-x^3}{R_Ax^2-x+1},\\
		F_H^{\pm}&=&\int_{0}^{1}dx\frac{-x^2(1-x)}{R_{H^\pm}x^2+(1-R_{H^\pm})x}.
	\end{eqnarray}
	where $R_{\phi}=m^2_\mu/M^2_{\phi}$ ($\phi=h,\,H,\,A,\,H^{\pm}$). While the dominant contribution at two-loops is induced by $A$ circulating inside the loop kind box, and is given by
	\begin{equation}
		\delta a_\mu^{2-loops}=\frac{\alpha^2}{8\pi^2 s_W^2}\frac{m_\mu^2 g_{A\mu\mu}}{m_W^2}\sum_{f}N_c^f Q_f^2 R_A \bar{F}_A g_{Af_i\bar{f}_j},
	\end{equation}
	where
	\begin{equation}
		\bar{F}_A=\int_{0}^{1}dy\frac{\log\Big(\frac{R_A}{y(1-y)}\Big)}{R_A-y(1-R_A)}
	\end{equation}
	\subsubsection{$\ell_i\to\ell_j\gamma$ decays}
\begin{figure}[!htb]
		\centering
		\includegraphics[width=6cm]{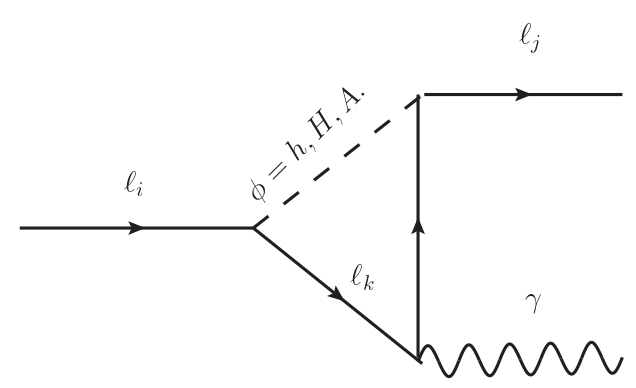}
		\caption{Feynman diagram of the process $\ell_i\to\ell_j\gamma$ at one-loop level induced by $\phi=h,\,H,\,A$.}\label{FDliljgamma}
	\end{figure}

	The effective Lagrangian for the $\ell_i\to\ell_j\gamma$ is given by
	\begin{equation}\label{EffLagrangian}
		\mathcal{L}_{\text{eff}}=C_L Q_{L\gamma} C_R Q_{R\gamma}+h.c.,
	\end{equation}
	where the dim-5 electromagnetic penguin operators read
	\begin{equation}\label{dim5Op}
		Q_{L\gamma,\,R\gamma}=\frac{e}{8\pi^2}(\bar{\ell}_j\sigma^{\alpha\beta}P_{L,\,R}\ell_i)F_{\alpha\beta},
	\end{equation}
	here $F_{\alpha\beta}$ is the electromagnetic field strength tensor. The Feynman diagram of the process $\ell_i\to\ell_j\gamma$ is depicted in Fig. \ref{FDliljgamma}.

	The Wilson coefficients $C_{L,\,R}$ receive contributions at one-loop level and an important contribution from the Barr-Zee two-loops level. For the particular case where $\ell_i=\tau$ and $\ell_j=\mu$, the approximations $g_{\phi\mu\mu}\ll g_{\phi\tau\tau}$ and $m_\mu\ll m_{\tau}\ll m_\phi$ are assumed. According to this assertion, the one-loop Wilson coefficients $C_{L,\,R}$ simplify as follows \cite{Harnik:2012pb, Blankenburg:2012ex}
	
    \begin{align}\label{WilsonCoe1loop}
    C_L^{1-loop} &\simeq \sum_{\phi}\frac{g_{\phi\tau\tau}g_{\phi\tau\mu}}{12m_\phi^2}\Bigg(-4+3\log\frac{m_\phi^2}{m_{\tau}^2}\Bigg),\notag \\
    C_R^{1-loop} &\simeq\sum_{\phi}\frac{g_{\phi\tau\tau}g_{\phi\tau\mu}}{12m_\phi^2}\Bigg(-4+3\log\frac{m_\phi^2}{m_{\tau}^2}\Bigg),
    \end{align}

	The numerical expressions for 2-loop contributions are given by
	
	\begin{align}\label{WilsonCoe2loop}
		C_L^{2-loops}&=\sum_{\phi=h,\,H,\,A}g_{\phi\tau\mu}^*(-0.082g_{\phi tt}+0.11)/(m_\phi\text{GeV})^2, \nonumber \\
		C_R^{2-loops}&=C_L^{2-loops}(g_{\phi\tau\mu}^*\to g_{\phi\tau\mu})
	\end{align}
	
	The rate for $\tau\to\mu\gamma$ is
	
	\begin{equation}\label{ratetaumugamma}
		\Gamma(\tau\to\mu\gamma)=\frac{\alpha m_{\tau}^2}{64\pi^4}(|C_L|^2+|C_R|^2).
	\end{equation}
	
	To obtain the corresponding width decay of the processes $\mu\to e\gamma$ and $\tau\to e\gamma$, the replacements $\tau\to \mu,\, \mu\to e$ for the former decay and $\mu\to e$ for the second process from \eqref{EffLagrangian} to \eqref{ratetaumugamma} are required.

	\subsubsection{$\ell_i\to\ell_j\ell_k\bar{\ell}_k$ decays}
\begin{figure}[!htb]
		\centering
		\includegraphics[width=5cm]{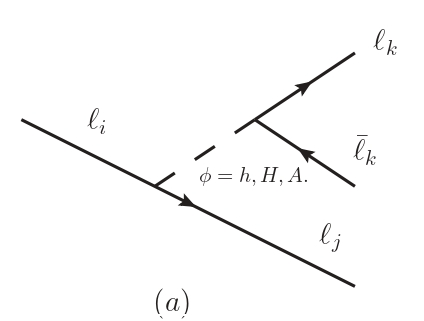}
		\includegraphics[width=5cm]{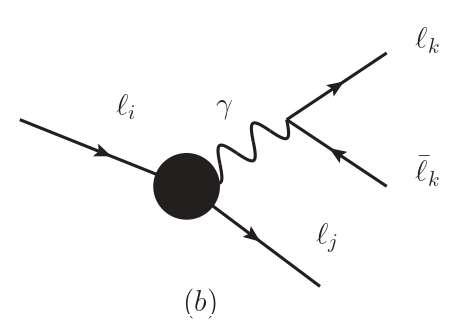}
		\caption{Feynman diagrams that contribute to the $\ell_i\to \ell_j\ell_k\bar{\ell}_k$, (a) tree-level and (b) one-loop level, where the black circle represents the loop as Feynman diagram of Fig. \ref{FDliljgamma}}\label{FDliljlklk}
	\end{figure}

	Within the 2HDM-III framework, these kinds of decays can occur at tree-level via the exchange of $h,\,H,\,A$, as shown in Fig. \ref{FDliljlklk}. Nevertheless, the process is suppressed by the flavor violating Yukawa couplings $Y_{\ell_i\ell_j}$ and by the flavor-conserving coupling $Y_{\ell_k\ell_k}$. However, there are higher order contributions: at one-loop and two-loops level. The corresponding flavor violating partial width decay reads
	\begin{equation}
		\Gamma(\tau\to3\mu)\sim \frac{\alpha m_{\tau}^5}{6(2\pi)^5}\Big|\log\frac{m_{\mu}^2}{m_{\tau}^2}-\frac{11}{4} \Big|(|C_L|^2+|C_R|^2),
	\end{equation}  
where the suppressed terms by the muon mass are neglected. The Wilson coefficients $C_{L,\,R}$ are given in Eqs. \eqref{WilsonCoe1loop}-\eqref{WilsonCoe2loop}. 
	
	\subsubsection{$h\to \ell_i \ell_j$ decays}\label{hlilj}
	
\begin{figure}[!htb]
		\centering
		\includegraphics[width=5cm]{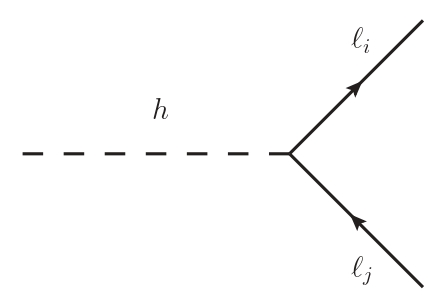}
		\caption{Feynman diagram of the process $h\to\ell_i\ell_j$ at tree level.}\label{FDhlilj}
	\end{figure}

	The LFV processes $h\to \ell_i\ell_j$ ($\ell_{i,\,j}=\ell_{i,\,j}^-\ell_{i,\,j}^+$) where $\ell_i\ell_j=e\mu,\,e\tau,\,\tau\mu$ can arise at tree level in many models that extend to the SM, as shown in Fig. \ref{FDhlilj}.
	
	The relevant interactions can be extracted from the Yukawa Lagrangian
	\begin{equation}
		\mathcal{L}_Y\supset -Y_{ij}\bar{\ell}_L^i \ell_R^jh+h.c.
	\end{equation}	 
	The corresponding full decay width of the $h\to \bar{f}_i f_j$ decays is given by:
	\begin{align}
		\Gamma(h\to \bar{f}_i f_j)=\frac{N_c g^2_{h \bar{f}_i f_j }m_h}{128\pi}\Bigg[ 4-(\sqrt{\tau_{f_i}} \notag \\
        +\sqrt{\tau_{f_j}})^2  \Bigg]^{3/2}\sqrt{4-(\sqrt{\tau_{f_i}}-\sqrt{\tau_{f_j}})^2},
	\end{align}
	where $g_{h \bar{f}_i f_j}$ is the $h\bar{f}_i f_j$ coupling coming from an extension of the SM, $Nc=3\,(1)$ is the color number for quarks (leptons), $m_h$ is the Higgs boson mass and $\tau_i=4m_i^2/m_h^2$. 
	
\begin{figure}[!htb]
		\centering
		\includegraphics[width=8.5cm]{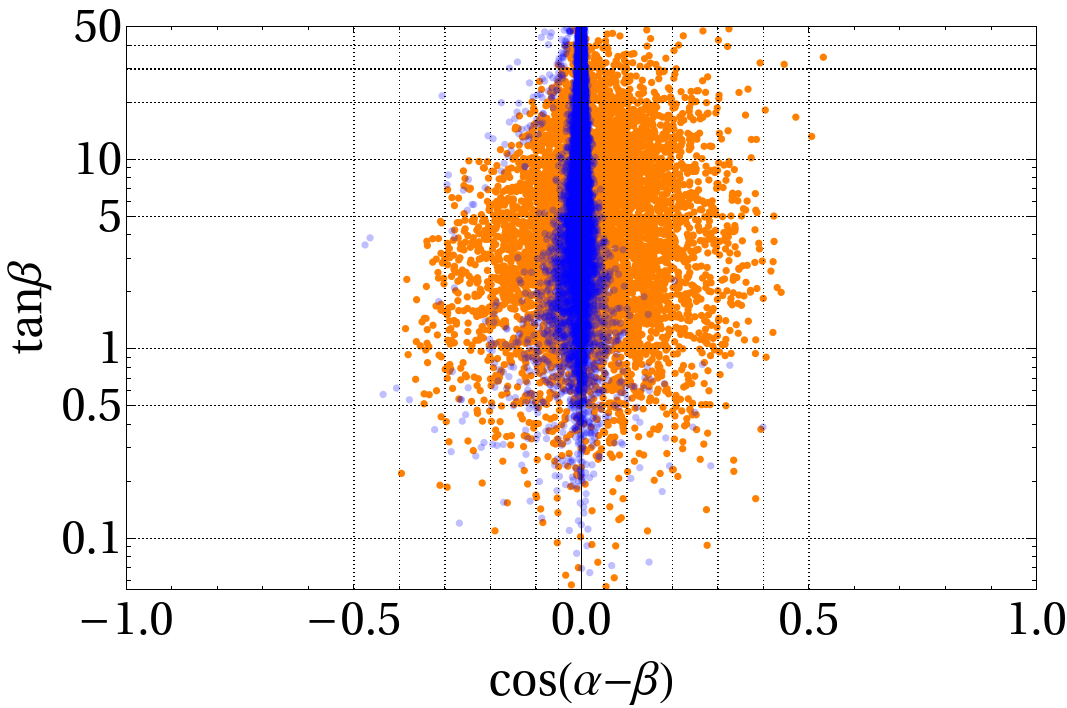}
		\caption{Blue (Orange) points stand for these allowed values by all the signal strength modifiers $\mathcal{\mu}_{X}$ (LFV processes). All the allowed values are presented in the $\cos(\alpha-\beta)-\tan\beta$ plane.}\label{HixBosonData}
	\end{figure}

	Once all the analytical expressions for calculating the BRs of the exposed observables have been given, we evaluate all the processes of Sec. \ref{HiggsConstraints} with the \texttt{Mathematica} package so-called \texttt{SpaceMath}\footnote{This software has implemented all the experimental constraints considered in our project.} \cite{Arroyo-Urena:2020qup} in order to find the values of the free parameters involved in our analysis. 
	In Fig. \ref{HixBosonData}, we present the plane $\cos(\alpha-\beta)-\tan\beta$, whose points represent the values allowed by experimental constraints. The orange points represent the experimental measurements (or upper limits) on branching ratios of the $B_{s,d}\to\mu^+\mu^-$ and the LFV processes, as described above. Meanwhile, the blue points stand for those that satisfy the LHC Higgs boson data. If one considers the region allowed by all the observables, it becomes possible to explore different scenarios to predict several observables. For illustration, two realistic scenarios are given by:
	\begin{itemize}
		\item $-0.04\lesssim \cos(\alpha-\beta)\lesssim 0.025$ for $1\lesssim \tan\beta \lesssim 10$,
		\item  $-0.01\lesssim \cos(\alpha-\beta)\lesssim 0.01$ for $1\lesssim \tan\beta \lesssim 50$.
	\end{itemize}
	
These bounds include to the decoupling limit $\alpha-\beta\to \pi/2$, it suppress Flavor-Violating SM-like Higgs boson decays, but at the same time $\cos(\alpha-\beta)\sim 0$ enhance those kind of processes for the heavy Higgs boson $H$ decays.
	We include in Appx. \ref{IndividualConstraints} individual allowed values of the model parameters associated with each observable.
	\subsection{Constraint on $M_H$, $M_A$  and $M_{H^\pm}$}\label{subsubsection:constraintoncharguedscalar}
\subsubsection{Collider constraints} 
\begin{figure}[!htb]
		\center{\includegraphics[scale=0.6]{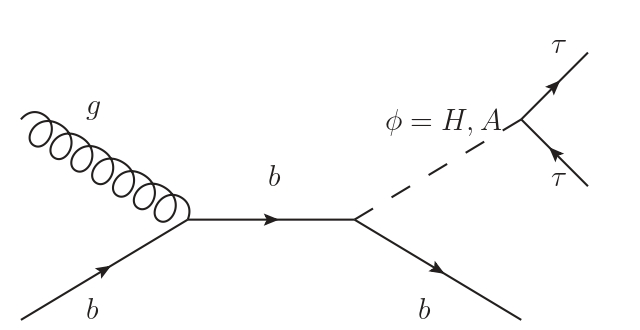}}
		\caption{Feynman diagram of the production of $\phi$ in association with a bottom quark at LHC, with a subsequent decay into a $\tau\tau$ pair.  \label{FeynmanDiagram2}}
	\end{figure}

Several searches for additional neutral Higgs bosons in different channels have been done by the ATLAS and CMS collaborations, one of them is through the di-tau chennel $gb\to\phi\to\tau\tau$, with $\phi=A,\,H$. Figure \ref{FeynmanDiagram2} shows the Feynman diagram of the process. However, no evidence of any additional Higgs boson was observed, but upper limits on the production cross-section $\sigma(gb\to\phi)$ times the branching ratio $\mathcal{BR}(\phi\to\tau\tau)$ were imposed.
	
\begin{figure}[!htb]\label{constMasses}
		\centering
		\subfigure[ ]{\includegraphics[scale=0.295]{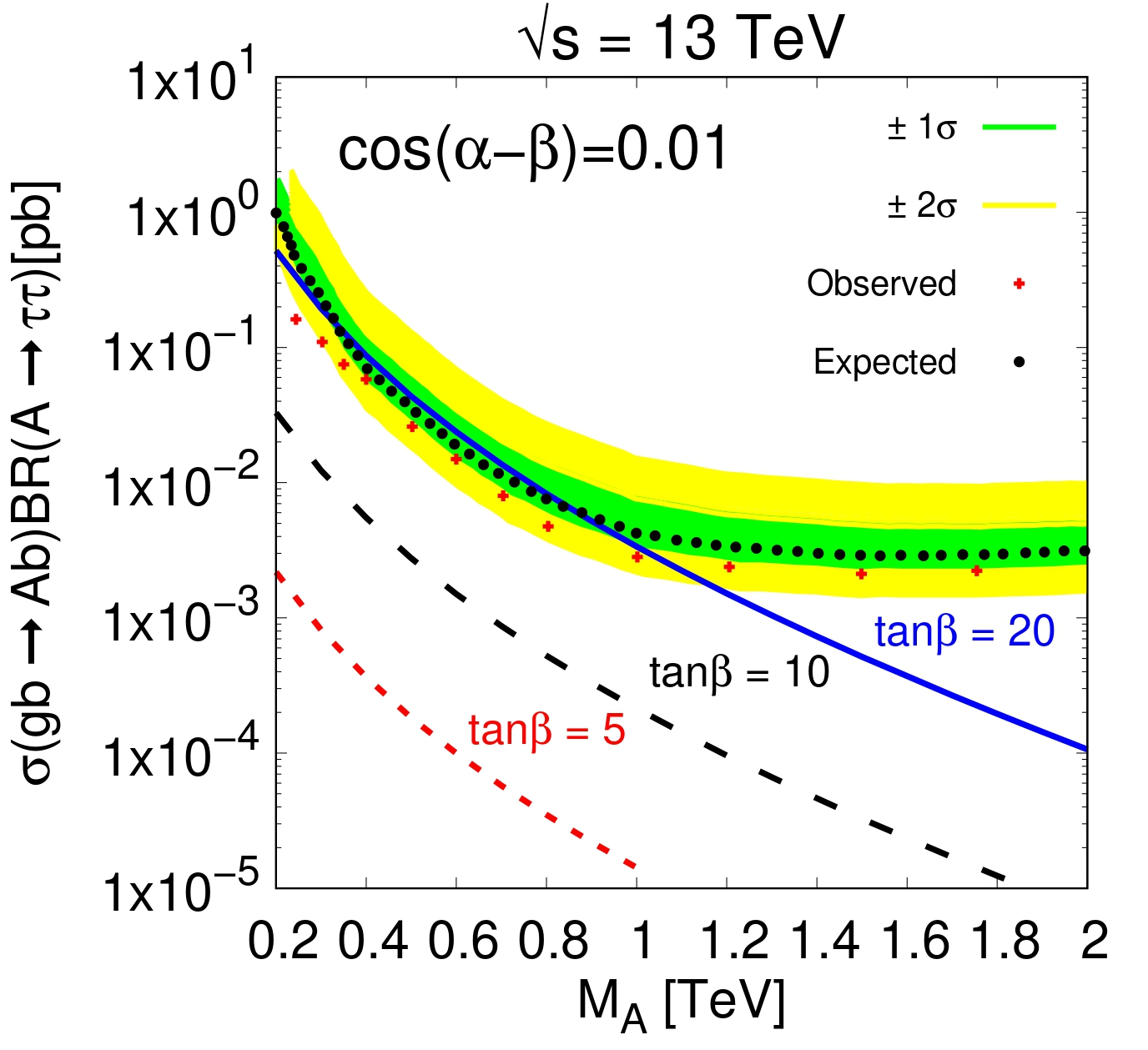}\label{XS_A_timesBRHtautau}}
		\subfigure[ ]{\includegraphics[scale=0.295]{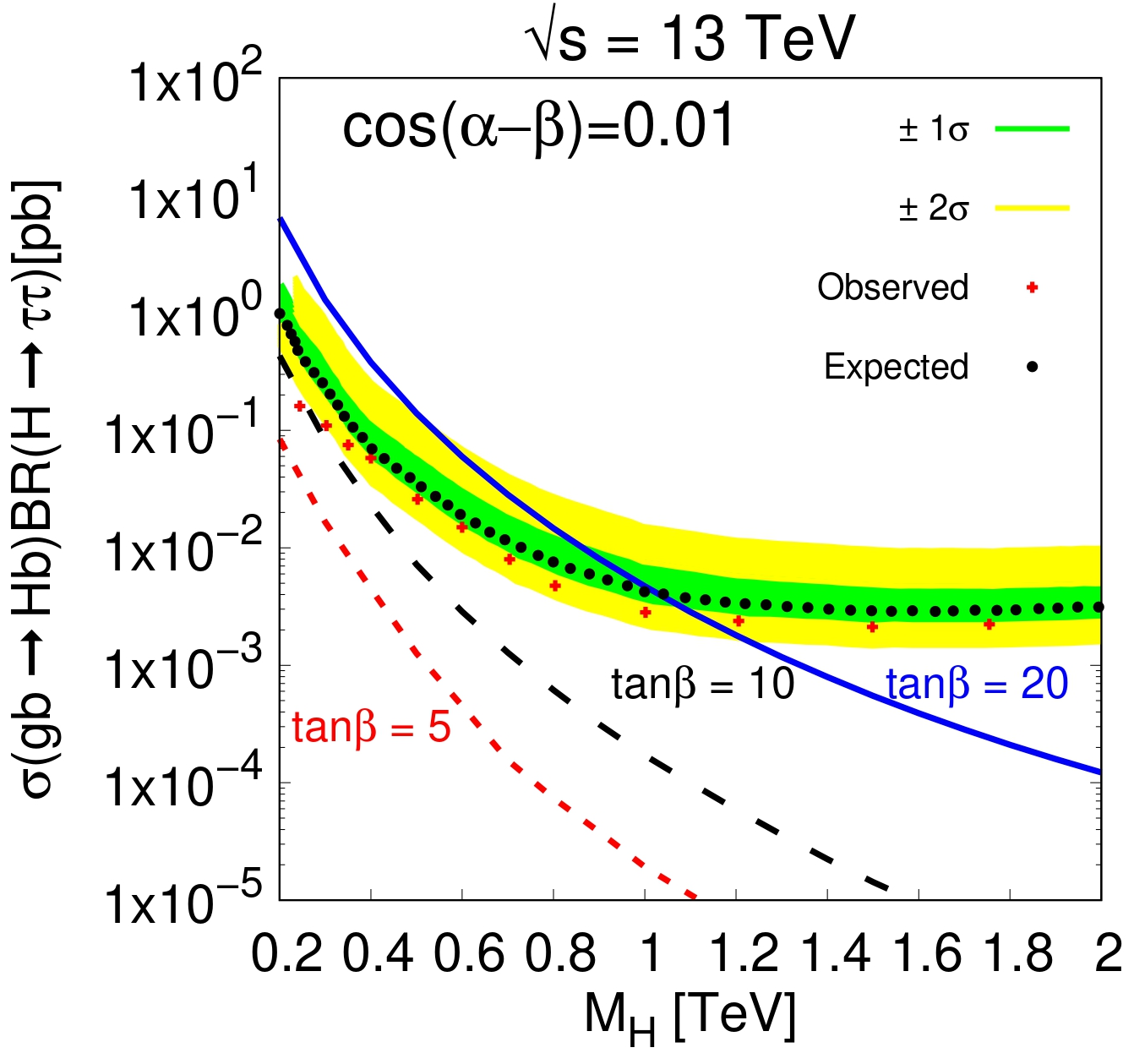}\label{XS_H_timesBRHtautau}}
		\caption{The observed and expected at 95$\%$ CL upper limits on the production cross-section times di-tau branching ratio for a scalar boson produced via $b$-associated production as a function of (a) $M_A$ (b) $M_H$. In both cases we consider $t_{\beta}=$ 5, 10, 20 and $c_{\alpha\beta}=0.01$.}
	\end{figure}

	Figure \ref{XS_A_timesBRHtautau} presents the $\sigma(gb\to Ab)\times\mathcal{BR}(A\to\tau\tau)$ as a function of $M_A$ for illustrative values of $\tan{\beta}=5,\,10,\,20$ and $\cos(\alpha-\beta)=0.01$, while Fig. \ref{XS_H_timesBRHtautau} shows the same plane but for $\phi=H$. In both plots, the black points and the red crosses represent the expected and observed values at 95$\%$ CL upper limits, respectively; while the green (yellow) band indicates the interval at $\pm 1 \sigma$ ($\pm 2 \sigma$) with respect to the expected value.  
	
\begin{figure}[!htb]
		\center{\includegraphics[scale=0.2]{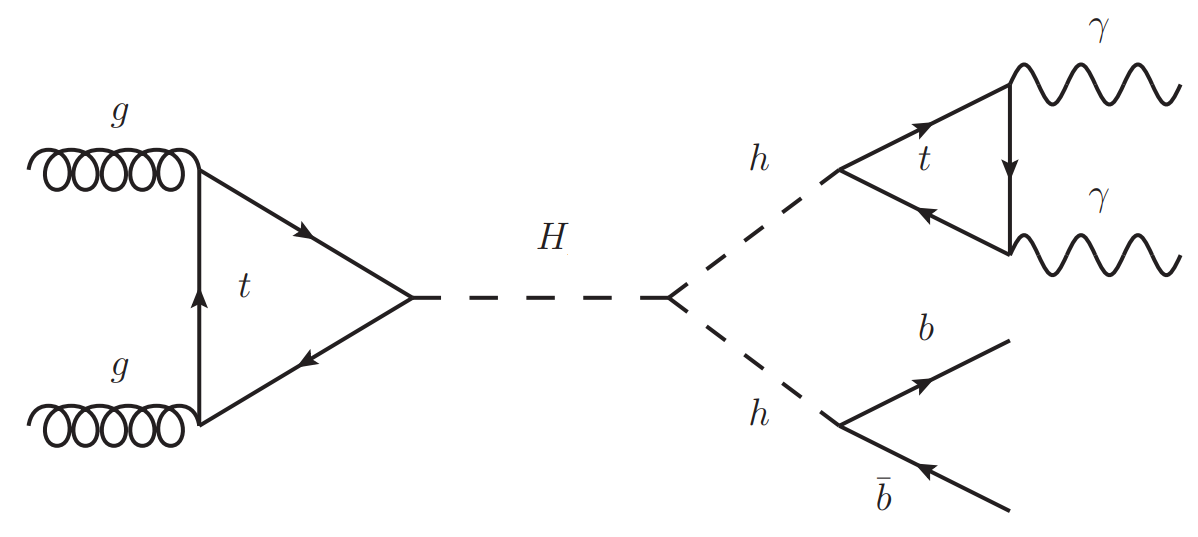}}
		\caption{Feynman diagram of the production of $H$ with its subsequent decay into $hh$.\label{FDHhh}}
	\end{figure}

\begin{figure}[!htb]
		\centering
		{\includegraphics[scale=0.23]{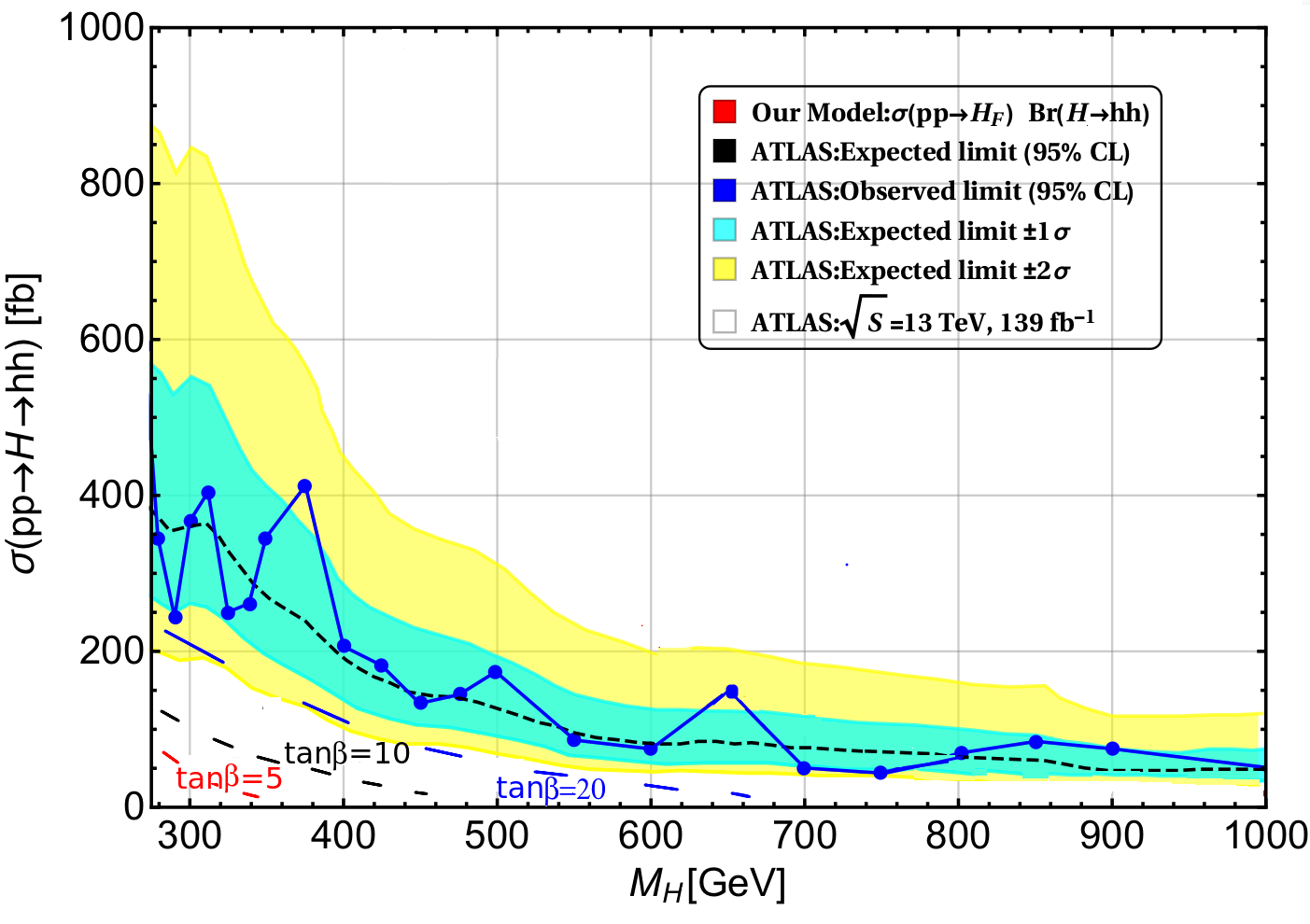}}
		\caption{The observed and expected at 95$\%$ CL upper limits on the production cross-section times di-Higgs branching ratio for a scalar boson produced via $pp$ collisions as a function of $M_H$. We consider $\tan{\beta}=$ 5, 10, 20 and $\cos(\alpha-\beta)=0.01$.}\label{Hhh}
	\end{figure}

	From Fig. \ref{XS_A_timesBRHtautau} we note that $M_A\lesssim 1$ TeV ($M_A\lesssim 1.2$ TeV) are excluded at 1$\sigma$ (2$\sigma$) for $\tan{\beta}= 20$, while for $\tan{\beta}\lesssim 5,\,10$ the upper limit on $\sigma(gb\to \phi b)\times\mathcal{BR}(\phi\to\tau\tau)$ is easily accomplished. Meanwhile, from Fig. \ref{XS_H_timesBRHtautau}, we find $M_H\lesssim 1.1$ TeV ($M_H\lesssim$ 1.3 GeV) are excluded at 1$\sigma$ (2$\sigma$) for $\tan{\beta}= 20$. For $\tan\beta\lesssim 15$, a wide range of masses are allowed $(300<M_A)$.  Another process used to constrain the mass of the heavy scalar $M_H$ is when it decays into a Higgs boson pair with their subsequent decays to $b\bar{b}$ and $\gamma\gamma$, i.e., $pp\to H\to hh, h\to b\bar{b}, h\to \gamma\gamma$, as shown in Fig. \ref{FDHhh}.
		 Then, we present in Fig. \ref{Hhh} the cross-section $\sigma(pp\to H\to hh, h\to b\bar{b}, h\to \gamma\gamma)$ as a function of $M_H$ including the upper limits at 1$\sigma$ (green band) and at 2$\sigma$ (yellow band), the expected limit (segmented black line) and the observed limit (blue line). These results were reported by the ATLAS collaboration \cite{ATLAS:2021ifb}. On the same plot, we also add the prediction of the 2HDM-III for $\tan\beta=5,\,10,\,20$ and $\cos(\alpha-\beta)=0.01$. We note that that the aforementioned limits on $M_H$ are easily met within the theoretical framework of the 2HDM-III for $\tan\beta\lesssim 18$. We also analyze the $H\rightarrow VV, (V=WZ)$ channel \cite{ATLAS:2020tlo,CMS:2019bnu}, however because of the coupling $g_{HVV}$ is proportional to $\cos(\alpha-\beta)$ these constraints are not enough restricted in the present model. Finally, we also analyzed the $t\bar{t}$ channel and compared it with experimental results for a model independent analysis~\cite{ATLAS:2024jja}. However, we found that the cross-section predicted by the 2HDM-III is highly supressed (4 to 8 orders of magnitude smaller) because it is inversely proportional to $\tan^4 \beta$.
		 
\subsubsection{$b\to s \gamma$ decay}	
\begin{figure}[!htb]
		\centering
		\subfigure[]{{\includegraphics[scale=0.25]{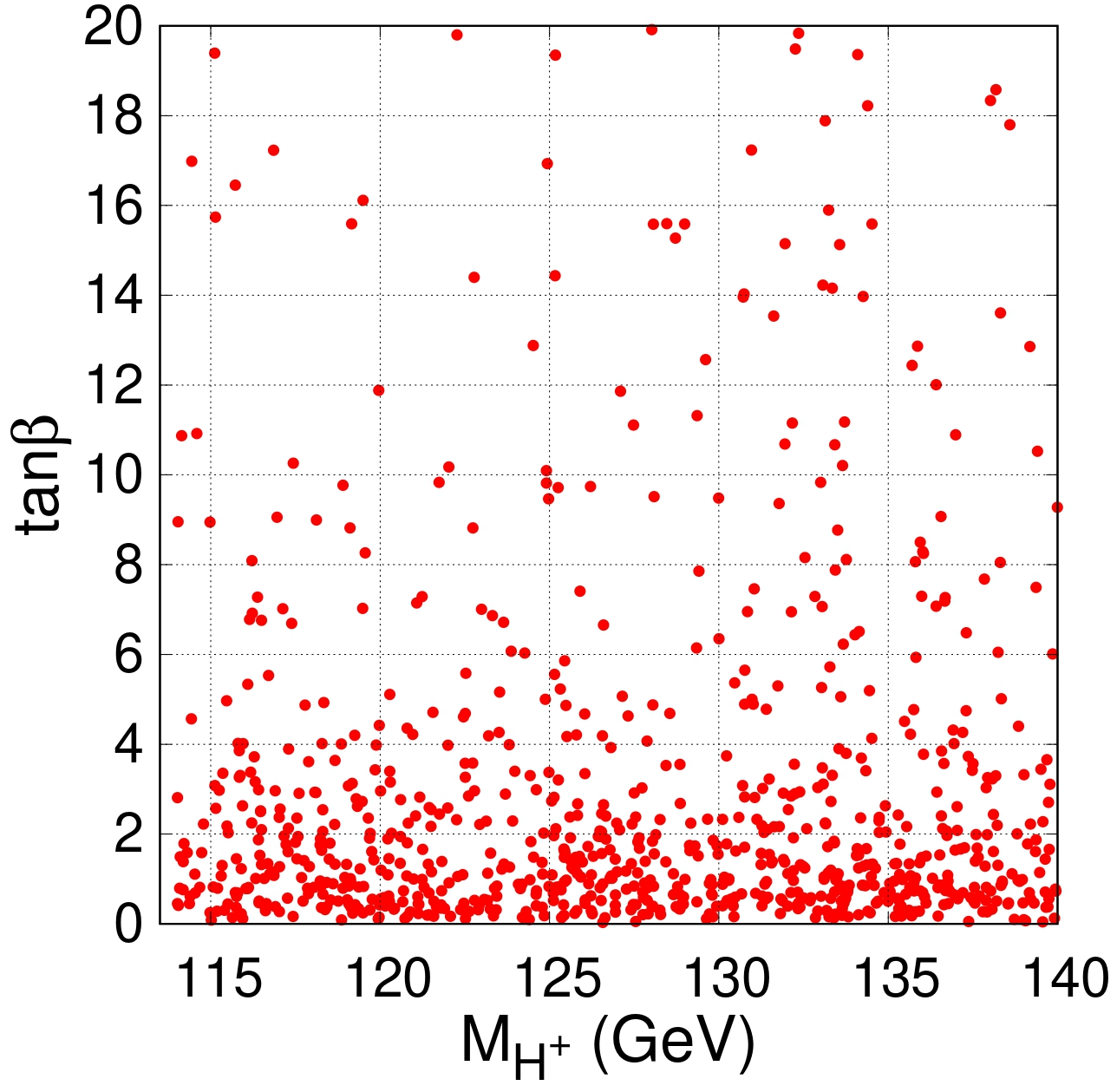}}}
		\subfigure[]{{\includegraphics[scale=0.25]{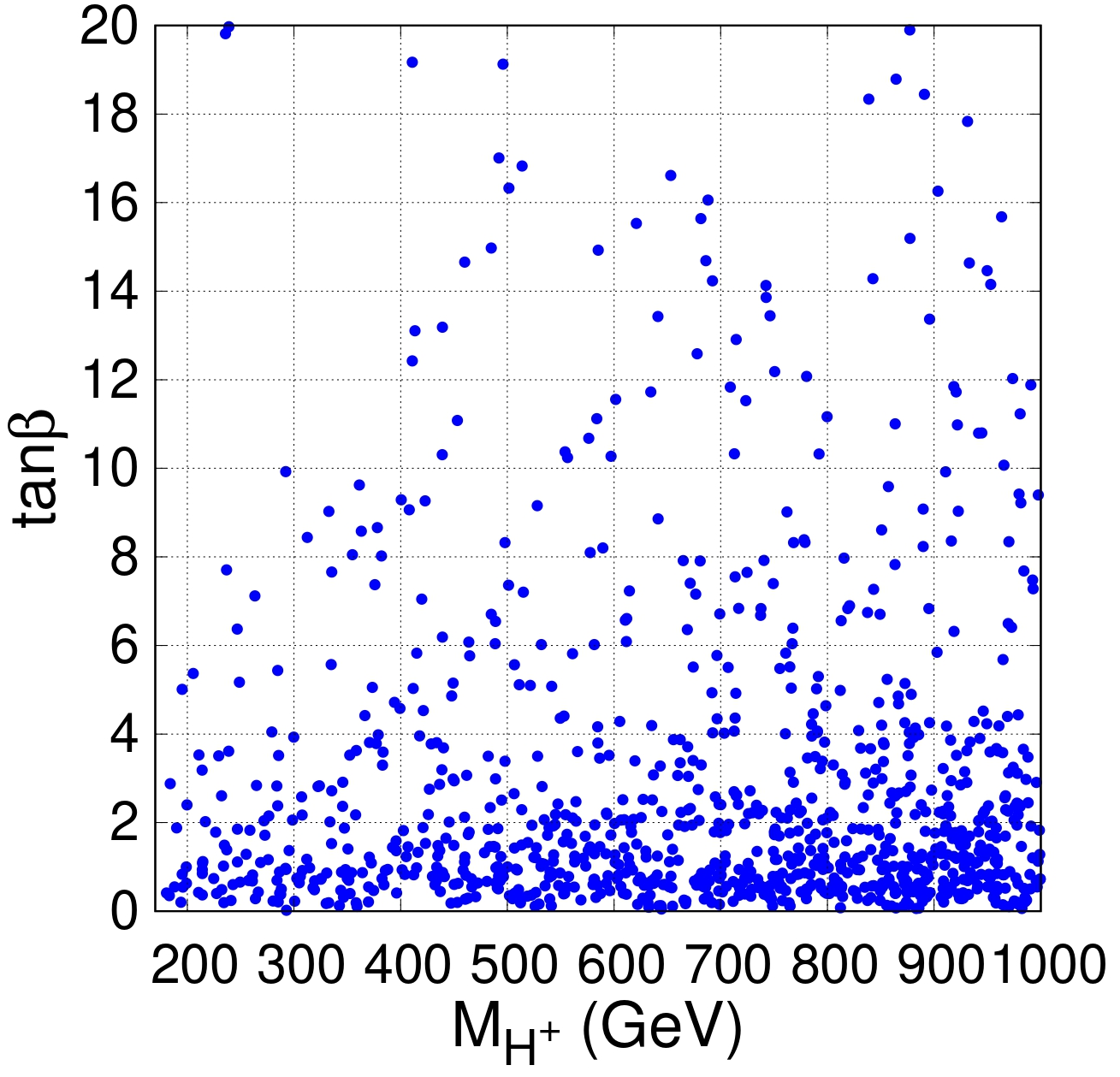}}}
		\caption{Scatter plot in the $M_{H^{\pm}}-\tan\beta$ plane. (a) Red points stand for these allowed by $R_{\rm quark}$ for the $114\leq M_{H^{\pm}}\leq 140$ GeV interval. (b) The same as in (a) but for the range of masses $m_t+m_b\leq M_{H^{\pm}}\leq 1000$. $R_{\rm quark}$ is defined in Eq. \eqref{eq:rquark}.}\label{mCH}
	\end{figure}

	Concerning to the charged scalar boson $H^{\pm}$, its detection would represent a clear signature of new physics. Constraints on its mass $M_{H^{\pm}}$ were obtained from collider searches for the $H^{\pm}$ production and its subsequent decay into a $\tau^{\pm}\nu_\tau$ pair~\cite{ATLAS:2018gfm}. However, we find that such processes are not a good way to constrain the charged scalar mass $M_{H^{\pm}}$ predicted in 2HDM-III. Nevertheless, the situation is opposite if one considers the decay $b\to s\gamma$ which imposes severe lower limits on $M_{H^\pm}$ because of the charged boson contribution~\cite{Ciuchini:1997xe, Misiak:2017bgg}.
	Figure~\ref{mCH}(a)-(b) shows a scatter plot in the $M_{H^\pm}$-$\tan\beta$ plane whose red and blue points stand for these allowed by the ratio $2.77\times 10^{-3}<R_{\rm quark}<3.67\times 10^{-3}$ ~\cite{Misiak:2017bgg}, defined as follows:
	\begin{equation}
		R_{\rm quark}=\frac{\Gamma(b\to X_s\gamma)}{\Gamma(b\to X_ce\nu_e)}.
		\label{eq:rquark}
	\end{equation} 
	
\begin{figure}[!htb]
		\centering
		\includegraphics[scale=0.15]{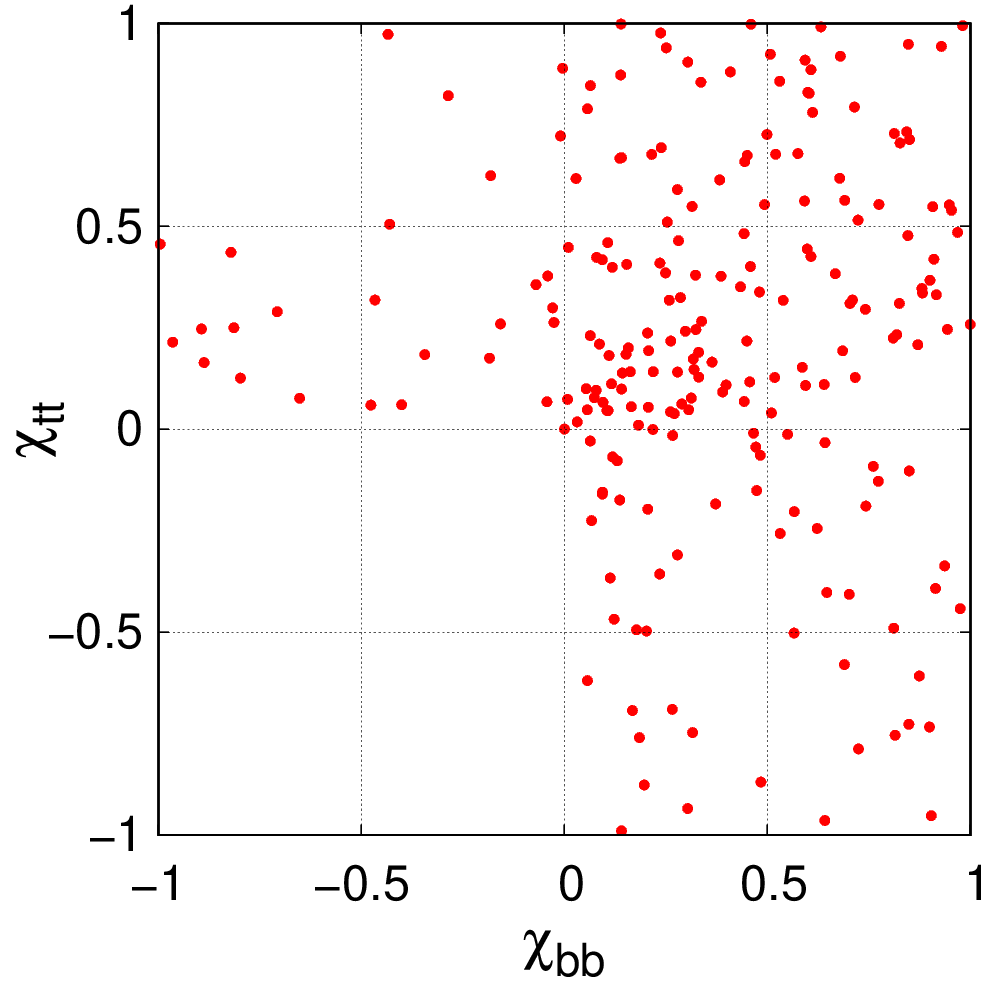}
		\caption{Scatter plot in the $\chi_{bb}-\chi_{tt}$ plane. The red points stand for these allowed by $R_{\rm quark}$ defined in Eq. \eqref{eq:rquark}.}\label{chitt-chibb}
	\end{figure}

	We note a preference for low $\tan\beta$ in both mass ranges, namely $114\leq M_{H^\pm}\leq 140$ GeV and $m_t+m_b\leq M_{H^\pm}\leq 1000$ GeV. This is due to the form of the contributions coming from 2HDM-III: $g_{\bar{u}dH^{-}}\sim 1/\tan\beta(1-\chi_{tt}/\sqrt{2})$ and $g_{\bar{d}uH^{+}}\sim \tan\beta(1-\chi_{bb}/\sqrt{2})$ of which we can observe that a large $\tan\beta$ would give dangerous contributions to $b\to s\gamma$.  The former scanned mass range is motivated by the current excess of events reported by the ATLAS collaboration \cite{ATLAS:2023bzb}. It is important to note that the parameters $\chi_{tt}$ and $\chi_{bb}$ have an attenuating effect on the size of the $g_{\bar{d}uH^{+}}$ and $g_{\bar{u}dH^{-}}$  couplings, allowing light masses for the charged scalar predicted in 2HDM-III, unlike in 2HDM-I,-II ($g_{\bar{u}dH^{-}}^{\rm I}=g_{\bar{d}uH^{+}}^{\rm I}= 1/\tan\beta$, $g_{\bar{u}dH^{-}}^{\rm II}= 1/\tan\beta$ and $g_{\bar{d}uH^{+}}^{\rm II}= \tan\beta$). 
		We show in Fig. \ref{chitt-chibb} the plane $\chi_{bb}-\chi_{tt}$, the red points correspond to those that meet the experimental results on $R_{\rm quark}$ in Eq. \eqref{eq:rquark}. 		
		Meanwhile, we present in Tab. \ref{scan2} the parameters scanned.
	\begin{table}
		
		\begin{centering}
			\caption{Scanned range of the parameters. We set $\chi_{ij}=1$, where $i,\,j$ stand for fermions not included in the Table (in general $i\neq j$). }\label{scan2}
			\begin{tabular}{|c|c|}
				\hline 
				Parameter & Scanned range\tabularnewline
				\hline 
				\hline 
				$\tan\beta$ & $[0.1,20]$\tabularnewline
				\hline 
				$\chi_{tt}$ & $[-1,1]$\tabularnewline
				\hline 
				$\chi_{bb}$ & $[-1,1]$\tabularnewline
				\hline 
				$M_{H^{\pm}}$ & $[114,140]$ GeV\tabularnewline
				\hline 
			\end{tabular}
			\par\end{centering}
	\end{table}
	
\begin{figure}[!h]
		\centering
		\subfigure[]{{\includegraphics[scale=0.25]{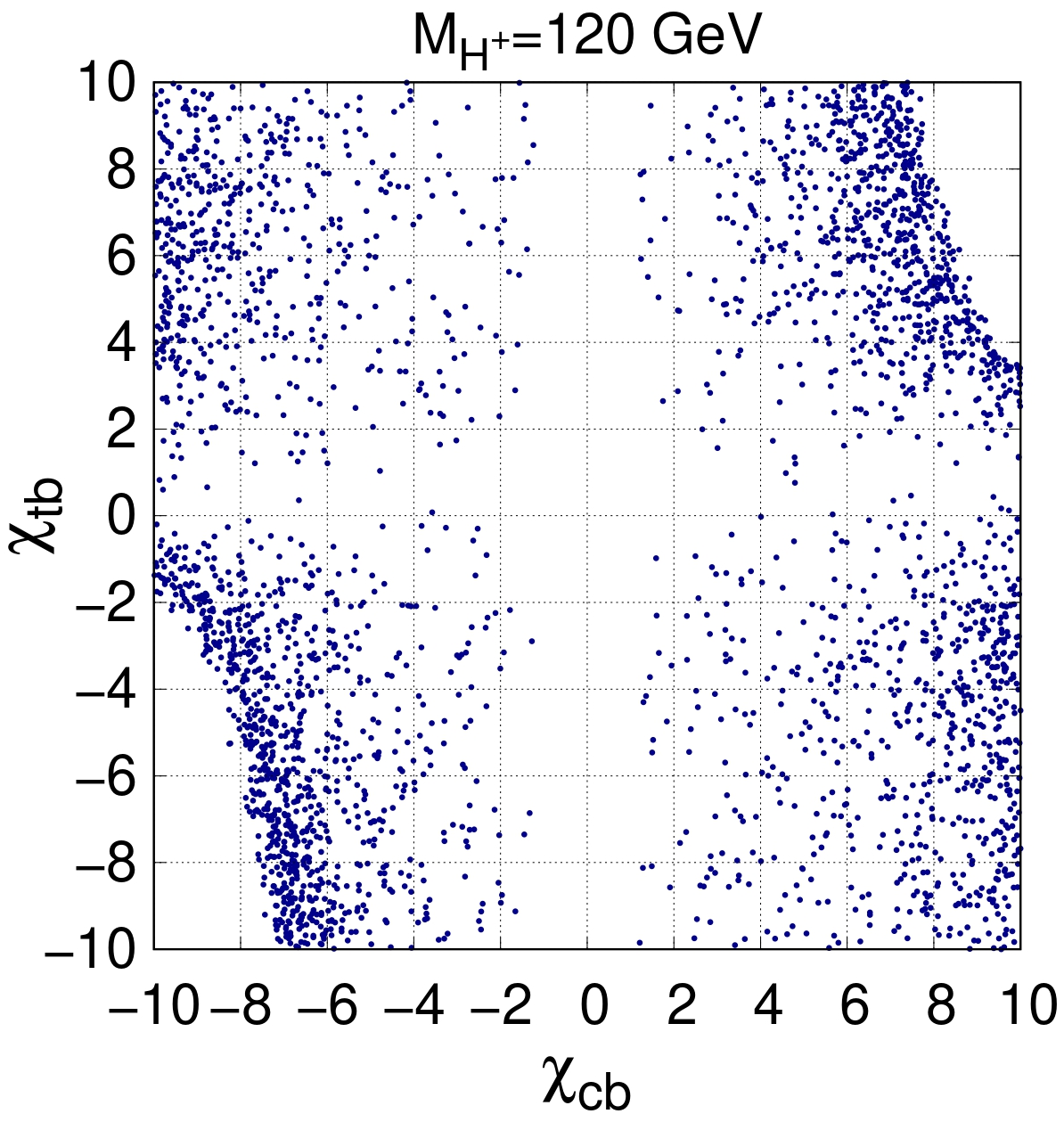}}}
		\subfigure[]{{\includegraphics[scale=0.25]{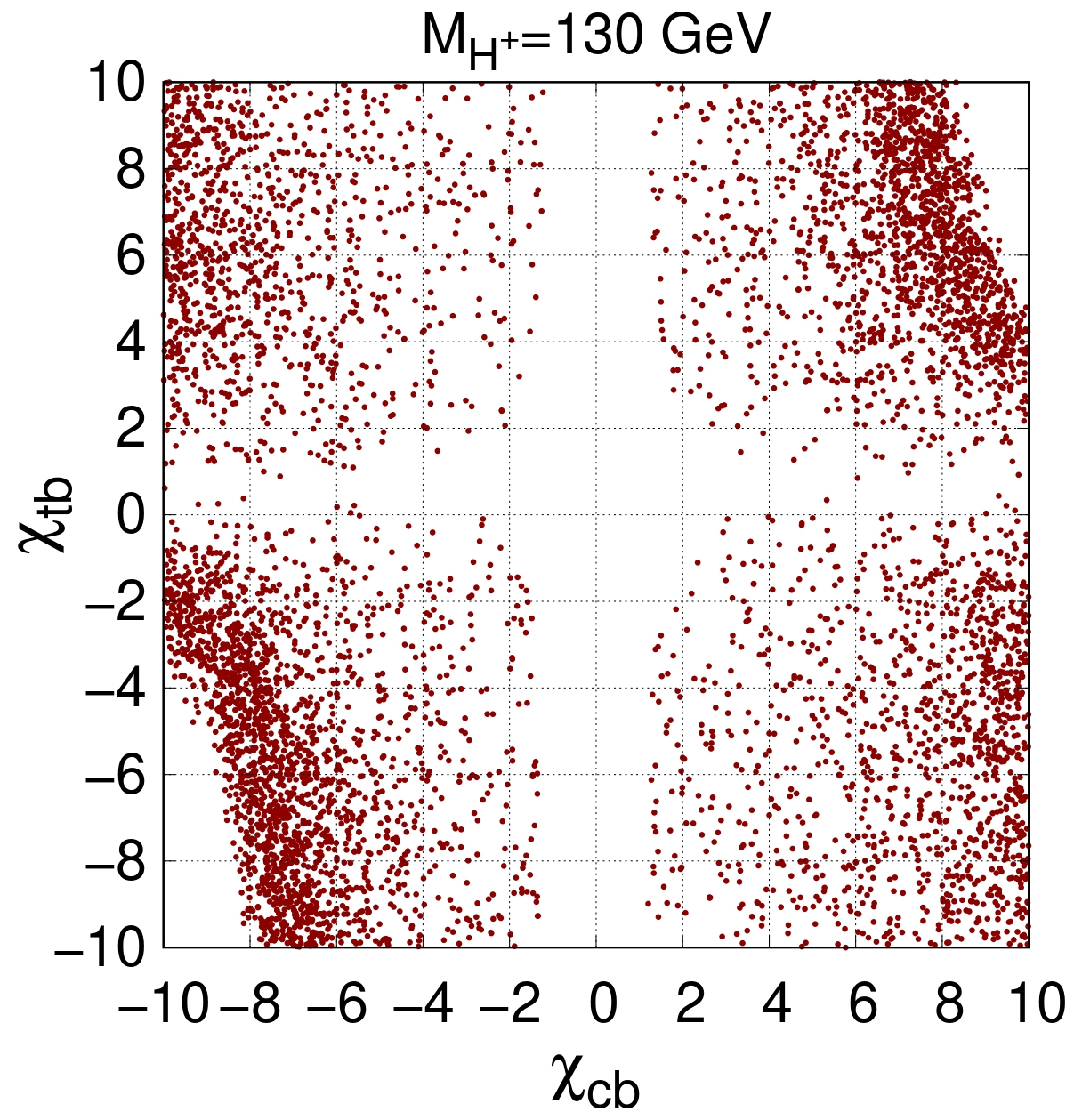}}}
		\subfigure[]{{\includegraphics[scale=0.25]{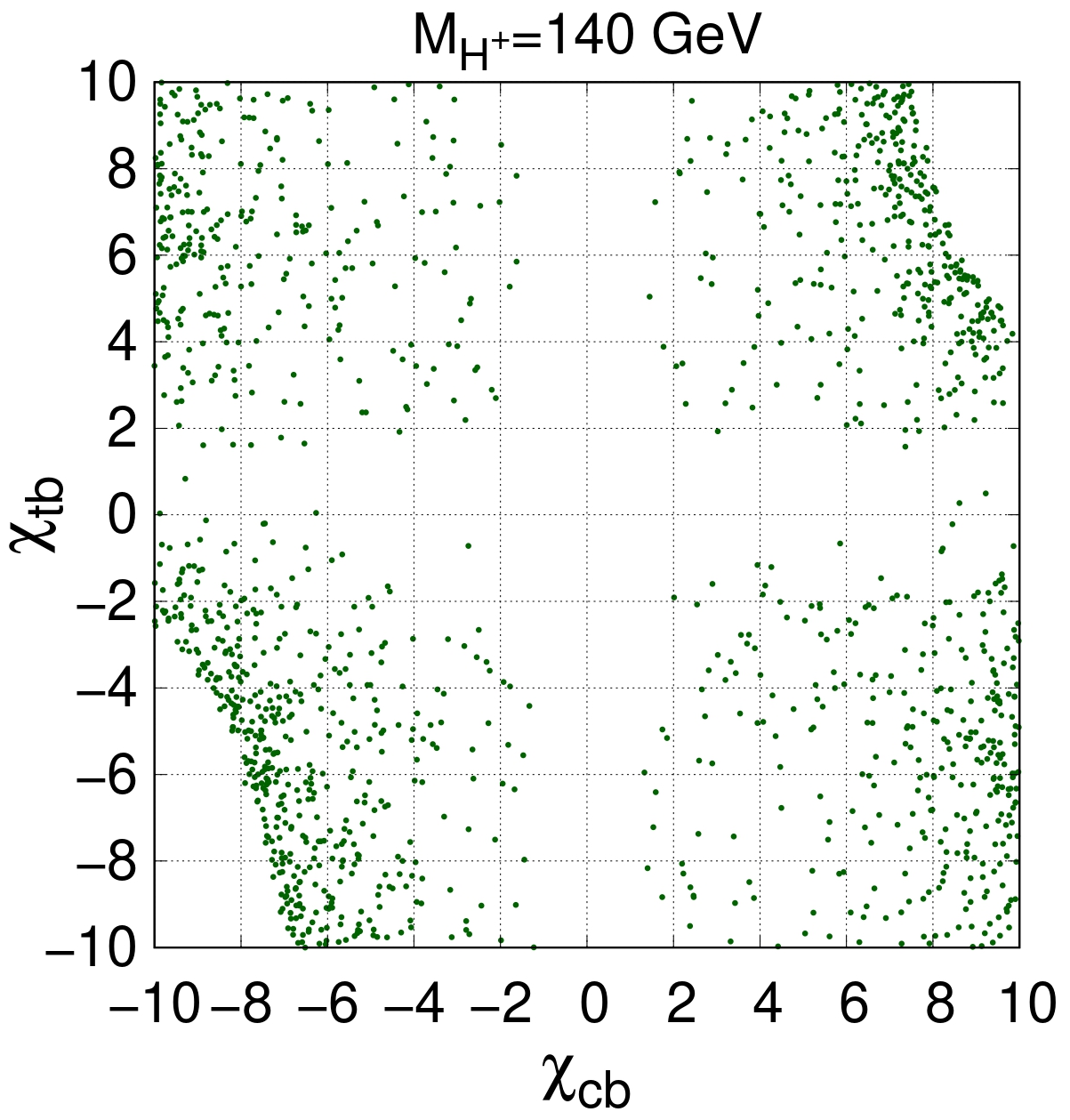}}}
		\caption{$\chi_{cb}-\chi_{tb}$ plane. The colored points indicate the values for $\chi$'s that accommodate the current excess of events reported by ATLAS collaboration \cite{ATLAS:2023bzb}. (a) Blue points represent these values that explain such an excess of events for $M_{H^{\pm}}=120$ GeV. (b) and (c) the same as in (a) but for $M_{H^{\pm}}=130$ GeV and $M_{H^{\pm}}=140$ GeV, respectively. }\label{excessMHp}
	\end{figure}

	Since our analysis focuses on the case of light charged scalar masses
	($M_{H^{\pm}}<m_t+m_b$), it is worth mentioning that the model under consideration may explain the recent slight excess in data with a significance of around $3\sigma$ for $M_{H^{\pm}}=130$ GeV reported by the ATLAS collaboration \cite{ATLAS:2023bzb}. This search used a data set of collisions $pp$ collected at a center-of-mass energy $\sqrt{s}=13$ TeV amounting to an integrated luminosity of 139 fb$^{-1}$. The analysis focuses on a data sample enriched in top-quark pair production, where one top quark decays into a leptonically decaying $W$ boson and a bottom quark, and the other top quark may decay into a $H^{\pm}$ boson and a bottom quark. The model-independent exclusion at 95\% confidence level on the product of branching ratios $\mathcal{BR}=\mathcal{BR}(t\to H^{\pm}b)\times \mathcal{BR}(H^{\pm}\to cb)$ was reported as a function of $M_{H^{\pm}}$.
	In Fig. \ref{excessMHp} we present the plane $\chi_{cb}-\chi_{tb}$ for $0.1<\tan\beta<20$, $\chi_{ij}=1$, $-10<\chi_{\mu\mu}<10$. Blue (Red, Green) points represent values that can accommodate the current excess for $M_{H^{\pm}}=120\,(130,\,140)$ GeV.
	
	We mainly note two characteristics, 
	\begin{enumerate}
		\item The density of red points is greater than other points (blue and green) because this red region represent values associated with the excess for $M_{H^{\pm}}=130$ GeV, which is the largest excess reported by ATLAS collaboration, opposite to the $M_{H^{\pm}}=140$ GeV case, as shown in Fig. 8 of Ref. \cite{ATLAS:2023bzb}.  
		\item  According to our scan over the model parameter space, as shown in Table \ref{scan1}, the impact of $\chi_{tb}$ and $\chi_{cb}$ to accommodate the current excess is transcendental because of the high sensitivity on these parameters, contrary to the 2HDM-I, II, Lepton Specific and Flipped, which lack these parameters. It is important to note that there are several parameters $\chi_{ij}$ associated with fermions $i$ and $j$ which are functions of the total width decay of the charged scalar boson $H^{\pm}$. In our analysis, we have set $\chi_{ij}=1$.    
	\end{enumerate}
	\begin{table}
		
		\begin{centering}
			\caption{Scanned range of the parameters. We set $\chi_{ij}=1$, where $i,\,j$ stand for fermions not included in the Table (in general $i\neq j$). }\label{scan1}
			\begin{tabular}{|c|c|}
				\hline 
				Parameter & Scanned range\tabularnewline
				\hline 
				\hline 
				$\tan\beta$ & $[0.1,20]$\tabularnewline
				\hline 
				$\chi_{tb}$ & $[-10,10]$\tabularnewline
				\hline 
				$\chi_{cb}$ & $[-10,10]$\tabularnewline
				\hline 
				$\chi_{\mu\mu}$ & $[-10,10]$\tabularnewline
				\hline 
				$M_{H^{\pm}}$ & $[114,140]$ GeV\tabularnewline
				\hline 
			\end{tabular}
			\par\end{centering}
	\end{table}

\subsubsection{Oblique parameters}
We require that the Peskin-Takeunchi parameters $S$, $T$ and $U$ are within the experimental results \cite{ParticleDataGroup:2024cfk}, 
	\begin{itemize}
		\item $S=-0.04\pm 0.1$,
		\item $T=0.01\pm 0.12$,
		\item $U=-0.01\pm 0.09$.   
	\end{itemize}
	  The corresponding expressions of the three parameters $S$, $T$ and $U$ are given in Appx. \ref{ObParam}. Considering a scenario in which $\cos(\alpha-\beta)=0.01$ and $0.1 \leq \tan\beta\leq 15$, the oblique parameters impose restrictive bounds on the mass difference of the scalars predicted in the model.
  
\begin{figure}[!htb]
  	\centering
  	{\includegraphics[scale=0.2]{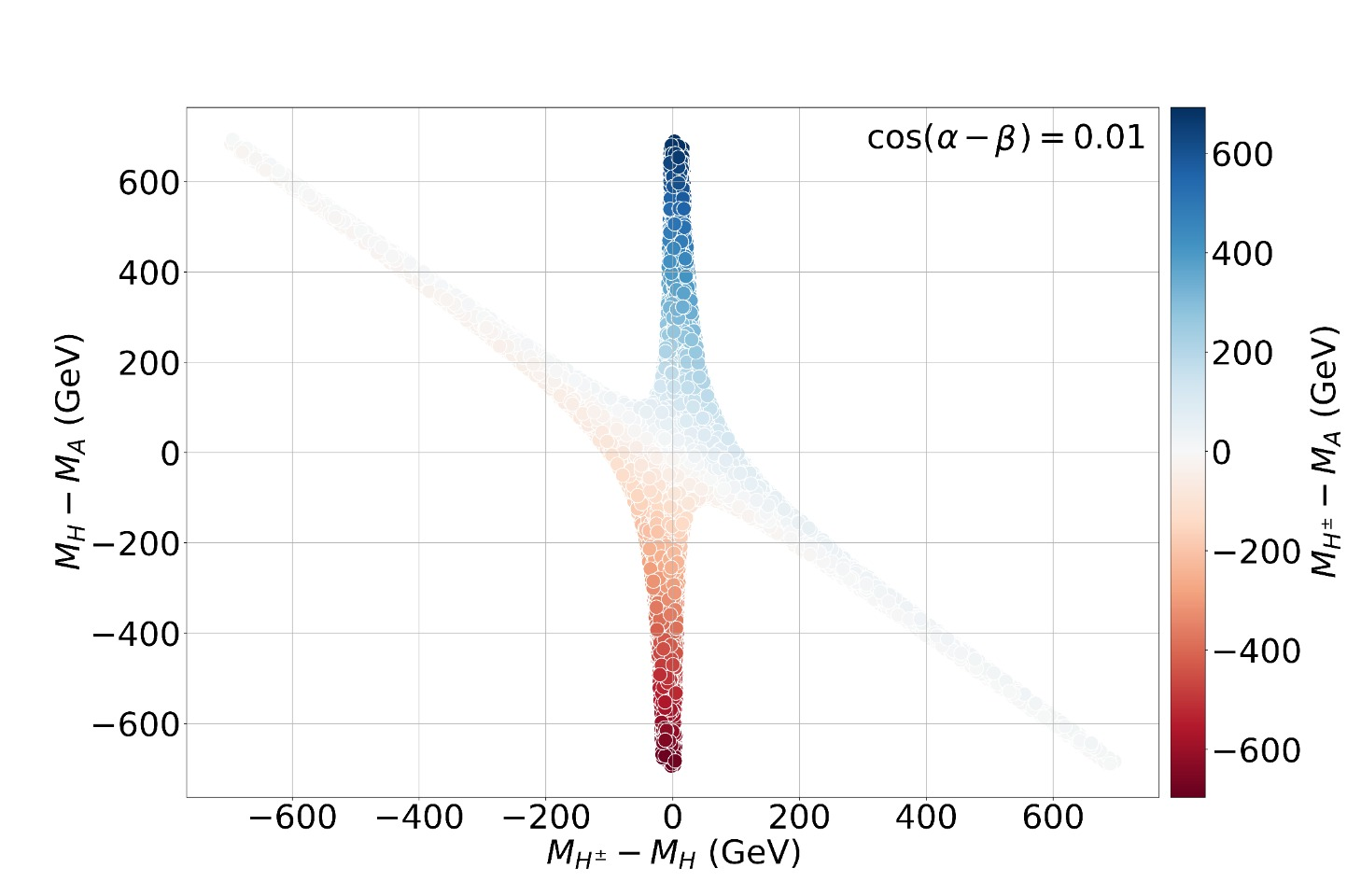}}
  	\caption{Allowed points (simultaneously) for the difference of scalar masses by oblique parameters.}\label{OP}
  \end{figure}

  We present in Fig. \ref{OP} a scattering plot whose points represent values for $M_H$, $M_A$ and $M_{H^+}$ such that they simultaneously satisfy the constraints of the oblique parameters $U$, $S$, $T$. We note that the main contribution for the production cross-section $\sigma(pp\to H^- H^+)$ is through the on-shell decay  $H\to H^+ H^-$. This could suppress the cross-section of the signal if one considers degeneration in the $M_{H^\pm}$, $M_H$, $M_A$ masses. To avoid it, we analyze cases such as $M_{H^\pm}-M_H\leq -250$ GeV. Then, we consider three masses for the neutral scalar boson $M_H=500,\,800,\,1000$ GeV. For the particular case $M_H=500$ GeV, the charged scalar mass is $M_{H^\pm}\leq 250$ GeV, it corresponds to $M_H-M_A\approx 250$ GeV and $M_{H^\pm}\approx M_A$. Thus, although the parameters $U$, $S$, and $T$ impose strong constraints on the mass difference, under the scenarios proposed below, we are free from these constraints. as well as restrictions from colliders and LFV processes.

 In conclusion, according to our analysis of the 2HDM-III parameter space, we present in Table \ref{scenarios} three realistic scenarios to be used in the simulations in the next section.
		 \begin{table}[!htb]
		 	
		 	\caption{Benchmark scenarios to be used in the subsequent calculations.}\label{scenarios}
		 	
		 	\begin{centering}
		 		\begin{tabular}{ccccccc}
		 			\hline 
		 			Scenario & $\tan\beta$&$\chi_{tt}$ & $\chi_{bc}$ & $\chi_{\mu\mu}$ & $\cos(\alpha-\beta)$ & $M_{H}$(GeV)\tabularnewline
		 			\hline 
		 			\hline 
		 			$S1$ & 1 &0.1& 10 & 10 & 0.01 & 500, 800, 1000\tabularnewline
		 			\hline 
		 			$S2$ & 5 &0.1& 5 & 5 & 0.01 & 500, 800, 1000\tabularnewline
		 			\hline 
		 			$S3$ & 10 &0.1& 1 & 1 & 0.01 & 500, 800, 1000\tabularnewline
		 			\hline 
		 		\end{tabular}
		 		\par\end{centering}
		 \end{table}
	 
		 	Through these simulations, we aim to explore the phenomenological implications of the model and identify potential signatures that could be observed at future collider experiments.


	\section{Collider analysis}\label{SecIV}

Let us first present the analytical expressions to evaluate the decay width of the processes $H^{\pm}\to u_i d_j$ and $H^{\pm}\to \ell^{\pm}\nu_\ell$, they are given by
\begin{eqnarray*}
	\Gamma(H^{\pm}\to u_{i}d_{j})&=&\frac{3G_{F}M_{H^{\pm}}}{\sqrt{2}4\pi}\Big(m_{u_{i}}^{2}|Y_{ij}|^{2}+m_{d_{j}}^{2}|X_{ij}|^{2}\Big),\\ \nonumber
	\Gamma(H^{\pm}\to \ell^{\pm}\nu_{\ell})&=&\frac{G_{F}M_{H^{\pm}}}{\sqrt{2}4\pi}\Big(m_{\ell}^{2}|Z_{ij}^\ell|^{2}\Big)\nonumber,
	\end{eqnarray*}
where 
	\begin{eqnarray*}
		X_{ij}&=&\sum_{l=1}^{3}(V_{\rm CKM})_{il}\Big[t_\beta \frac{m_{d_l}}{m_{d_j}}\delta_{l j}-\frac{\sqrt{1+t^2_\beta}}{\sqrt{2}}\sqrt{\frac{m_{d_l}}{m_{d_j}}}\chi_{lj}^d \Big],\\
		Y_{ij}&=&\sum_{l=1}^{3}\Big[1/t_\beta\delta_{il}-\frac{\sqrt{1+1/t^2_\beta}}{\sqrt{2}}\sqrt{\frac{m_{u_l}}{m_{u_j}}}\chi_{\l j}^u\Big](V_{\rm CKM})_{lj},\\
		Z_{ij}^\ell&=&\Big[t_\beta \frac{m_{\ell_i}}{m_{\ell_j}}\delta_{i j}-\frac{\sqrt{1+t_\beta^2}}{\sqrt{2}}\sqrt{\frac{m_{\ell_i}}{m_{\ell_j}}}\chi_{ij}^\ell \Big],
		\end{eqnarray*}	
where $t_\beta\equiv\tan\beta$.

\begin{figure}[!h]
	\centering
	\subfigure[]{{\includegraphics[scale=0.3]{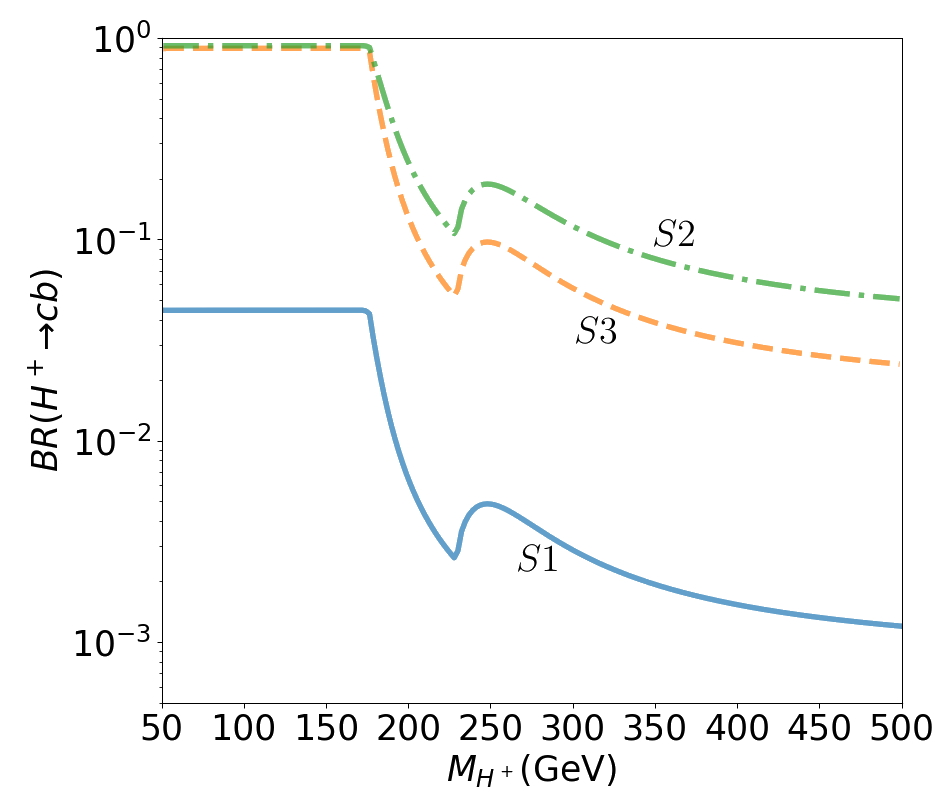}}}
	\subfigure[]{{\includegraphics[scale=0.3]{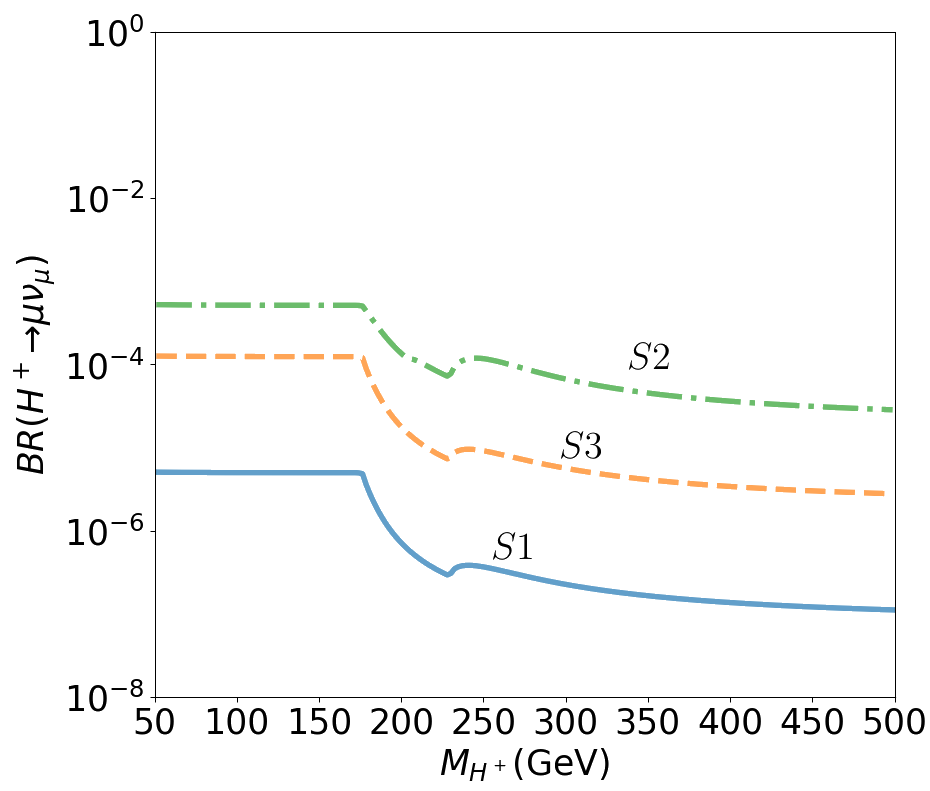}}}
	\caption{ Branching ratios (a) $BR(H^+\to cb)$ (b) $BR(H^+\to cb)$ as a function of the charged scalar mass $M_{H^+}$.}\label{BRs}
\end{figure}

We present in Fig. \ref{BRs} the branching ratios of the processes $H^{+}\to cb$ and $H^+\to \mu{\nu_\mu}$ as a function of the charged scalar mass $M_{H^+}$ for the scenarios $S1,\,S2,\,S3$.
	\subsection{Signal and background}
	The signal and background processes coming from $pp$ collisions at the LHC ...
	
	\begin{itemize}
		\item \textbf{SIGNAL:} We focus on the search for a specific final state $cb\mu\nu_{\mu}$. This final state arises from the pair production of charged Higgs bosons in proton-proton collisions, i.e., $pp\to H^- H^+ \to \mu^+\nu_{\mu}\bar{c}b+\mu^-\bar{\nu}_{\mu}c\bar{b}\,(\equiv \mu\nu_\mu cb)$. We consider a $b$ tagging efficiency $\epsilon_b=80\%$, the probability that a $c$-jet is mistagged as a $b$-jet is $\epsilon_c=10\%$, while the probability that any other jet is $\epsilon_j=1\%$. The relevant contributions coming from the 2HDM-III are depicted in Fig. \ref{FDsignal}.
		\begin{figure}[!htb]
			\centering
			{\includegraphics[scale=0.3]{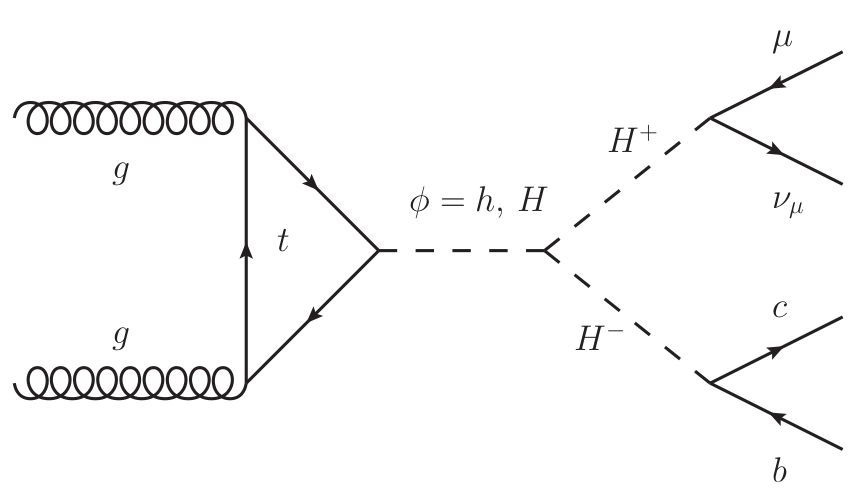}
                \includegraphics[scale=0.3]{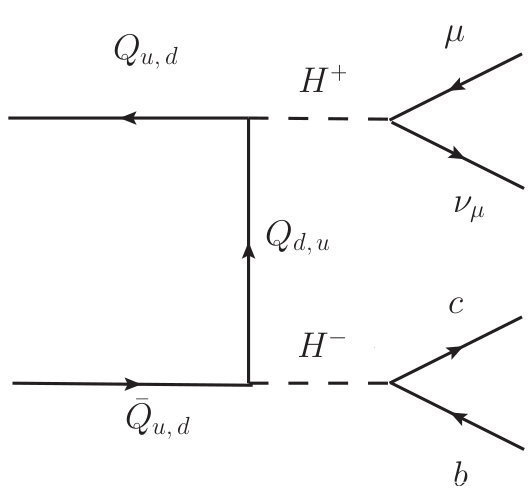}}
			\caption{Feynman diagrams of the production cross-section of the signal $pp\to H^+H^-\to \mu\nu_\mu cb$.}\label{FDsignal}
		\end{figure}
		\item \textbf{BACKGROUND:} The main SM background comes from the final state of $bj\ell\nu_\ell$, whose source arises from
		\begin{itemize}
			\item $Wjj+Wb\bar{b}$,
			\item $tb+tj$,
			\item $t\bar{t}$.
		\end{itemize}
	\end{itemize}
	In the last background process ($t\bar{t}$) either one of the two leptons is missed in the semi-leptonic top quark decays or two of the four jets are missed when one of the top quarks decays semi-leptonically. The numerical cross-section and branching ratios associated to the signal for $M_H=500,\,800,\,1000$ GeV and scenario $S2$ are presented in Table \ref{XSsignal}, while the corresponding cross-sections of the dominant SM background processes are shown in Table \ref{XSSMBGD}. 
		\begin{widetext}
	\begin{center}
		\begin{table}[!htb]
			
			\caption{Dominant contribution for the cross-section and branching ratios for
				scenario $S2$.}\label{XSsignal}
			
			\centering{}%
			\begin{tabular}{|c|c|c|c|c|c|c|}
				\hline 
				$M_{H}=500$(GeV)\tabularnewline
				\hline 
				$M_{H^{\pm}}$(GeV) & $\sigma(gg\to H)$&$\mathcal{BR}(H\to H^{-}H^+)$ & $\mathcal{BR}(H^{-}\to b\bar{c})$ & $\mathcal{BR}(H^{+}\to\mu^{+}\nu_{\mu})$ & $\sigma(pp\to\mu^{+}\nu_{\mu}b\bar{c})$& Events (3000 fb$^{-1}$) \tabularnewline
				\hline 
				100 & 1200 fb & 0.32 &0.944 & $5.27\times10^{-4}$ & 0.189 fb&567\tabularnewline
				\hline 
				150 & 1200 fb &0.54 &0.943 & $5.26\times10^{-4}$ & 0.324 fb&972\tabularnewline
				\hline 
				\hline
				\hline 
				$M_{H}=800$(GeV)\tabularnewline
				\hline 
				
				$M_{H^{\pm}}$(GeV) & $\sigma(gg\to H)$&$\mathcal{BR}(H\to H^{-}H^+)$ & $\mathcal{BR}(H^{-}\to b\bar{c})$ & $\mathcal{BR}(H^{+}\to\mu^{+}\nu_{\mu})$ & $\sigma(pp\to\mu^{+}\nu_{\mu}b\bar{c})$& Events (3000 fb$^{-1}$)\tabularnewline

				\hline 
				100 & 151 fb & 0.27 &0.944 & $5.27\times10^{-4}$ & 0.02 fb&59\tabularnewline
				\hline 
				150 & 151 fb &0.40 &0.943 & $5.26\times10^{-4}$ & 0.03 fb&100\tabularnewline
				\hline 
				\hline 
				\hline 
				$M_{H}=1000$(GeV) \tabularnewline
				\hline 
				
				$M_{H^{\pm}}$(GeV) & $\sigma(gg\to H)$&$\mathcal{BR}(H\to H^{-}H^+)$ & $\mathcal{BR}(H^{-}\to b\bar{c})$ & $\mathcal{BR}(H^{+}\to\mu^{+}\nu_{\mu})$ & $\sigma(pp\to\mu^{+}\nu_{\mu}b\bar{c})$& Events (3000 fb$^{-1}$)\tabularnewline

				\hline 
				100 & 42 fb & 1.67$\times 10^{-3}$&0.944 & $5.27\times10^{-4}$ & 0.000035 fb&0.1\tabularnewline
				\hline 
				150 & 42 fb &2.83$\times 10^{-3}$ &0.943 & $5.26\times10^{-4}$ & 0.000059 fb&0.2\tabularnewline
				\hline 
			\end{tabular}
		\end{table}
		\par\end{center}
		\begin{table}[!htb]
		\caption{Cross-section of the dominant SM background processes.}\label{XSSMBGD}
		\begin{centering}
			\begin{tabular}{|c|c|c|}
				\hline 
				SM backgrounds & Cross-section {[}fb{]}&Events (3000 fb$^{-1}$)\tabularnewline
				\hline 
				\hline 
				$pp\to Wjj+Wb\bar{b}\,(W\to\ell\nu_{\ell})$ & $3745960$&$\mathcal{O}(10^{10})$\tabularnewline
				\hline 
				$pp\to tb+tj\,(t\to\ell\nu_{\ell}b)$ & $1734$&$\mathcal{O}(10^{6})$\tabularnewline
				\hline 
				$pp\to t\bar{t}\,(t\to\ell\nu_{\ell}b,\,t\to q_{i}q_{j}b)$ & $431001$&$\mathcal{O}(10^{9})$\tabularnewline
				\hline 
			\end{tabular}
			\par\end{centering}
	\end{table}
	\end{widetext}		
	We present in Fig. \ref{Brs} an overview of the production cross-section as a function of charged scalar mass $M_{H^{\pm}}$ for the scenarios previously defined ($S1,\,S2,\,S3$) and $M_H=500$ GeV. Meanwhile, Fig. \ref{XS8001000} shows the same plane but considering $M_H=800,\,1000$ GeV only for scenario $S2$.
	\begin{figure}[!h]
		\centering
		{\includegraphics[scale=0.2]{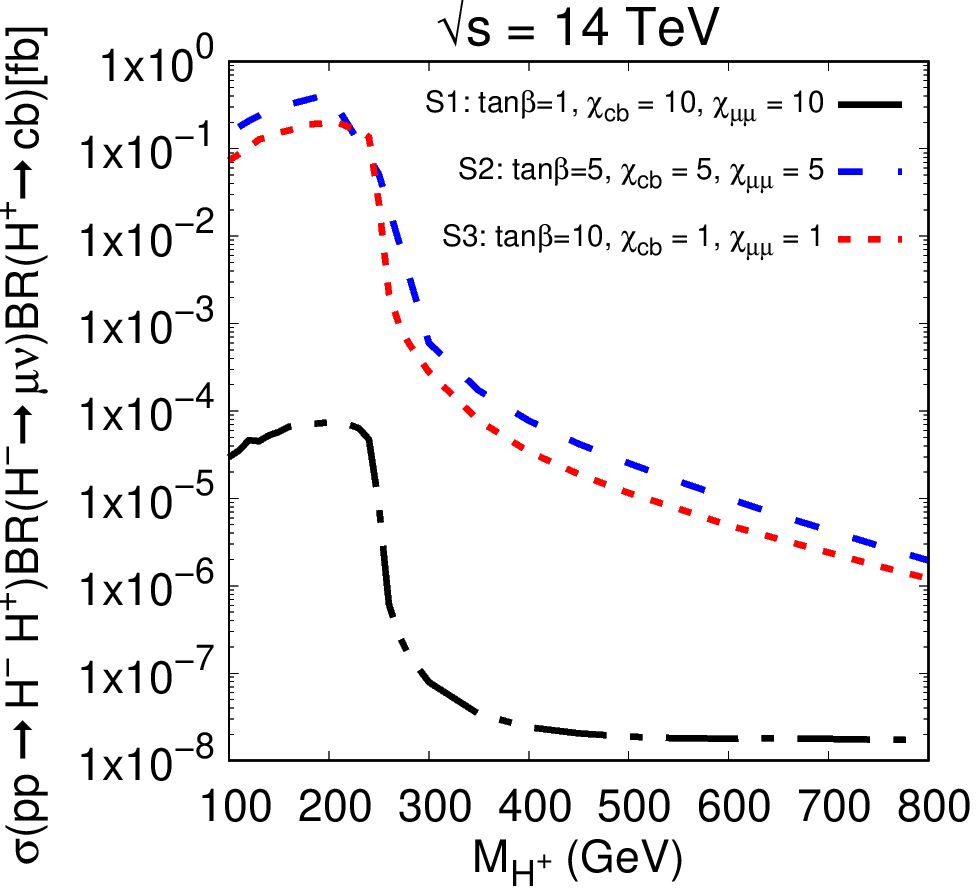}}
		\caption{Production cross-section of a charged Higgs pair with their subsequent decays into $\mu\nu_{\mu} cb$ for $M_H=500$ GeV. }\label{Brs}
	\end{figure}
	
	We observe for the three scenarios that the cross-sections are higher for masses in the range $100\leq M_{H^{\pm}} \leq 250$ GeV. This is because the dominant contribution comes from the neutral heavy Higgs boson $(M_H=500\,\rm GeV)$ which is capable of producing two real charged Higgs bosons. Once $M_{H^\pm}\geq 250\,\rm GeV$, the mediation of $H$ is virtual, which suppresses the cross-section as $M_{H^\pm}$ increases. From Tables \ref{XSsignal} and \ref{XSSMBGD}, for $M_{H^{\pm}}=100$ GeV and $M_H=500$ GeV, we note a difference of up to 7 orders of magnitude between the signal cross-section compared with the corresponding background production cross-section. This can be translated to a difference of $10^8$ number of events, assuming an integrated luminosity of $3000$ fb$^{-1}$. Even considering the maximum  integrated luminosity that the HL-LHC will reach (3000 fb$^{-1}$), in the case $M_H=1000$ GeV any event can be produced, which excludes a massive scalar for such integrated luminosity. However, according to our analysis, despite having a handful of events for the case when $M_H=800$ GeV, the HL-LHC may have the potential to scrutinize the possible evidence of the signal we propose, as will be seen later.
	
	   	\begin{figure}[!htb]
		\centering
		{\includegraphics[scale=0.24]{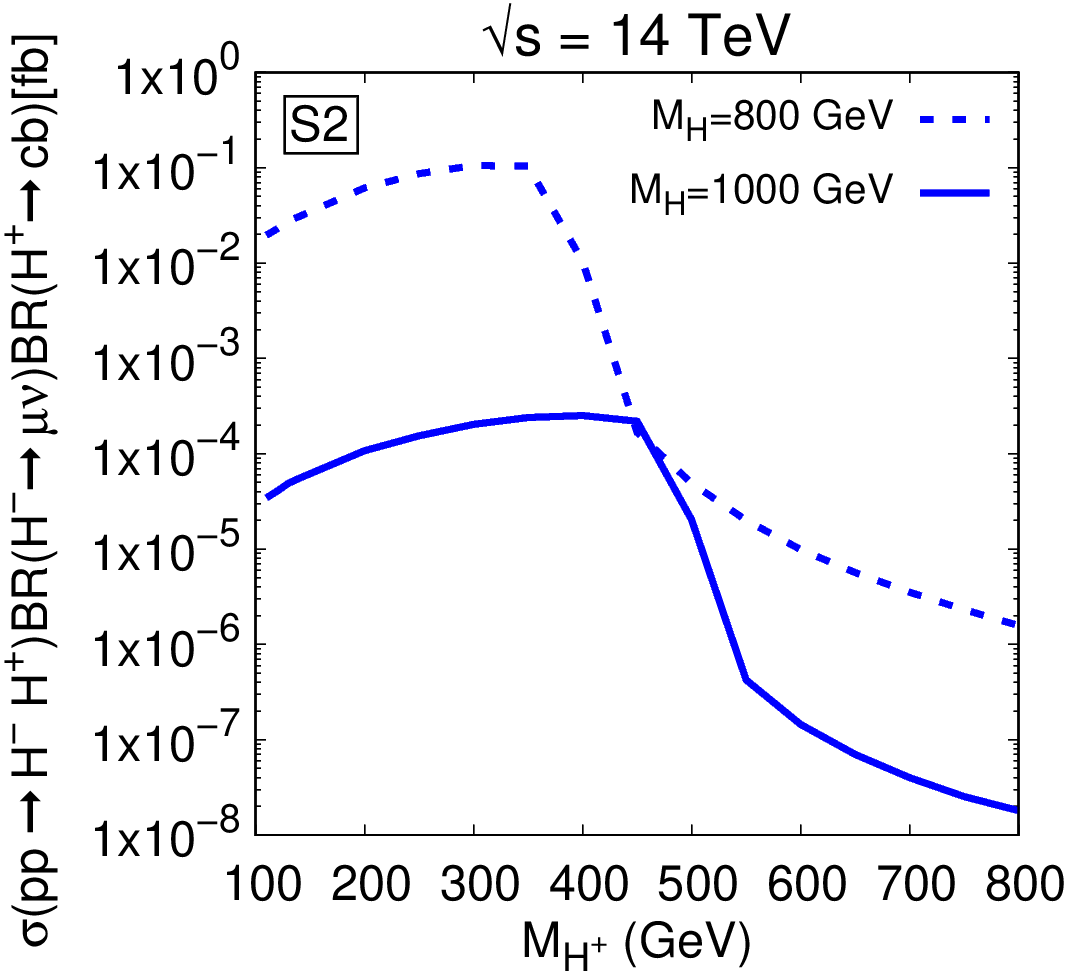}}
		\caption{Production cross-section of a charged Higgs pair with their subsequent decays into $\mu\nu_{\mu} cb$ for $M_H=800,\,1000$ GeV and scenario $S2$. }\label{XS8001000}
	\end{figure}
	As far as our computation scheme is concerned, we first implement the full model via $\texttt{FeynRules}$ \cite{Alloul:2013bka} for $\texttt{MadGraph5}$ \cite{MadGraphNLO}, later it is interfaced with \texttt{Pythia8} \cite{Sjostrand:2008vc} and \texttt{Delphes 3} \cite{delphes} for detector simulations. Concerning the jet reconstruction, the finding package \texttt{FastJet} \cite{Cacciari:2011ma} and the \texttt{anti-$k_t$} algorithm \cite{Cacciari:2008gp} were used.

	\subsection{Signal significance}
	The traditional strategy based in kinematic cuts on observables often rejects a significant number of background events, however also refuse an important number of signal events. This situation can be improved by using a Multivariate Analysis (MVA). In this way, we compute a Boosted Decision Tree (BDT) training \cite{Woodruff2018} on the set variables shown in Table \ref{VarBDT} of the Appx. \ref{VI}. We choose the relevant hyperparameters as follows: Number of trees \texttt{NTree}=50, maximum depth of the decision tree \texttt{MaxDepth}=5, maximum number of leaves \texttt{MaxLeaves}=8; other parameters are set to its default values.
	
\begin{figure}[!h]
		\centering
		\includegraphics[scale=0.35]{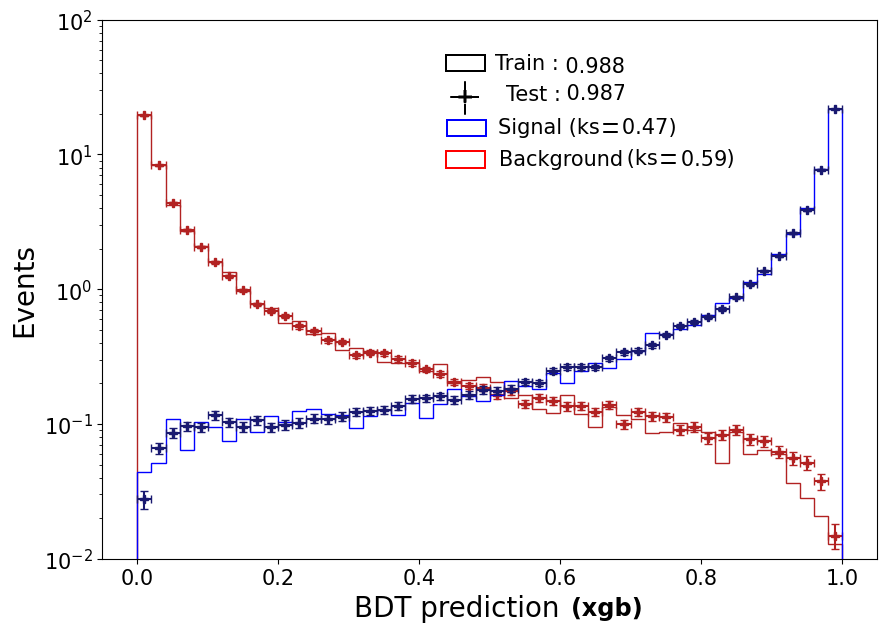}
		\caption{Plot of the discriminant for signal and background data.}\label{Discriminat}
	\end{figure}

	We present in Fig. \ref{Discriminat} the discriminant for the signal and background. The goodness of fit is checked with the Kolmogorov-Smirnov (KS) test. We observed that the KS value is within the permissible [0,1] interval and it is 0.47 (0.59) for the signal (background). 
Once the classifier was trained, it exports the output in terms of a single variable $(\textbf{xgb})$ which separate the signal events from the background ones, as shown in Fig. \ref{Discriminat}. Then, the signal-to-background ratio is optimised.	
	
According to our analysis, we find that the best ranked observables are i) transverse moment of the $b-jet$, ii) missing energy transverse $\slashed{E}_T$, iii) transverse moment of the $c-jet$, iv) transverse mass of the muon $M_{T}[\mu]$ and v) invariant mass of one of the charged Higgs decaying into a $bc$-jets pair $M_{inv}[bc]$. Considering the scenario $S2$, $M_{H^{\pm}}=110$ GeV and $M_H=500$ GeV, we present in Fig. \ref{distributions} the aforementioned distributions. From the $M_{T}[\mu]$ and $M_{inv}[bc]$, we observe a remarkable fact, it is the difference between the behavior of the background and signal processes. In both observables there is a resonant peak (around $M_{H^{\pm}}=110$ GeV) that identifies the charged scalar bosons, one of them decaying into a $bc$ pair and the another decaying into $\mu+\slashed{E}_T$; these are the clearest signatures of our signal and one could directly impose cuts on these variables as follows,
\begin{enumerate}
	\item $M_{H^{\pm}}-20<M_T[\mu]<M_{H^{\pm}}+20$ GeV,
	\item $M_{H^{\pm}}-30<M_{inv}[bc]<M_{H^{\pm}}+20$ GeV,\\
	and additionally,	
	\item $P_T[b]>50$ GeV,
	\item $\rm MET>20$ GeV,
	\item $P_T[j_1]>30$ GeV.
\end{enumerate}    
With the previous kinematic cuts (we really make sure of this via the $\texttt{MadAnalysis5}$ package \cite{Conte:2012fm}), we can obtain a good signal significance. Nevertheless, instead, we choose to impose cuts on the variable $\textbf{xgb}$, as described below.
	\begin{figure}[!htb]
		\subfigure[]{{\includegraphics[scale=0.3]{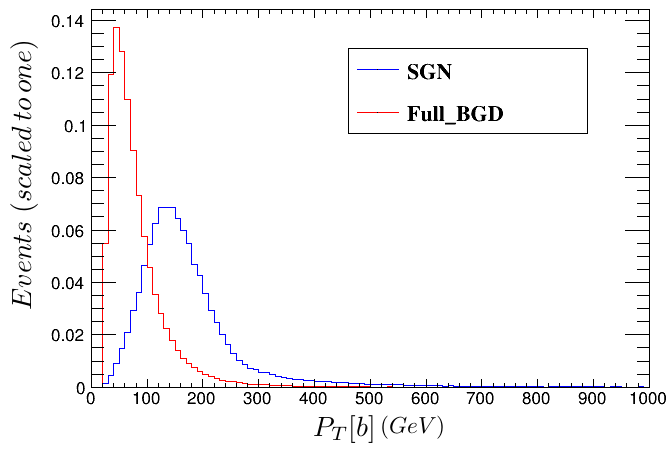}}\label{a}} \\
		\subfigure[]{{\includegraphics[scale=0.3]{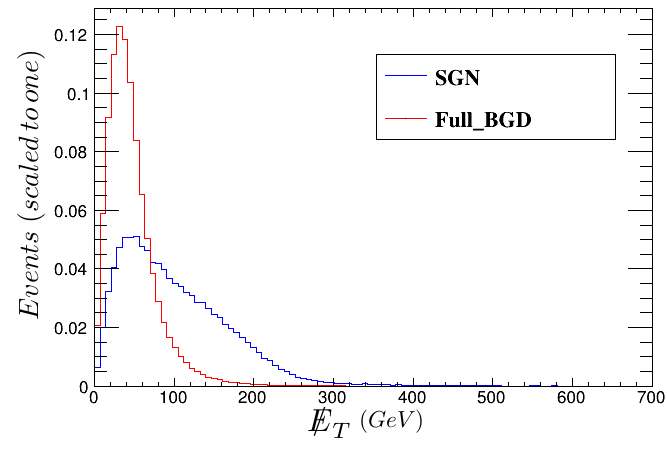}}\label{b}} \\
			\subfigure[]{{\includegraphics[scale=0.3]{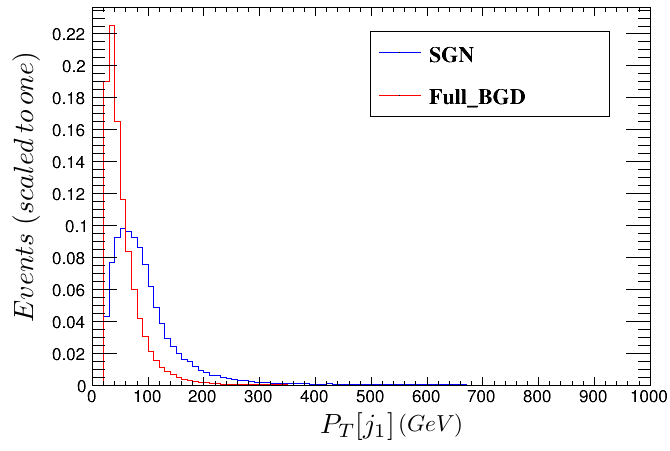}}\label{c}} \\
		\subfigure[]{{\includegraphics[scale=0.3]{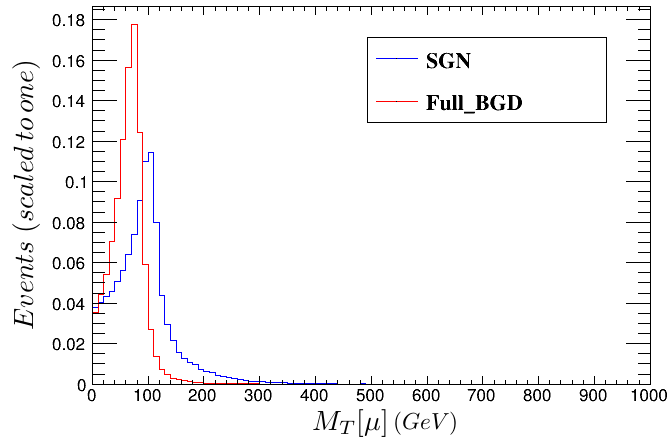}}\label{d}}
		\end{figure}
	
	\begin{figure}[!htb]
		\subfigure[]{{\includegraphics[scale=0.24]{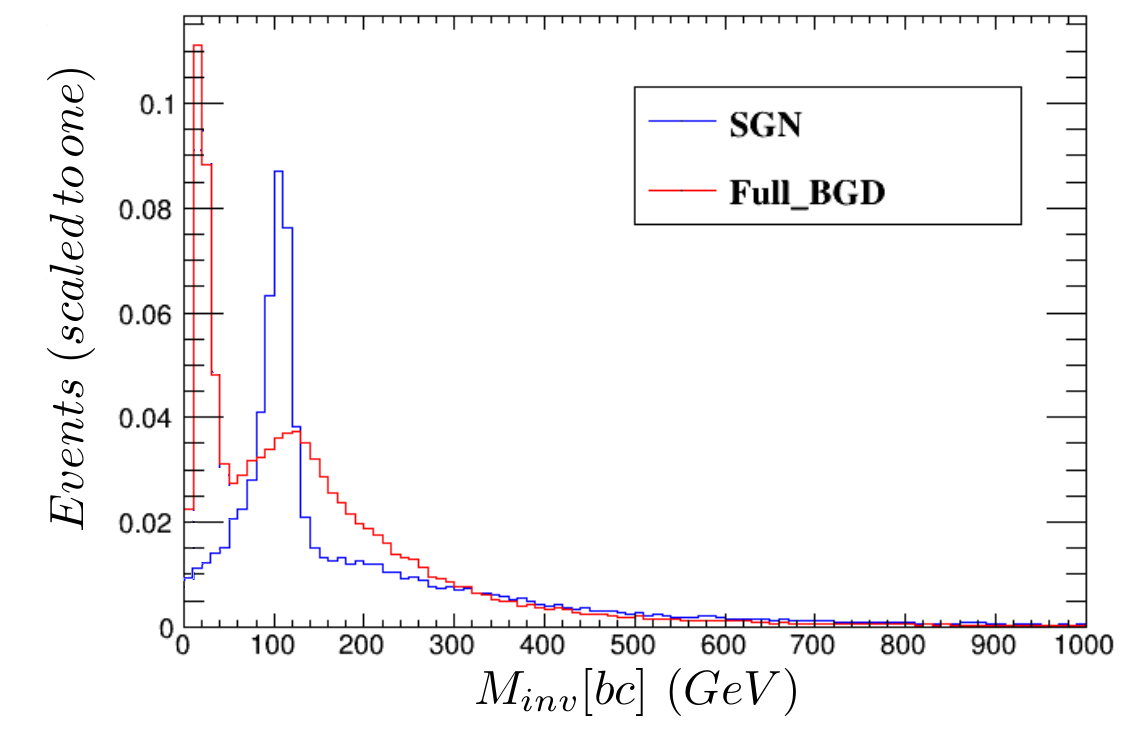}}\label{e}} 
		\caption{Kinematic distributions of the signal and SM background processes: (a) transverse moment of the $b-jet$, (b) missing energy transverse $\slashed{E}_T$, (c) transverse moment of the $c-jet$, (d) transverse mass of the muon $M_{T}[\mu]$ and (e) invariant mass of one of the charged Higgs decaying into a $bc$-jets pair $M_{inv}[bc]$. In all graphs, \texttt{SGN} labels the signal, while \texttt{Full$\_$BGD} labels the full background.}\label{distributions}
	\end{figure}

\begin{figure}[!htb]
		\centering
		\subfigure[]{{\includegraphics[scale=0.2]{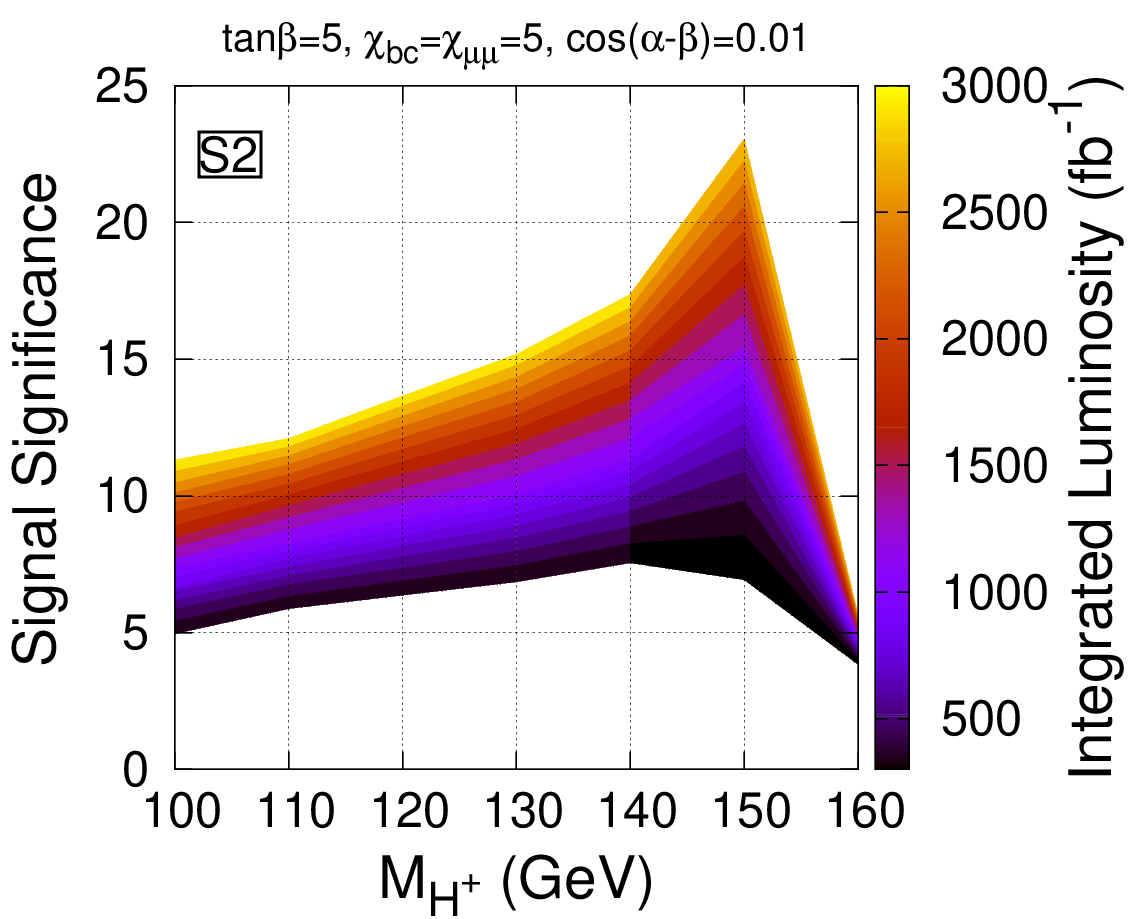}}\label{b}}
		\subfigure[]{{\includegraphics[scale=0.2]{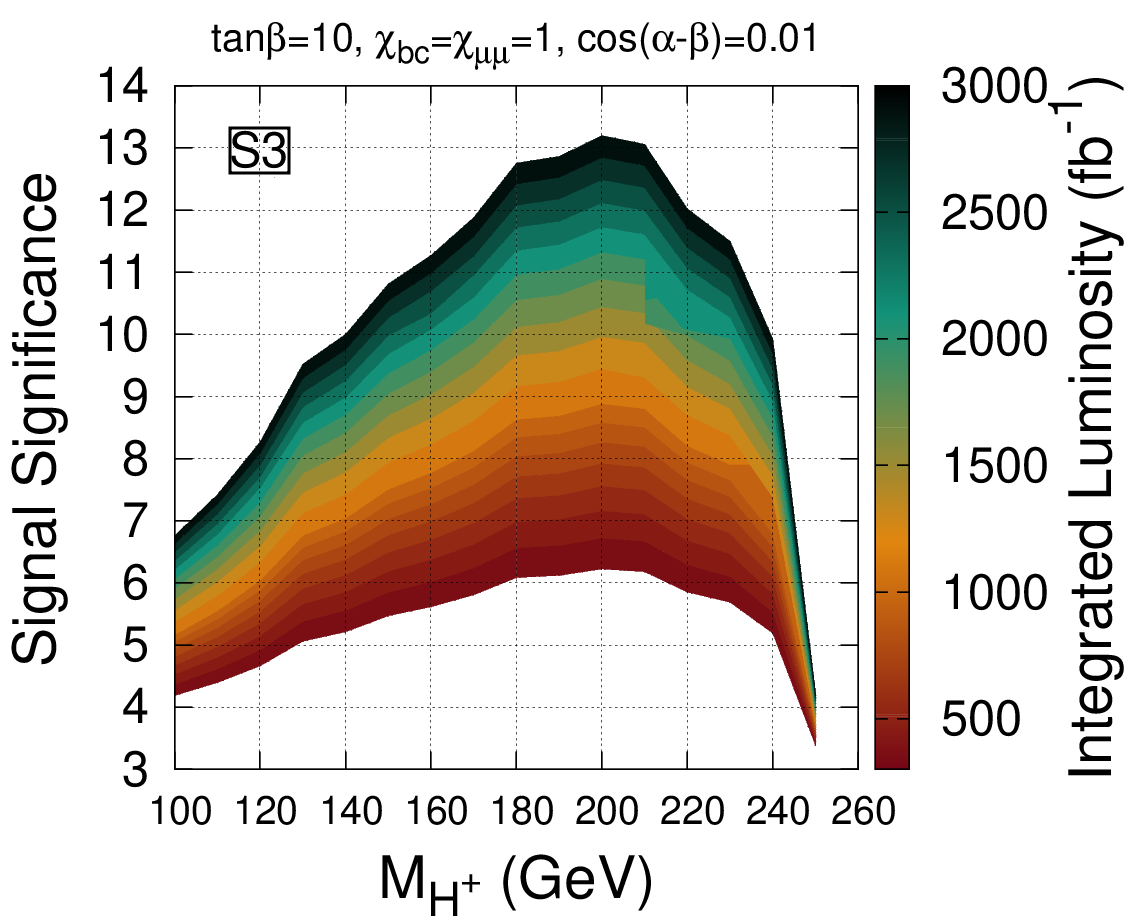}}\label{c}}
		\subfigure[]{{\includegraphics[scale=0.2]{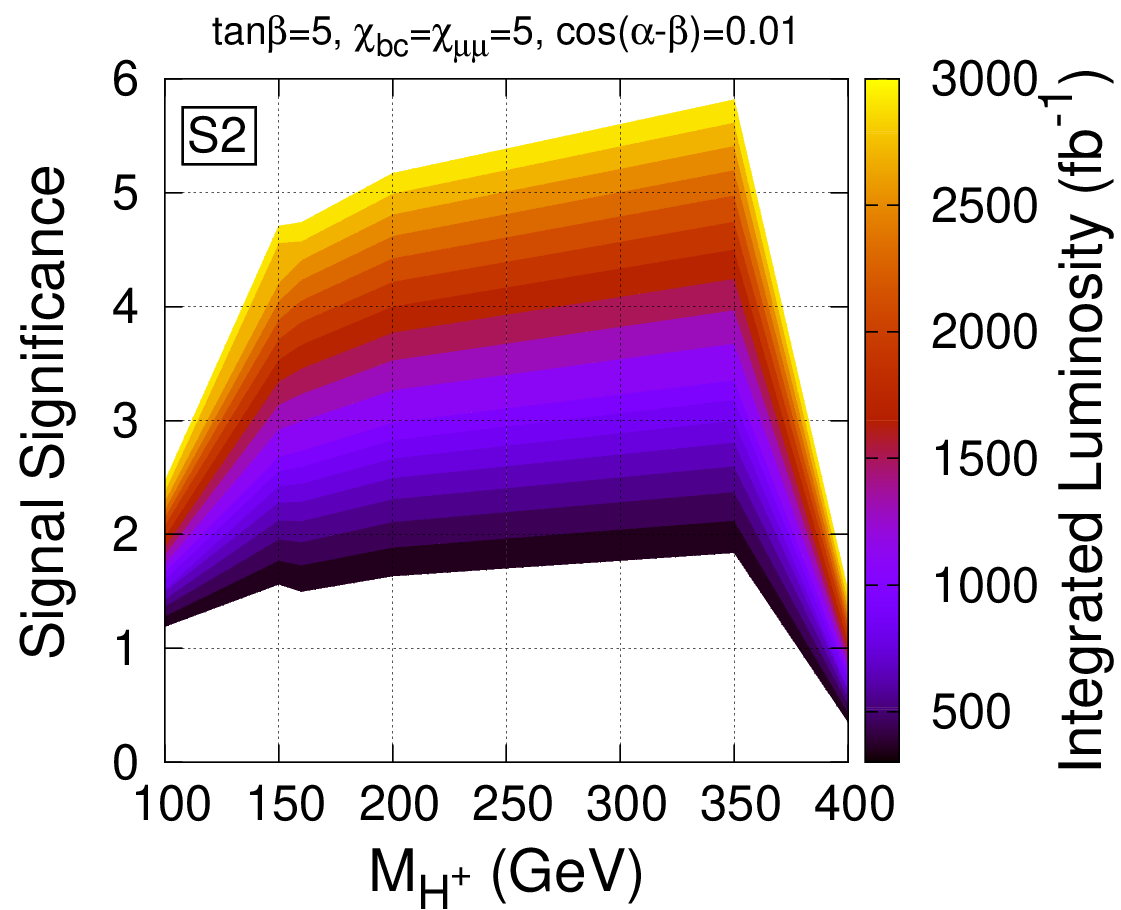}}\label{d}}
		\caption{Signal significance as function of the charged Higgs boson mass and the integrated luminosity for (a) $S2$, (b) $S3$ ($M_H=500$ GeV; (c) $S2$ ($M_H=800$ GeV).}\label{SignalSignificance}
	\end{figure}

Our procedure is to take the maximum signal significance, defined as the ratio $\mathcal{N}_S/\sqrt{\mathcal{N}_S+\mathcal{N}_B}$, where $\mathcal{N}_S\,(\mathcal{N}_B)$ is the number of signal (background) events. Such a signal significance is achieved by making a cut on the BDT output $\textbf{xgb}>0.95$, which is clearly visible in Fig. \ref{Discriminat}.
	 Therefore, Fig. \ref{SignalSignificance} displays contour plots of the signal significance as function of the charged Higgs boson mass $M_{H^{\pm}}$ and the integrated luminosity for the scenarios $S2$\footnote{We explore up to $M_{H^{\pm}}=200$ GeV, however we found a tiny signal significance.} and $S3$ for (a) and (b): $M_H=500$ GeV, and (c): $M_H=800$ GeV. We find that the scenario $S1$ presents difficulties to be tested in the experiment, which is why it is not included.

	 We find that the best scenario is $S2$, which predicts a statistical significance of $5\sigma$ in the $100\leq M_{H^\pm}\leq 160$ GeV interval for an integrated luminosity in the range $250-300$fb$^{-1}$. This scenario could be explored even at the LHC and subsequently tested at the HL-LHC. Concerning scenario $S3$, it also has the potential to be brought to experimental scrutiny, even explore up to $M_{H^{\pm}}\sim 250$ GeV. We foretell a statistical signal of $5\sigma$ for $100\lesssim M_{H^{\pm}}\lesssim 250$ GeV once an amount of data corresponding to an integrated luminosity in the range $220-1000$ fb$^{-1}$ is accumulated.

	\section{Conclusions}\label{SecV}
	
	In this work, we have explored the production and possible detection of a charged scalar boson pair predicted in an extension of the Standard Model, the so-called Two-Higgs Doublet Model of type III. The production channel studied occurs via proton-proton collisions at the LHC and the HL-LHC, subsequently, the charged scalars decay to the final state $\mu\nu_\mu cb$, i.e. $pp\to H^- H^+ \to \mu\nu_\mu cb$. Through a deep analysis of the model parameter space, we identify highly promising benchmarks that could be tested in the forthcoming HL-LHC. Specifically, two of the three scenarios defined in the article present valuable opportunities to be analyzed experimentally: $S2:\,\tan\beta=\chi_{\mu\mu}=\chi_{cb}=5,\,\cos(\alpha-\beta)=0.01$ and $S3:\,\tan\beta=10,\,\chi_{\mu\mu}=\chi_{cb}=1,\,\cos(\alpha-\beta)=0.01$. We also note that the dominant contribution is through the production of an on-shell neutral scalar boson $H$ (predicted in the model) that decays to the charged Higgs boson pair. Using the scenarios $S2$ and $S3$ in Monte Carlo simulations and the Boosted Decision Trees algorithm, it is possible to discriminate a significant number of background events that obscure the proposed signal, achieving a substantial improvement with respect to traditional straight kinematic cuts. In this way, once the HL-LHC reaches integrated luminosities $\mathcal{L}{\rm int}\geq 300$ fb$^{-1}$, we predict a signal significance at the level of $\geq 5\sigma$ for the $110\lesssim M{H^{\pm}}\lesssim 250$ GeV interval, corresponding to $M_H=500$ GeV. Meanwhile, for a neutral scalar boson mass of $M_H=800$ GeV, we need integrated luminosities $\mathcal{L}{\rm int}\geq 2400$ fb$^{-1}$ for the range $180\lesssim M{H^{\pm}}\lesssim 360$ GeV.
	\section*{Acknowledgments}
    We are grateful to Jhovanny Andres Mejia Guisao for valuable discussions and his helpful guide and advice with the software used in this study. The work of M. A. Arroyo-Ure\~na is supported by ``Estancias posdoctorales por M\'exico (CONAHCYT)" and ``Sistema Nacional de Investigadores" (SNI-CONAHCYT). S. Rosado-Navarro thanks to Vicerrector\'ia de Investigaci\'on y Estudios de Posgrado through ``Centro Interdisciplinario de Investigaci\'on y ense\~nanza de la Ciencia". E. A. Herrera-Chac\'on acknowledges support from CONAHCYT.	
	\appendix
	\section*{Appendix V}
	\subsubsection*{Constraints} \label{IndividualConstraints}	
	We present individual allowed regions by different constraints for each observable described in Sec. \ref{SecIII}, namely,
	\begin{itemize}
		\item Signal strengths
	\end{itemize}
	We present in Fig. \ref{senalesIntensidadIND} individual  $\cos(\alpha-\beta)-\tan\beta$ planes associated to each $\mu_X$. The colored points correspond to those allowed by $\mu_X$. While Tab. \ref{Scan_muX} shows the range of scanned parameters involved in the evaluation of the signal strengths $\mu_X$.
\begin{figure}[H]
	\centering
	\subfigure[]{\centering{\includegraphics[scale=0.2]{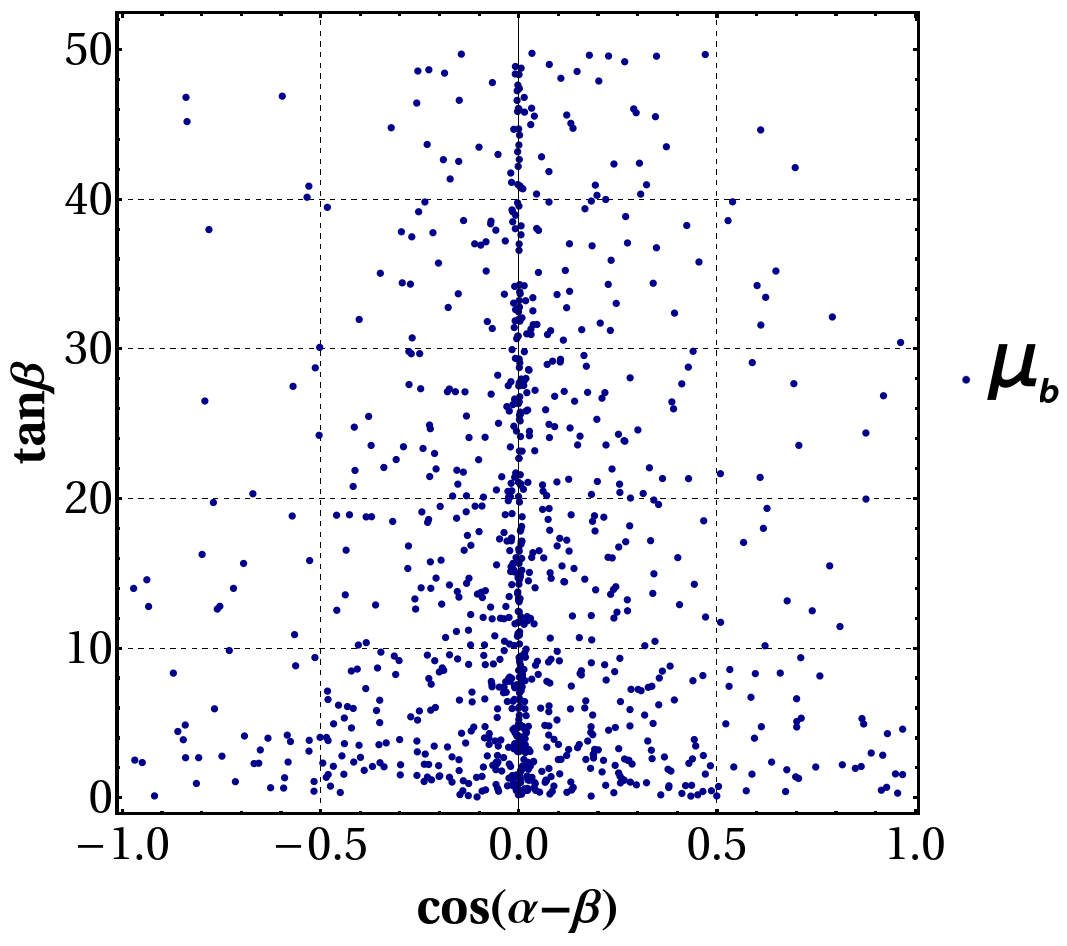}}}
	\subfigure[]{\centering{\includegraphics[scale=0.2]{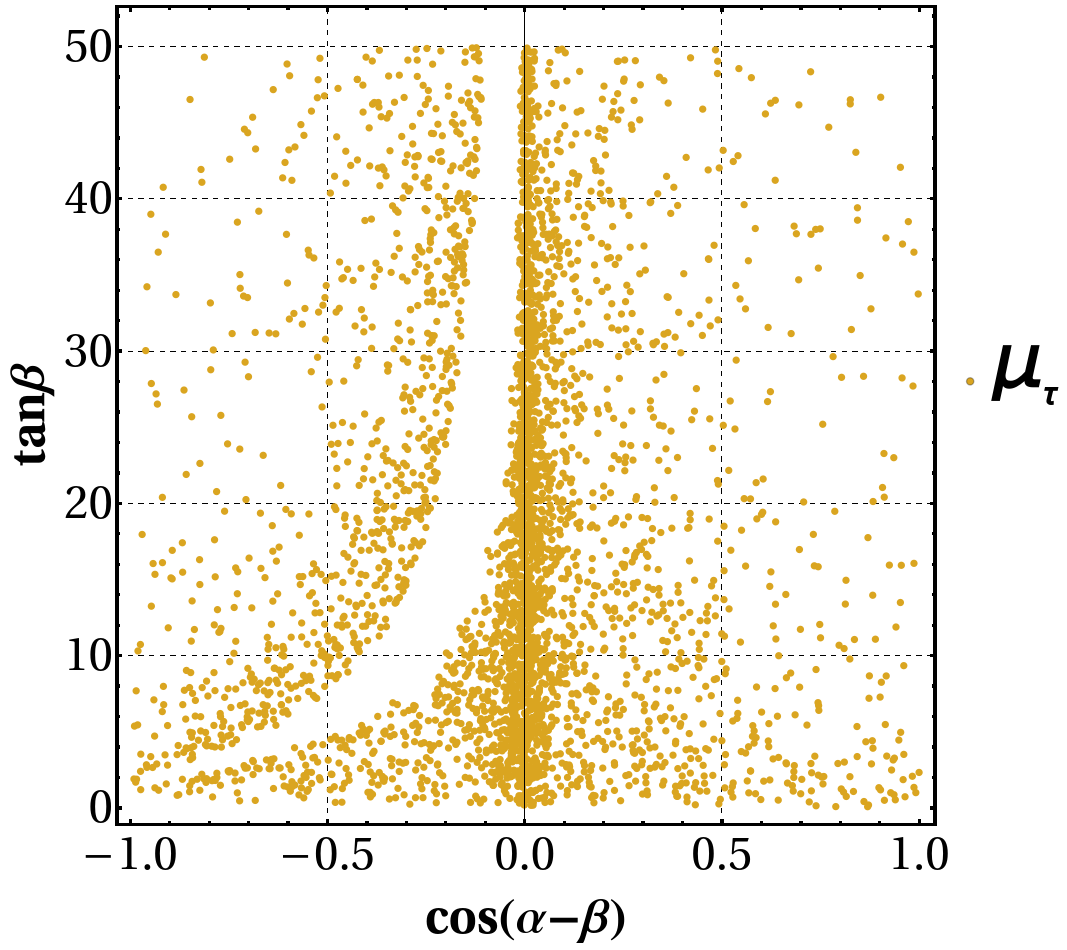}}}
	\subfigure[]{\centering{\includegraphics[scale=0.2]{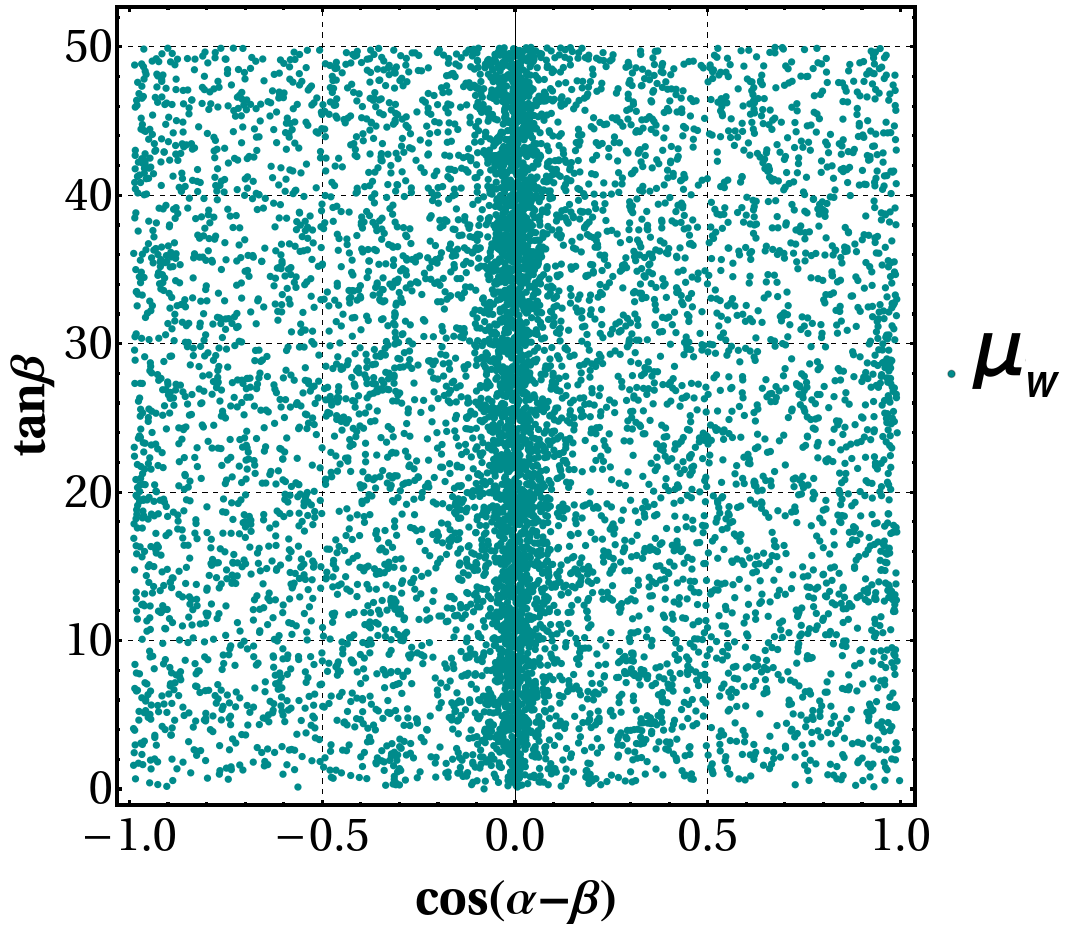}}}
	\subfigure[]{\centering{\includegraphics[scale=0.2]{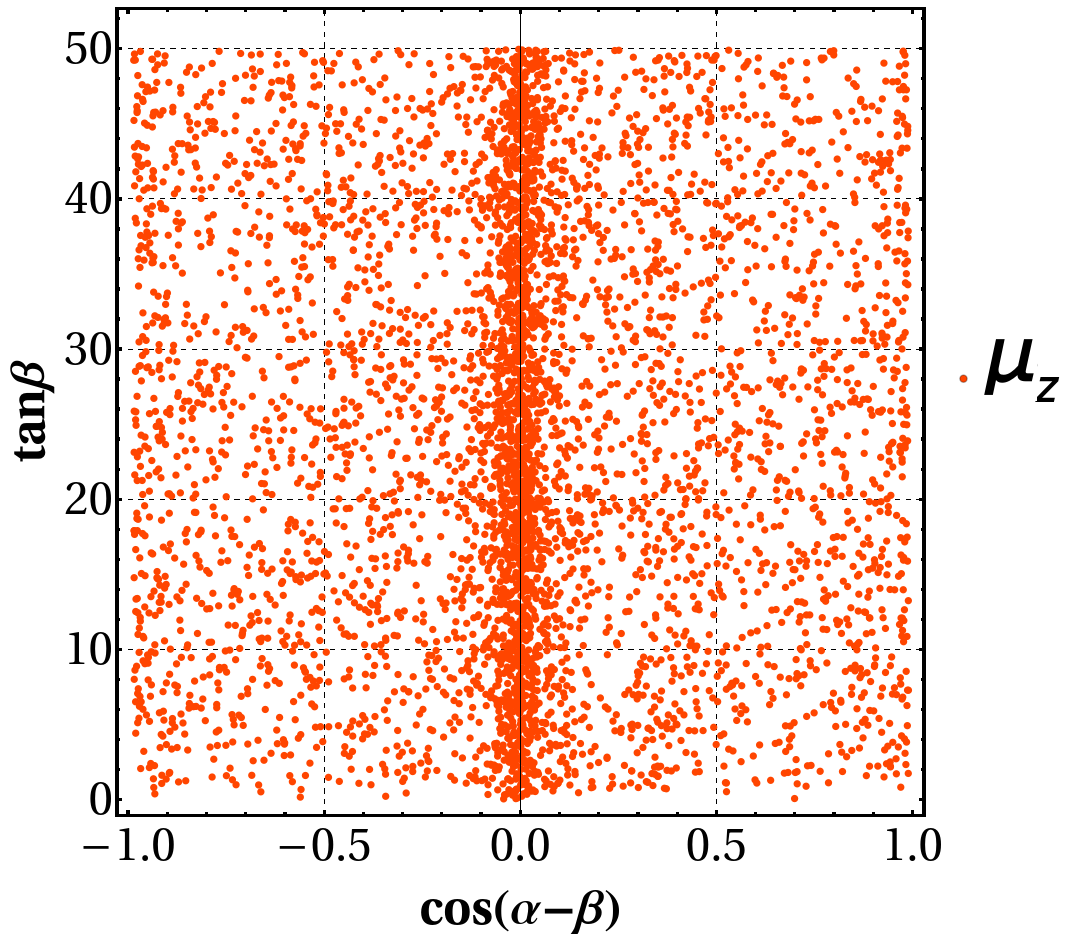}}}
\end{figure}

\begin{figure}[H]
	\centering
	\subfigure[]{\centering{\includegraphics[scale=0.2]{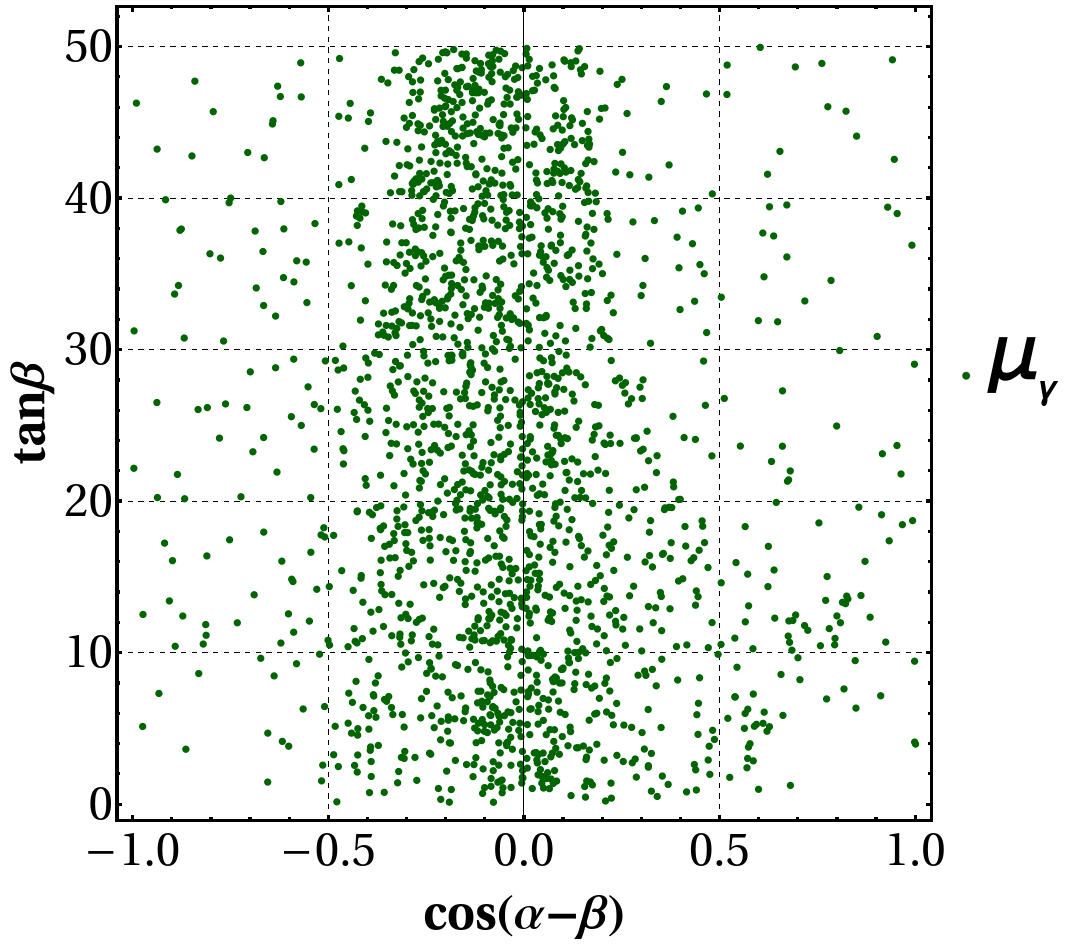}}}
	\subfigure[]{\centering{\includegraphics[scale=0.2]{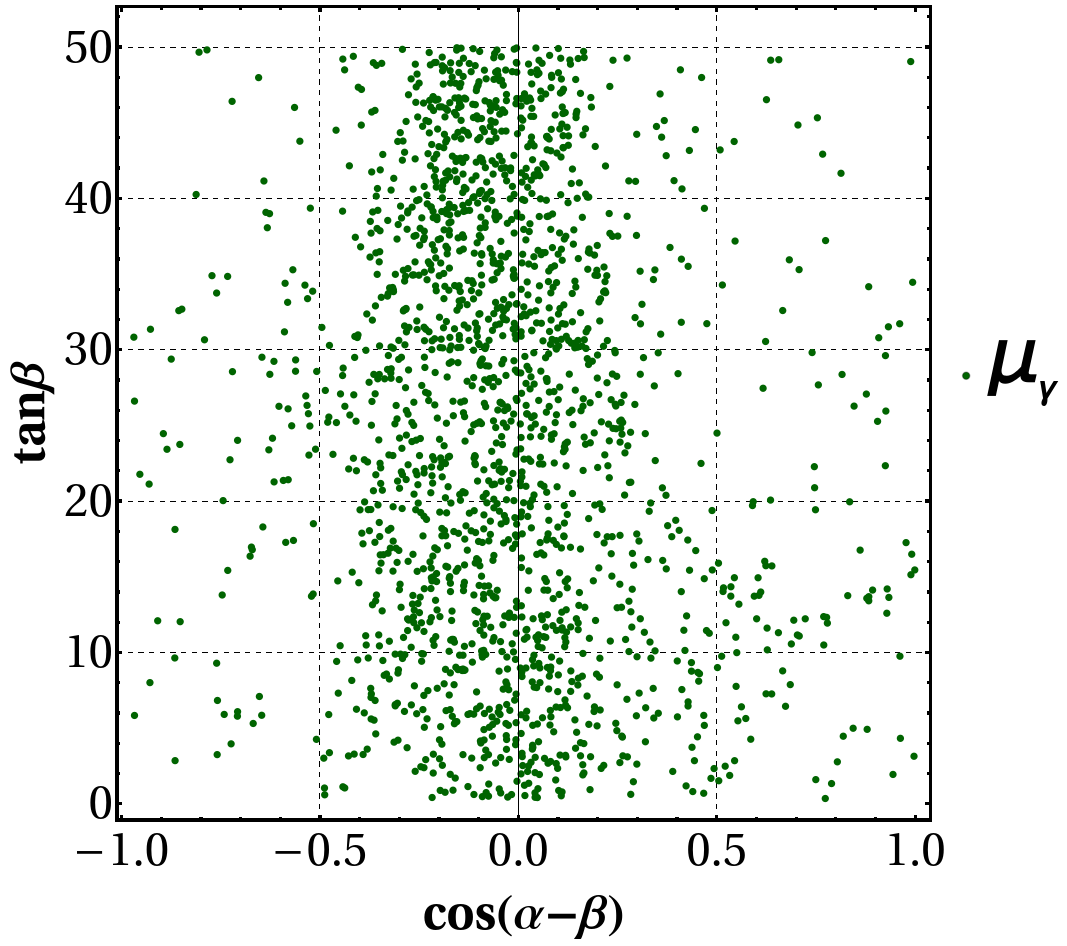}}}
	\caption{Signal strengths: (a) $\mu_b$, (b) $\mu_{\tau}$, (c) $\mu_W$, (d) $\mu_Z$, (e) $\mu_{\gamma}$ ($114\leq M_{H^{\pm}}\leq 140$), (f) $\mu_{\gamma}$ ($m_t + m_b\leq M_{H^{\pm}}\leq 1000$). In all cases we generate $50$K random points.}\label{senalesIntensidadIND}
\end{figure}
	
\begin{figure}[!htb]
		\centering
		\includegraphics[width=6cm]{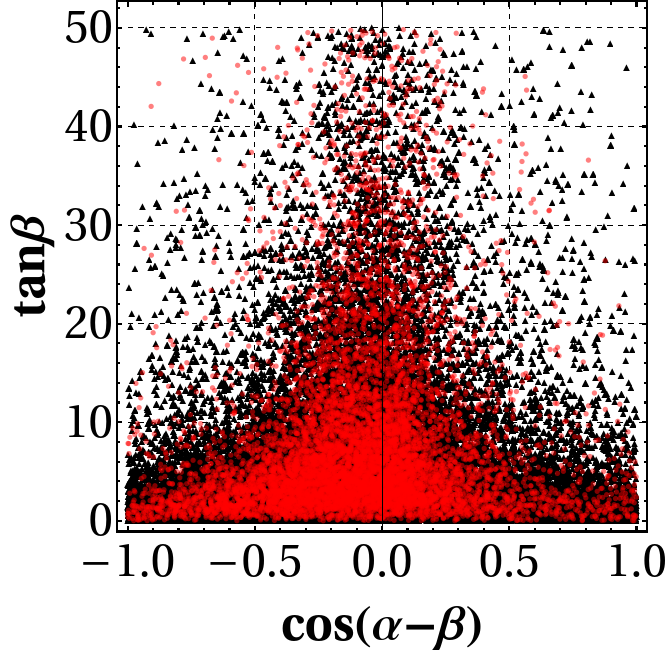}
		\caption{Allowed values by the measurement on $\mathcal{BR}(B_d^0\to\mu^+\mu^-)$ (black triangles) and by the upper limit on $\mathcal{BR}(B_s^0\to\mu^+\mu^-)$ (light red circles). In both cases we generate $50$K random points.}\label{INDBs}
	\end{figure}

	\begin{table}[!htb]
		
		\caption{Range of scanned parameters for the signal strengths $\mu_{X}.$}\label{Scan_muX}
		
		\begin{centering}
			\begin{tabular}{|c|c|}
				\hline 
				Parameter & Range\tabularnewline
				\hline 
				\hline 
				$\tan\beta$ & $[0,50]$\tabularnewline
				\hline 
				$\cos(\alpha-\beta)$ & $[-1,1]$\tabularnewline
				\hline 
				$\chi_{bb}$ & $[-1,1]$\tabularnewline
				\hline 
				$\chi_{tt}$ & $[-100,100]$\tabularnewline
				\hline 
				$\chi_{\tau\tau}$ & $[-1,1]$\tabularnewline
				\hline 
				$M_{H^{\pm}}$ (GeV) & $[114,140]$\tabularnewline
				\hline 
				$M_{H^{\pm}}$ (GeV) & $[m_{t}+m_{b},1000]$\tabularnewline
				\hline 
			\end{tabular}
			\par\end{centering}
	\end{table}
	\begin{itemize}
		\item Decays $B_{s,d}^0\to \mu^+\mu^-$
	\end{itemize}
	Fig. \ref{INDBs} presents the allowed points by experimental measurements on $\mathcal{BR}(B_{s,d}^0\to \mu^+\mu^-)$. The range of scanned parameters is displayed in Table \ref{ScanB2mumu}.
	
	\begin{table}[!htb]
		
		\caption{Range of scanned parameters. On the left: $B_{d}^{0}\to\mu^{+}\mu^{-}$,
			on the right: $B_{s}^{0}\to\mu^{+}\mu^{-}$ }\label{ScanB2mumu}
		
		\begin{centering}
			\begin{tabular}{cc}
				\hline 
				Parameter & Range\tabularnewline
				\hline 
				\hline 
				$\tan\beta$ & $[0,50]$\tabularnewline
				\hline 
				$\cos(\alpha-\beta)$ & $[-1,1]$\tabularnewline
				\hline 
				$\chi_{\mu\mu}$ & $[-1,1]$\tabularnewline
				\hline 
				$\chi_{bd}$ & $[-1,1]$\tabularnewline
				\hline 
				$\chi_{db}$ & $[-1,1]$\tabularnewline
				\hline 
				$M_{A}$ (GeV) & $[100,1000]$\tabularnewline
				\hline 
				$M_{H}$ (GeV) & $[300,1000]$\tabularnewline
				\hline 
			\end{tabular}$\qquad$%
			\begin{tabular}{cc}
				\hline 
				Parameter & Range\tabularnewline
				\hline 
				\hline 
				$\tan\beta$ & $[0,50]$\tabularnewline
				\hline 
				$\cos(\alpha-\beta)$ & $[-1,1]$\tabularnewline
				\hline 
				$\chi_{\mu\mu}$ & $[-1,1]$\tabularnewline
				\hline 
				$\chi_{bs}$ & $[-1,1]$\tabularnewline
				\hline 
				$\chi_{sb}$ & $[-1,1]$\tabularnewline
				\hline 
				$M_{A}$ (GeV) & $[100,1000]$\tabularnewline
				\hline 
				$M_{H}$ (GeV) & $[300,1000]$\tabularnewline
				\hline 
			\end{tabular}
			\par\end{centering}
	\end{table}
	
\begin{figure}[!htb]
		\centering
		\subfigure[]{{\includegraphics[scale=0.2]{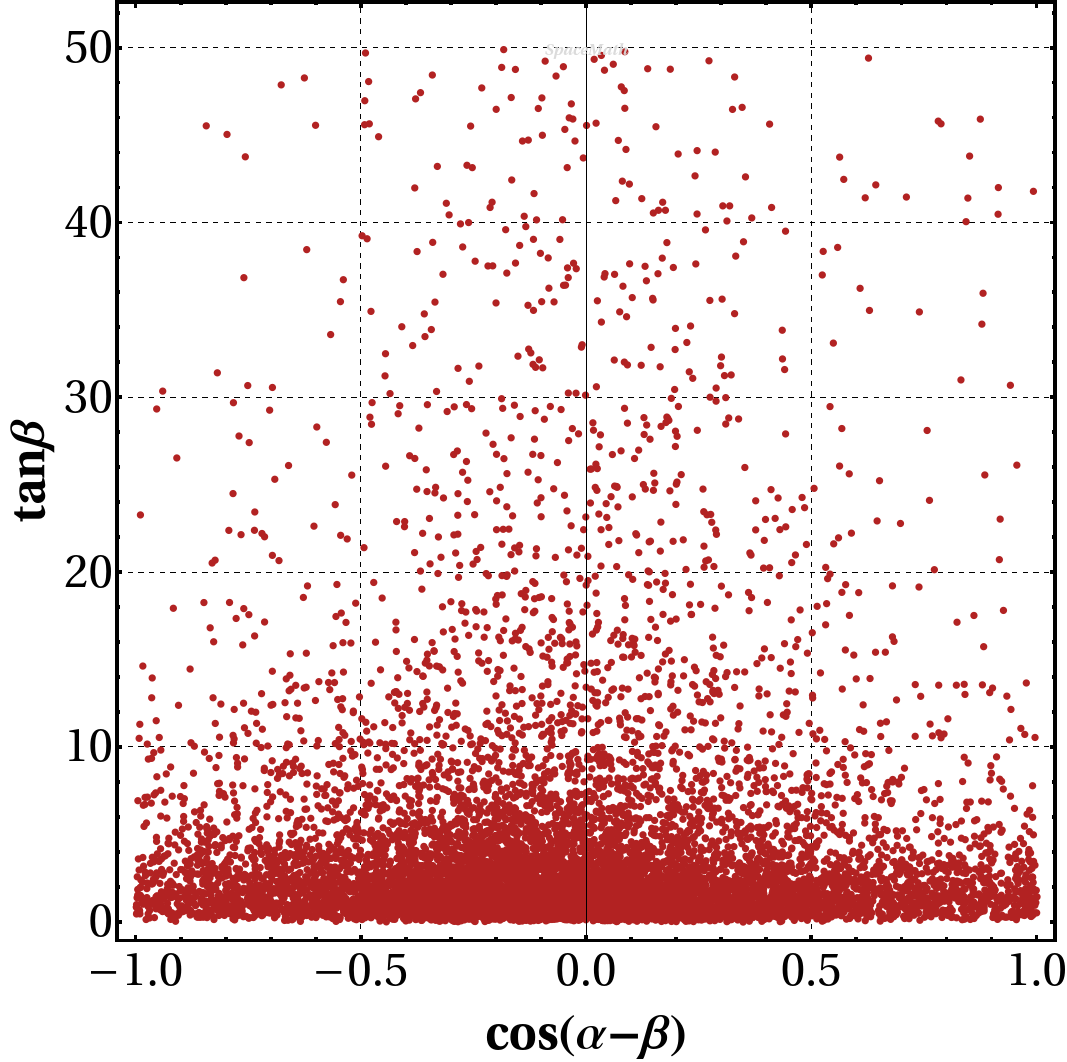}}}
		\subfigure[]{{\includegraphics[scale=0.2]{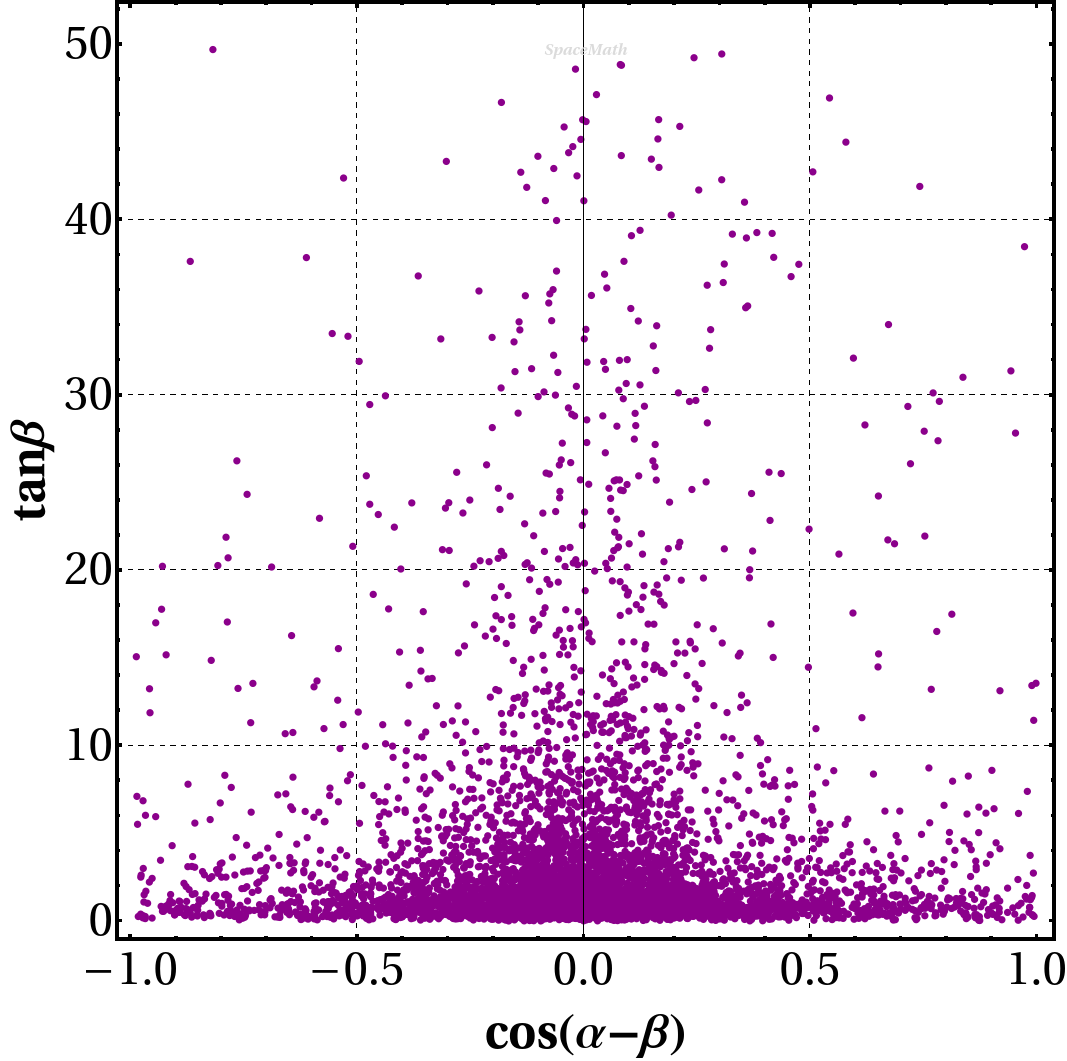}}}
		\subfigure[]{{\includegraphics[scale=0.2]{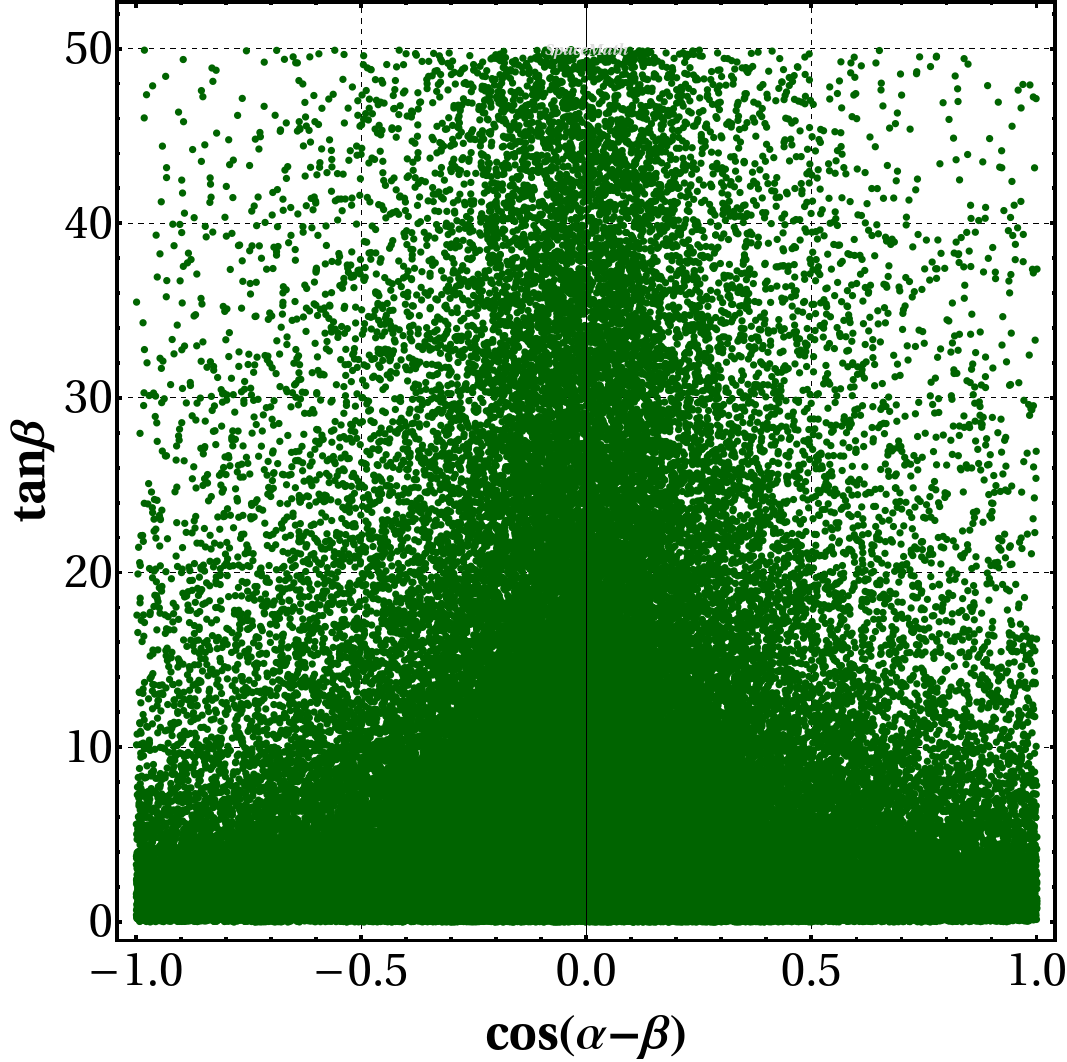}}}
		\caption{Decays $\ell_i\to \ell_j \gamma$: (a) $\mu\to e\gamma$, (b) $\tau\to \mu\gamma$, (c) $\tau\to e\gamma$. In all cases we generate $100$K random points.}\label{RADdecays}
	\end{figure}

	\begin{itemize}
		\item $\ell_i\to \ell_j\gamma$
	\end{itemize}
	Fig. \ref{RADdecays} shows the corresponding allowed points that meet upper limits on $\mathcal{BR}(\ell_i\to \ell_j \gamma)$. The range of scanned parameters is displayed in Table \ref{Scan_li-lj_Gamma}.
	\begin{table}[!htb]
	\caption{Range of scanned parameters. On the left: $\mu \to e \gamma$, on the right: $\tau \to \mu \gamma$, and below: $\tau \to e \gamma$}\label{Scan_li-lj_Gamma}
	\begin{centering}
		\begin{tabular}{cc}
			\hline 
			Parameter & Range\tabularnewline
			\hline 
			\hline 
			$\tan\beta$ & $[0,50]$\tabularnewline
			\hline 
			$\cos(\alpha-\beta)$ & $[-1,1]$\tabularnewline
			\hline 
			$\chi_{\mu\mu}$ & $[-1,1]$\tabularnewline
			\hline 
			$\chi_{\mu e}$ & $[-1,1]$\tabularnewline
			\hline 
			$\chi_{tt}$ & $[-100,100]$\tabularnewline
			\hline 
			$M_{A}$ (GeV) & $[100,1000]$\tabularnewline
			\hline 
			$M_{H}$ (GeV) & $[300,1000]$\tabularnewline
			\hline 
			$M_{H^{\pm}}$ (GeV) & $[110,1000]$\tabularnewline
			\hline
		\end{tabular}$\qquad$%
		\begin{tabular}{cc}
			\hline 
			Parameter & Range\tabularnewline
			\hline 
			\hline 
			$\tan\beta$ & $[0,50]$\tabularnewline
			\hline 
			$\cos(\alpha-\beta)$ & $[-1,1]$\tabularnewline
			\hline 
			$\chi_{\tau\mu}$ & $[-100,100]$\tabularnewline
			\hline 
			$\chi_{\tau\tau}$ & $[-1,1]$\tabularnewline
			\hline 
			$\chi_{tt}$ & $[-100,100]$\tabularnewline
			\hline 
			$M_{A}$ (GeV) & $[100,1000]$\tabularnewline
			\hline 
			$M_{H}$ (GeV) & $[300,1000]$\tabularnewline
			\hline
			$M_{H^{\pm}}$ (GeV) & $[110,1000]$\tabularnewline
			\hline 
		\end{tabular}
	\par\end{centering}
	\begin{centering}
		\begin{tabular}{cc}
			\hline 
			Parameter & Range\tabularnewline
			\hline 
			\hline 
			$\tan\beta$ & $[0,50]$\tabularnewline
			\hline 
			$\cos(\alpha-\beta)$ & $[-1,1]$\tabularnewline
			\hline 
			$\chi_{\tau e}$ & $[-1,1]$\tabularnewline
			\hline 
			$\chi_{\tau\tau}$ & $[-1,1]$\tabularnewline
			\hline 
			$\chi_{tt}$ & $[-100,100]$\tabularnewline
			\hline 
			$M_{A}$ (GeV) & $[100,1000]$\tabularnewline
			\hline 
			$M_{H}$ (GeV) & $[300,1000]$\tabularnewline
			\hline
			$M_{H^{\pm}}$ (GeV) & $[110,1000]$\tabularnewline
			\hline 
		\end{tabular}
	\par\end{centering}
\end{table}

\begin{figure}[!h]
		\centering
		\subfigure[]{{\includegraphics[scale=0.2]{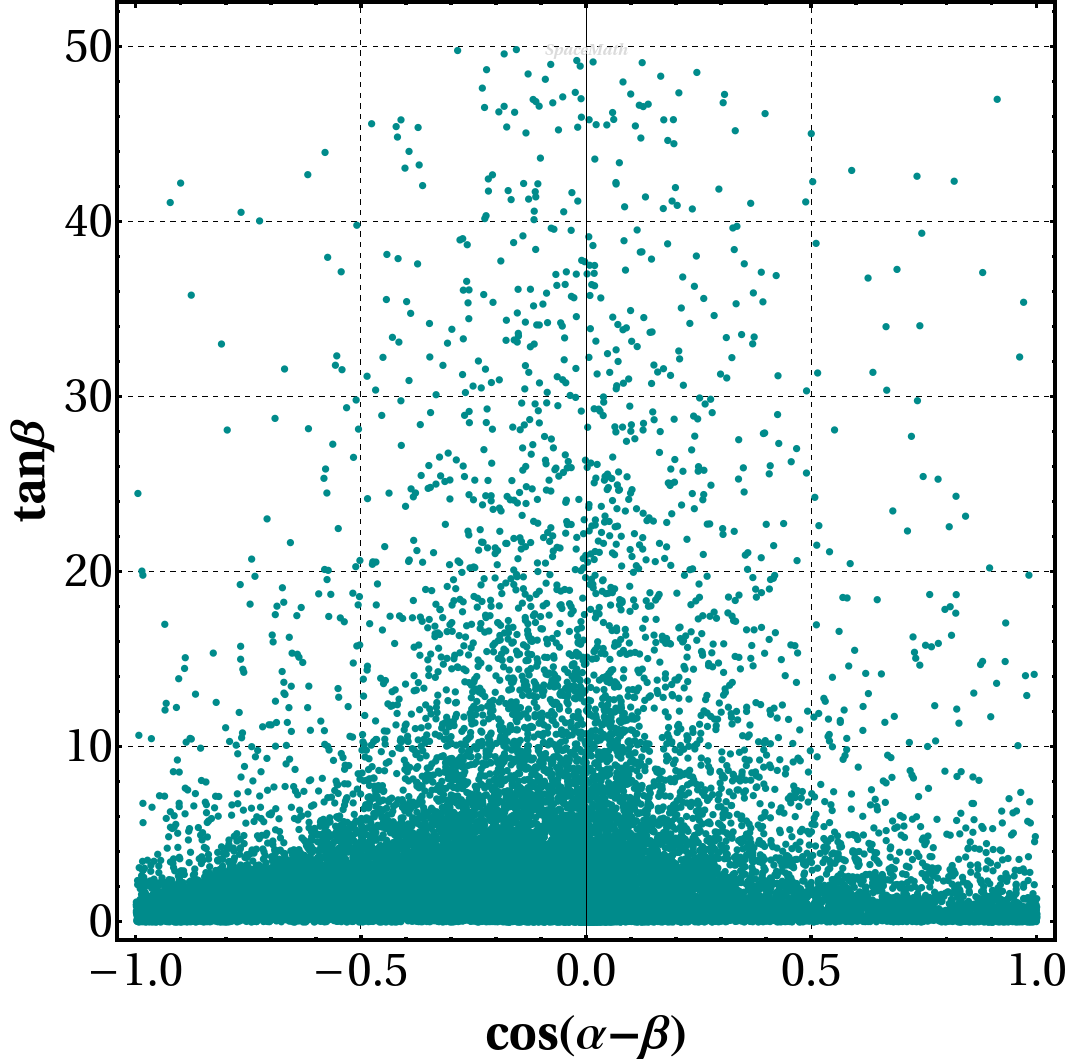}}}
		\subfigure[]{{\includegraphics[scale=0.2]{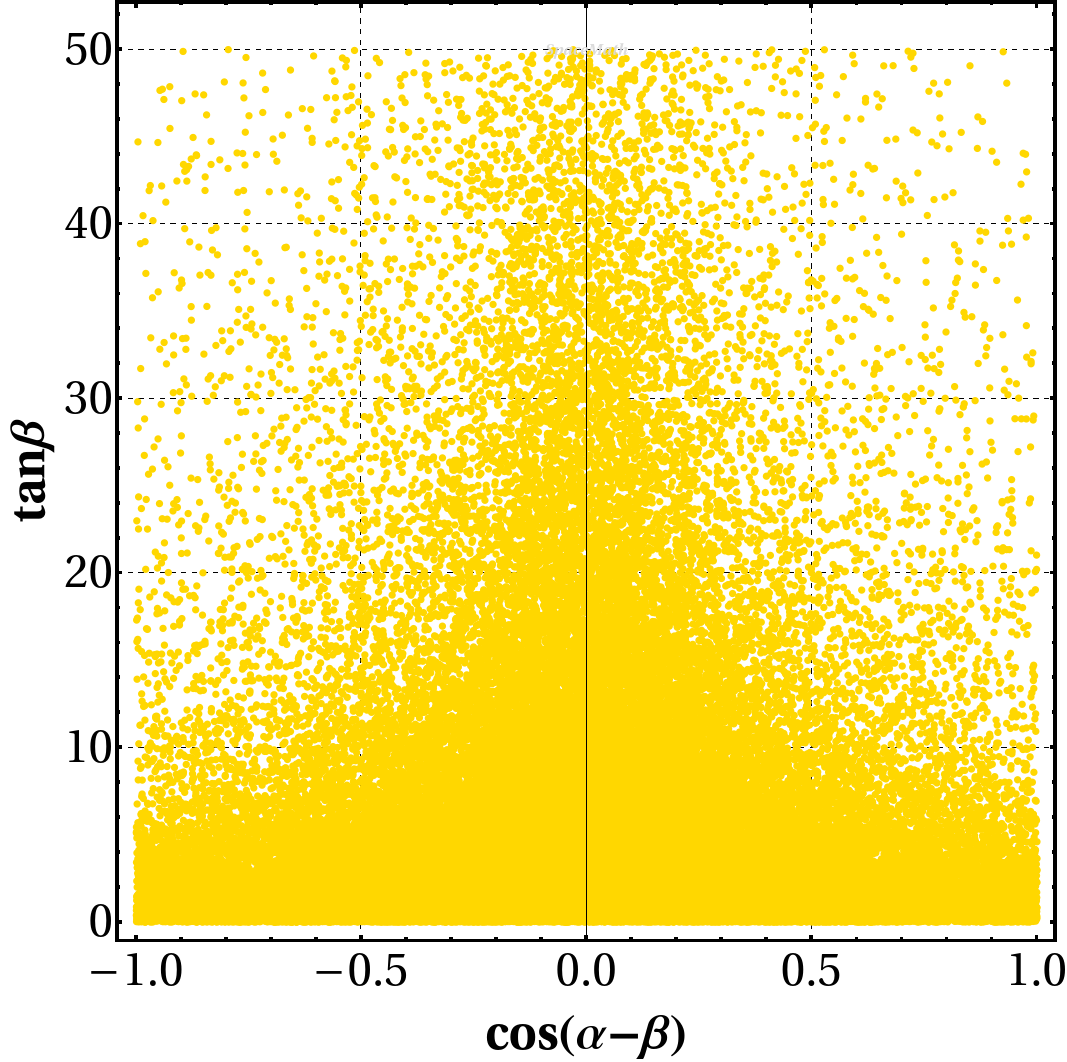}}}
		\caption{Decays $\ell_i\to \ell_j \ell_k \bar{\ell}_k$: (a) $\mu \to 3e$, (b) $\tau\to 3 \mu$. In all cases we generate $100$K random points.}\label{liljlklkDecays}
	\end{figure}

\begin{figure}[H]\label{h-tau_mu}
		\centering
		\includegraphics[width=6cm]{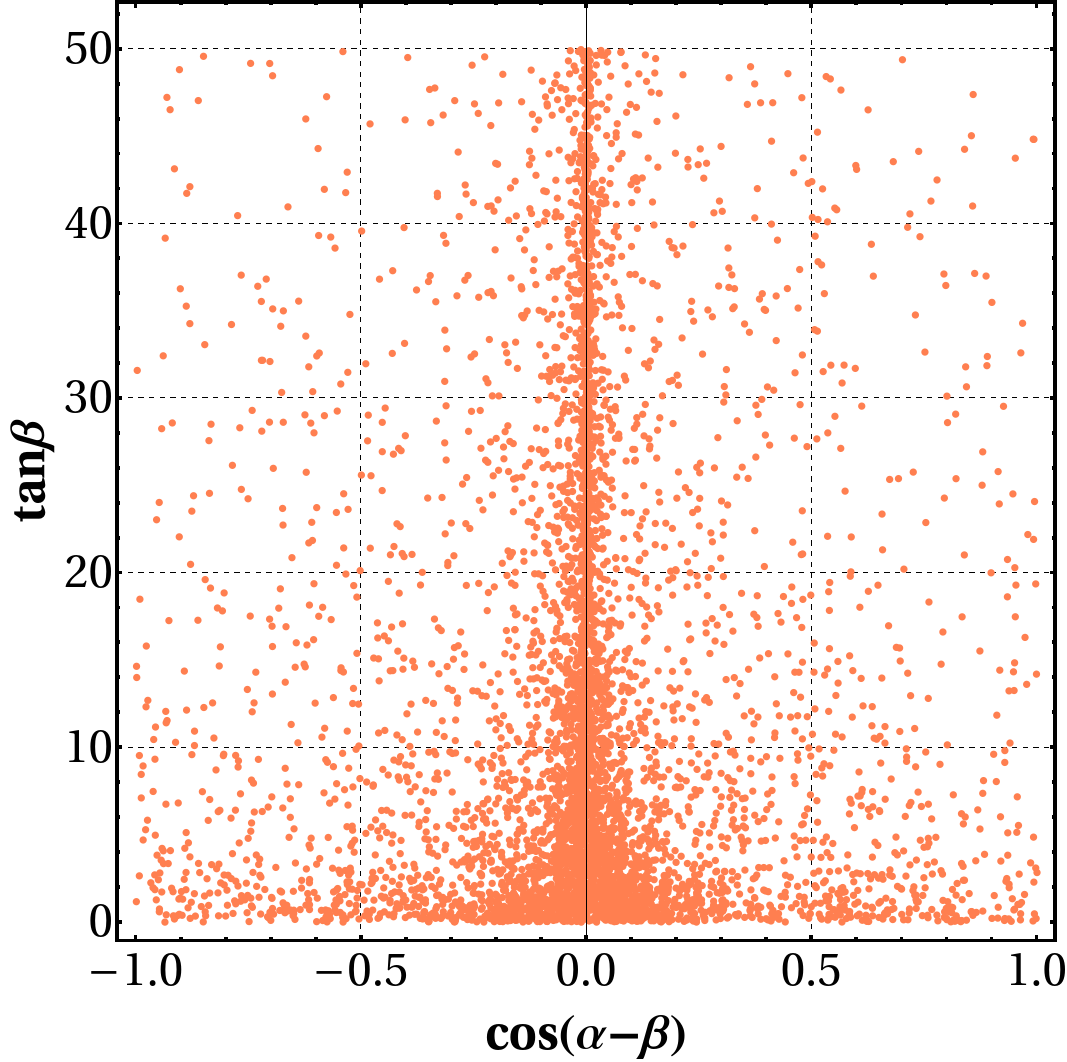}
		\caption{Allowed values by the upper limit on $\mathcal{BR}(h\to\tau\mu)$. We generate $50$K random points.}\label{INDhtaumu}
	\end{figure}

	\begin{itemize}
		\item Decays $\ell_i\to \ell_j \ell_k \bar{\ell}_k$
	\end{itemize}
	As far as the decays $\ell_i\to \ell_j \ell_k \bar{\ell}_k$ are concerned, we present in Fig. \ref{liljlklkDecays} the points that comply with the constraints on $\mathcal{BR}(\ell_i\to \ell_j \ell_k \bar{\ell}_k)$. And in Table \ref{Scanliljlklk} the intervals of the scan on the parameters are presented. 
	\begin{table}[!htb]
		\caption{Range of scanned parameters. On the left: $\mu \to 3e$, on the right: $\tau \to 3\mu$.}\label{Scanliljlklk}
		\begin{centering}
			\begin{tabular}{cc}
				\hline 
				Parameter & Range\tabularnewline
				\hline 
				\hline 
				$\tan\beta$ & $[0,50]$\tabularnewline
				\hline 
				$\cos(\alpha-\beta)$ & $[-1,1]$\tabularnewline
				\hline 
				$\chi_{ee}$ & $[-1,1]$\tabularnewline
				\hline 
				$\chi_{\mu e}$ & $[-1,1]$\tabularnewline
				\hline 
				$M_{A}$ (GeV) & $[100,1000]$\tabularnewline
				\hline 
				$M_{H}$ (GeV) & $[300,1000]$\tabularnewline
				\hline 
			\end{tabular}$\qquad$%
			\begin{tabular}{cc}
				\hline 
				Parameter & Range\tabularnewline
				\hline 
				\hline 
				$\tan\beta$ & $[0,50]$\tabularnewline
				\hline 
				$\cos(\alpha-\beta)$ & $[-1,1]$\tabularnewline
				\hline 
				$\chi_{\mu\mu}$ & $[-1,1]$\tabularnewline
				\hline
				$\chi_{\tau\mu}$ & $[-100,100]$\tabularnewline
				\hline 
				$M_{A}$ (GeV) & $[100,1000]$\tabularnewline
				\hline 
				$M_{H}$ (GeV) & $[300,1000]$\tabularnewline
				\hline
			\end{tabular}
			\par\end{centering}
	\end{table}
	\begin{itemize}
		\item Decays $h\to \ell_i \ell_j$
	\end{itemize}
	We only present the allowed region coming from $h\to\tau\mu$ because the process $h\to e\mu$ imposes very weak bounds to $\tan\beta$ and $\cos(\alpha-\beta)$. Fig. \ref{INDhtaumu} presents the scatter plot in the $\cos(\alpha-\beta)-\tan\beta$ plane whose points are the allowed by the upper limit on $\mathcal{BR}(h\to\tau\mu)$. 	
	The range of scanned parameters is displayed in Table \ref{Scan_h-tau_mu}.
	\begin{table}[!htb]
		\caption{Range of scanned parameters associated to the decay $h\to\tau\mu$. }\label{Scan_h-tau_mu}
		\begin{centering}
			\begin{tabular}{cc}
				\hline 
				Parameter & Range\tabularnewline
				\hline 
				\hline 
				$\tan\beta$ & $[0,50]$\tabularnewline
				\hline 
				$\cos(\alpha-\beta)$ & $[-1,1]$\tabularnewline
				\hline 
				$\chi_{\tau\mu}$ & $[-10,10]$\tabularnewline
				\hline 
			\end{tabular}
			\par\end{centering}
	\end{table}

\section*{Appendix VI}\label{VI}
The variables used to train and test the signal and background processes are shown in Table \label{VarBDT}.	

\begin{widetext}

\begin{table}[!htb]
	
	\caption{List of the variables used to train and test the signal and background
		events. The transverse mass is defined as $M_{T}[\ell]=\sqrt{2P_{T}^{\ell}E_{T}(1-\cos\phi_{\ell \slashed{E}_T}).}$}\label{VarBDT}
	\begin{centering}
		\begin{tabular}{|c|c|c|}
			\hline 
			Rank & Variable & Description\tabularnewline
			\hline 
			\hline 
			1 & $P_{T}[b]$ & $b-jet$ transverse momentum\tabularnewline
			\hline 
			2 & $\slashed{E}_T$ & Missing Energy Transverse\tabularnewline
			\hline 
			3 & $P_{T}[j_{1}]$ & Jet with the largest transverse momentum ($c-jet$)\tabularnewline
			\hline 
			4 & $M_{T}[\mu]$ & Transverse mass between the $\slashed{E}_T$ and the muon.\tabularnewline
			\hline 
			5 & $M_{inv}[b,c]$ & Invariant mass of the \textbf{$b-jet$ }and\textbf{ $c-jet$}.\tabularnewline
			\hline
			6 &  $\sum P_{T}$ & Sum of the moduli of the transverse momentum of the $b-jet$, leading
			jet and the muon\tabularnewline
			\hline 
			7 & $\eta_{b}$ & Pseudorapidity of the \textbf{$b-jet$}\tabularnewline
			\hline 
			8 & $P_{T}^{all}[jet]$ & Scalar sum of the transverse momentum of all jets.\tabularnewline
			\hline 
			9 & $\Delta\eta[j_{1},j_{2}]$ & Absolute value of the pseudorapidity separation between two jets\tabularnewline
			\hline 
			10 & $\eta[j_{1}]$ & Pseudorapidity of the leading jet\tabularnewline
			\hline 
			11 & $\slashed{E}_{tot}$ & Total transverse energy in the detector\tabularnewline
			\hline 
			12 & $\eta[j_{1}]\times\eta[j_{2}]$ & Product of the pseudorapidities of the two jets\tabularnewline
			\hline 
		\end{tabular}
		\par\end{centering}
\end{table}
\end{widetext}
\section*{Oblique parameters}\label{ObParam}
The $S,\,T,\,U$ oblique parameters in the theoretical framework of 2HDMs are given by
\begin{widetext}
\begin{align}
	\begin{array}{lll}
		S & = & \frac{1}{\pi m_{Z}^{2}}\left\{ \sin^{2}\left(\alpha-\beta\right)\left[\mathcal{B}_{22}\left(m_{Z}^{2};m_{Z}^{2},m_{h}^{2}\right)-M_{Z}^{2}\mathcal{B}_{0}\left(m_{Z}^{2};m_{Z}^{2},m_{h}^{2}\right)+\mathcal{B}_{22}\left(m_{Z}^{2};M_{H}^{2},M_{A}^{2}\right)\right]\right.\\
		& + & \cos^{2}\left(\alpha-\beta\right)\left[\mathcal{B}_{22}\left(m_{Z}^{2};m_{Z}^{2},M_{H}^{2}\right)-m_{Z}^{2}\mathcal{B}_{0}\left(m_{Z}^{2};m_{Z}^{2},M_{H}^{2}\right)+\mathcal{B}_{22}\left(m_{Z}^{2};m_{h}^{2},M_{A}^{2}\right)\right]\\
		& - & \left.\mathcal{B}_{22}\left(m_{Z}^{2};M_{H^{\pm}}^{2},M_{H^{\pm}}^{2}\right)-\mathcal{B}_{22}\left(m_{Z}^{2};m_{Z}^{2},m_{h}^{2}\right)+m_{Z}^{2}\mathcal{B}_{0}\left(m_{Z}^{2};m_{Z}^{2},m_{h}^{2}\right)\right\} ,
	\end{array}
\end{align}

\begin{align}
	\begin{array}{lll}
		T & = & \frac{1}{16\pi m_{W}^{2}s_{W}^{2}}\left\{ \sin^{2}\left(\alpha-\beta\right)\left[\mathcal{F}\left(M_{H^{\pm}}^{2},M_{H}^{2}\right)-\mathcal{F}\left(M_{H}^{2},M_{A}^{2}\right)+3\mathcal{F}\left(m_{Z}^{2},m_{h}^{2}\right)-3\mathcal{F}\left(m_{W}^{2},m_{h}^{2}\right)\right]\right.\\
		& + & \cos^{2}\left(\alpha-\beta\right)\left[\mathcal{F}\left(M_{H^{\pm}}^{2},m_{h}^{2}\right)-\mathcal{F}\left(m_{h}^{2},M_{A}^{2}\right)+3\mathcal{F}\left(m_{Z}^{2},M_{H}^{2}\right)-3\mathcal{F}\left(m_{W}^{2},M_{H}^{2}\right)\right]\\
		& + & \left.\mathcal{F}\left(M_{H^{\pm}}^{2},M_{A}^{2}\right)-3\mathcal{F}\left(m_{Z}^{2},m_{h}^{2}\right)+3\mathcal{F}\left(m_{W}^{2},m_{h}^{2}\right)\right\} ,
	\end{array}
\end{align}

\begin{align}
	\begin{array}{lll}
		U & = & \mathcal{H}\left(m_{W}^{2}\right)-\mathcal{H}\left(m_{Z}^{2}\right)\\
		& + & \frac{1}{\pi m_{W}^{2}}\left\{ \cos^{2}\left(\alpha-\beta\right)\mathcal{B}_{22}\left(m_{W}^{2};M_{H^{\pm}}^{2},m_{h}^{2}\right)+\sin^{2}\left(\alpha-\beta\right)\mathcal{B}_{22}\left(m_{W}^{2};M_{H^{\pm}}^{2},M_{H}^{2}\right)\right.\\
		& + & \left.\mathcal{B}_{22}\left(m_{W}^{2};M_{H^{\pm}}^{2},M_{A}^{2}\right)-2\mathcal{B}_{22}\left(m_{W}^{2};M_{H^{\pm}}^{2},M_{H^{\pm}}^{2}\right)\right\} \\
		& - & \frac{1}{\pi m_{Z}^{2}}\left\{ \cos^{2}\left(\alpha-\beta\right)\mathcal{B}_{22}\left(m_{Z}^{2};m_{h}^{2},M_{A}^{2}\right)+\sin^{2}\left(\alpha-\beta\right)\mathcal{B}_{22}\left(m_{Z}^{2};M_{H}^{2},M_{A}^{2}\right)\right.\\
		& - & \left.\mathcal{B}_{22}\left(m_{Z}^{2};M_{H^{\pm}}^{2},M_{H}^{2}\right)\right\} ,
	\end{array}
\end{align}	

where
\begin{align}
	\begin{array}{lll}
		\mathcal{H}\left(m_{V}^{2}\right) & \equiv & \frac{1}{\pi m_{V}^{2}}\left\{ \sin^{2}\left(\alpha-\beta\right)\left[\mathcal{B}_{22}\left(m_{V}^{2};m_{V}^{2},m_{h}^{2}\right)-m_{V}^{2}\mathcal{B}_{0}\left(m_{V}^{2};m_{V}^{2},m_{h}^{2}\right)\right]\right.\\
		& + & \cos^{2}\left(\alpha-\beta\right)\left[\mathcal{B}_{22}\left(m_{V}^{2};m_{V}^{2},M_{H}^{2}\right)-m_{V}^{2}\mathcal{B}_{0}\left(m_{V}^{2};m_{V}^{2},M_{H}^{2}\right)\right]\\
		& - & \left.\mathcal{B}_{22}\left(m_{V}^{2};m_{V}^{2},m_{h}^{2}\right)+m_{V}^{2}\mathcal{B}_{0}\left(m_{V}^{2};m_{V}^{2},m_{h}^{2}\right)\right\} ,
	\end{array}
\end{align}	

\begin{equation}\label{key}
	\mathcal{F}\left(m_{1}^{2},m_{2}^{2}\right)=\frac{1}{2}\left(m_{1}^{2}+m_{2}^{2}\right)-\frac{m_{1}^{2}m_{2}^{2}}{m_{1}^{2}-m_{2}^{2}}\log\left(\frac{m_{1}^{2}}{m_{2}^{2}}\right),
\end{equation}

\begin{equation}\label{key}
	\mathcal{B}_{22}\left(q^{2};m_{1}^{2},m_{2}^{2}\right)\equiv B_{22}\left(q^{2};m_{1}^{2},m_{2}^{2}\right)-B_{22}\left(0;m_{1}^{2},m_{2}^{2}\right),
\end{equation}

\begin{equation}\label{key}
	\mathcal{B}_{0}\left(q^{2};m_{1}^{2},m_{2}^{2}\right)\equiv B_{0}\left(q^{2};m_{1}^{2},m_{2}^{2}\right)-B_{0}\left(0;m_{1}^{2},m_{2}^{2}\right),
\end{equation}

\begin{align}
	\begin{array}{lll}
		B_{22}\left(q^{2};m_{1}^{2},m_{2}^{2}\right) & = & \frac{1}{6}\left[A_{0}\left(m_{2}^{2}\right)+2m_{1}^{2}B_{0}\left(q^{2};m_{1}^{2},m_{2}^{2}\right)+\left(m_{1}^{2}-m_{2}^{2}+q^{2}\right)B_{1}\left(q^{2};m_{1}^{2},m_{2}^{2}\right)\right.\\
		& - & \left.\frac{q^{2}}{3}+m_{1}^{2}+m_{2}^{2}\right],
	\end{array}
\end{align}
in which $A_0,\,B_0,\,B_1$ are scalar Passarino-Veltman functions.
\end{widetext}

\bibliography{aipsamp}

\begin{thebibliography}{43}%
\makeatletter
\providecommand \@ifxundefined [1]{%
 \@ifx{#1\undefined}
}%
\providecommand \@ifnum [1]{%
 \ifnum #1\expandafter \@firstoftwo
 \else \expandafter \@secondoftwo
 \fi
}%
\providecommand \@ifx [1]{%
 \ifx #1\expandafter \@firstoftwo
 \else \expandafter \@secondoftwo
 \fi
}%
\providecommand \natexlab [1]{#1}%
\providecommand \enquote  [1]{``#1''}%
\providecommand \bibnamefont  [1]{#1}%
\providecommand \bibfnamefont [1]{#1}%
\providecommand \citenamefont [1]{#1}%
\providecommand \href@noop [0]{\@secondoftwo}%
\providecommand \href [0]{\begingroup \@sanitize@url \@href}%
\providecommand \@href[1]{\@@startlink{#1}\@@href}%
\providecommand \@@href[1]{\endgroup#1\@@endlink}%
\providecommand \@sanitize@url [0]{\catcode `\\12\catcode `\$12\catcode
  `\&12\catcode `\#12\catcode `\^12\catcode `\_12\catcode `\%12\relax}%
\providecommand \@@startlink[1]{}%
\providecommand \@@endlink[0]{}%
\providecommand \url  [0]{\begingroup\@sanitize@url \@url }%
\providecommand \@url [1]{\endgroup\@href {#1}{\urlprefix }}%
\providecommand \urlprefix  [0]{URL }%
\providecommand \Eprint [0]{\href }%
\providecommand \doibase [0]{https://doi.org/}%
\providecommand \selectlanguage [0]{\@gobble}%
\providecommand \bibinfo  [0]{\@secondoftwo}%
\providecommand \bibfield  [0]{\@secondoftwo}%
\providecommand \translation [1]{[#1]}%
\providecommand \BibitemOpen [0]{}%
\providecommand \bibitemStop [0]{}%
\providecommand \bibitemNoStop [0]{.\EOS\space}%
\providecommand \EOS [0]{\spacefactor3000\relax}%
\providecommand \BibitemShut  [1]{\csname bibitem#1\endcsname}%
\let\auto@bib@innerbib\@empty
\bibitem [{\citenamefont {Aad}\ \emph {et~al.}(2023{\natexlab{a}})\citenamefont
  {Aad} \emph {et~al.}}]{ATLAS:2023ofo}%
  \BibitemOpen
  \bibfield  {author} {\bibinfo {author} {\bibfnamefont {G.}~\bibnamefont
  {Aad}} \emph {et~al.} (\bibinfo {collaboration} {ATLAS}),\ }\bibfield
  {title} {\bibinfo {title} {Search for a new pseudoscalar decaying into a pair
  of muons in events with a top-quark pair at s=13\,tev with the atlas
  detector},\ }\href {https://doi.org/10.1103/PhysRevD.108.092007} {\bibfield
  {journal} {\bibinfo  {journal} {Phys. Rev. D}\ }\textbf {\bibinfo {volume}
  {108}},\ \bibinfo {pages} {092007} (\bibinfo {year} {2023}{\natexlab{a}})},\
  \Eprint {https://arxiv.org/abs/2304.14247} {arXiv:2304.14247 [hep-ex]}
  \BibitemShut {NoStop}%
\bibitem [{\citenamefont {Aad}\ \emph {et~al.}(2012)\citenamefont {Aad} \emph
  {et~al.}}]{ATLAS:2012nhc}%
  \BibitemOpen
  \bibfield  {author} {\bibinfo {author} {\bibfnamefont {G.}~\bibnamefont
  {Aad}} \emph {et~al.} (\bibinfo {collaboration} {ATLAS}),\ }\bibfield
  {title} {\bibinfo {title} {Search for charged higgs bosons decaying via
  $h^{+} \to \tau \nu$ in top quark pair events using $pp$ collision data at
  $\sqrt{s}=7$ tev with the atlas detector},\ }\href
  {https://doi.org/10.1007/JHEP06(2012)039} {\bibfield  {journal} {\bibinfo
  {journal} {JHEP}\ }\textbf {\bibinfo {volume} {06}},\ \bibinfo {pages}
  {039}},\ \Eprint {https://arxiv.org/abs/1204.2760} {arXiv:1204.2760 [hep-ex]}
  \BibitemShut {NoStop}%
\bibitem [{\citenamefont {Chatrchyan}\ \emph {et~al.}(2012)\citenamefont
  {Chatrchyan} \emph {et~al.}}]{CMS:2012fgz}%
  \BibitemOpen
  \bibfield  {author} {\bibinfo {author} {\bibfnamefont {S.}~\bibnamefont
  {Chatrchyan}} \emph {et~al.} (\bibinfo {collaboration} {CMS}),\ }\bibfield
  {title} {\bibinfo {title} {Search for a light charged higgs boson in top
  quark decays in $pp$ collisions at $\sqrt{s}=7$ tev},\ }\href
  {https://doi.org/10.1007/JHEP07(2012)143} {\bibfield  {journal} {\bibinfo
  {journal} {JHEP}\ }\textbf {\bibinfo {volume} {07}},\ \bibinfo {pages}
  {143}},\ \Eprint {https://arxiv.org/abs/1205.5736} {arXiv:1205.5736 [hep-ex]}
  \BibitemShut {NoStop}%
\bibitem [{\citenamefont {Aad}\ \emph {et~al.}(2023{\natexlab{b}})\citenamefont
  {Aad} \emph {et~al.}}]{ATLAS:2023bzb}%
  \BibitemOpen
  \bibfield  {author} {\bibinfo {author} {\bibfnamefont {G.}~\bibnamefont
  {Aad}} \emph {et~al.} (\bibinfo {collaboration} {ATLAS}),\ }\bibfield
  {title} {\bibinfo {title} {Search for a light charged higgs boson in $t
  \rightarrow h^{\pm}b$ decays, with $h^{\pm} \rightarrow cb$, in the
  lepton+jets final state in proton-proton collisions at $\sqrt{s}=13$ tev with
  the atlas detector},\ }\href {https://doi.org/10.1007/JHEP09(2023)004}
  {\bibfield  {journal} {\bibinfo  {journal} {JHEP}\ }\textbf {\bibinfo
  {volume} {09}},\ \bibinfo {pages} {004}},\ \Eprint
  {https://arxiv.org/abs/2302.11739} {arXiv:2302.11739 [hep-ex]} \BibitemShut
  {NoStop}%
\bibitem [{\citenamefont {Hern\'andez-S\'anchez}\ \emph
  {et~al.}(2013)\citenamefont {Hern\'andez-S\'anchez}, \citenamefont {Moretti},
  \citenamefont {Noriega-Papaqui},\ and\ \citenamefont
  {Rosado}}]{HernandezSanchez:2012eg}%
  \BibitemOpen
  \bibfield  {author} {\bibinfo {author} {\bibfnamefont {J.}~\bibnamefont
  {Hern\'andez-S\'anchez}}, \bibinfo {author} {\bibfnamefont {S.}~\bibnamefont
  {Moretti}}, \bibinfo {author} {\bibfnamefont {R.}~\bibnamefont
  {Noriega-Papaqui}},\ and\ \bibinfo {author} {\bibfnamefont {A.}~\bibnamefont
  {Rosado}},\ }\bibfield  {title} {\bibinfo {title} {Off-diagonal terms in
  yukawa textures of the type-iii 2-higgs doublet model and light charged higgs
  boson phenomenology},\ }\href {https://doi.org/10.1007/JHEP07(2013)044}
  {\bibfield  {journal} {\bibinfo  {journal} {JHEP}\ }\textbf {\bibinfo
  {volume} {1307}},\ \bibinfo {pages} {044}},\ \Eprint
  {https://arxiv.org/abs/1212.6818} {arXiv:1212.6818 [hep-ph]} \BibitemShut
  {NoStop}%
\bibitem [{\citenamefont {F\'elix-Beltr\'an}\ \emph {et~al.}(2015)\citenamefont
  {F\'elix-Beltr\'an}, \citenamefont {Gonz\'alez-Canales}, \citenamefont
  {Hern\'andez-S\'anchez}, \citenamefont {Moretti}, \citenamefont
  {Noriega-Papaqui},\ and\ \citenamefont {Rosado}}]{Felix-Beltran:2013tra}%
  \BibitemOpen
  \bibfield  {author} {\bibinfo {author} {\bibfnamefont {O.}~\bibnamefont
  {F\'elix-Beltr\'an}}, \bibinfo {author} {\bibfnamefont {F.}~\bibnamefont
  {Gonz\'alez-Canales}}, \bibinfo {author} {\bibfnamefont {J.}~\bibnamefont
  {Hern\'andez-S\'anchez}}, \bibinfo {author} {\bibfnamefont {S.}~\bibnamefont
  {Moretti}}, \bibinfo {author} {\bibfnamefont {R.}~\bibnamefont
  {Noriega-Papaqui}},\ and\ \bibinfo {author} {\bibfnamefont {A.}~\bibnamefont
  {Rosado}},\ }\bibfield  {title} {\bibinfo {title} {Analysis of the quark
  sector in the 2hdm with a four-zero yukawa texture using the most recent data
  on the ckm matrix},\ }\href {https://doi.org/10.1016/j.physletb.2015.02.003}
  {\bibfield  {journal} {\bibinfo  {journal} {Phys. Lett. B}\ }\textbf
  {\bibinfo {volume} {742}},\ \bibinfo {pages} {347} (\bibinfo {year}
  {2015})},\ \Eprint {https://arxiv.org/abs/1311.5210} {arXiv:1311.5210
  [hep-ph]} \BibitemShut {NoStop}%
\bibitem [{\citenamefont {Hern\'andez-S\'anchez}\ \emph
  {et~al.}(2020)\citenamefont {Hern\'andez-S\'anchez}, \citenamefont
  {Honorato}, \citenamefont {Moretti},\ and\ \citenamefont
  {Rosado-Navarro}}]{Hernandez-Sanchez:2020vax}%
  \BibitemOpen
  \bibfield  {author} {\bibinfo {author} {\bibfnamefont {J.}~\bibnamefont
  {Hern\'andez-S\'anchez}}, \bibinfo {author} {\bibfnamefont {C.~G.}\
  \bibnamefont {Honorato}}, \bibinfo {author} {\bibfnamefont {S.}~\bibnamefont
  {Moretti}},\ and\ \bibinfo {author} {\bibfnamefont {S.}~\bibnamefont
  {Rosado-Navarro}},\ }\bibfield  {title} {\bibinfo {title} {Charged higgs
  boson production via $cb$-fusion at the large hadron collider},\ }\href
  {https://doi.org/10.1103/PhysRevD.102.055008} {\bibfield  {journal} {\bibinfo
   {journal} {Phys. Rev. D}\ }\textbf {\bibinfo {volume} {102}},\ \bibinfo
  {pages} {055008} (\bibinfo {year} {2020})},\ \Eprint
  {https://arxiv.org/abs/2003.06263} {arXiv:2003.06263 [hep-ph]} \BibitemShut
  {NoStop}%
\bibitem [{\citenamefont {Arroyo-Ure\~na}\ \emph {et~al.}(2020)\citenamefont
  {Arroyo-Ure\~na}, \citenamefont {Valencia-P\'erez}, \citenamefont {Gait\'an},
  \citenamefont {Montes De~Oca},\ and\ \citenamefont
  {Fern\'andez-T\'ellez}}]{Arroyo-Urena:2020mgg}%
  \BibitemOpen
  \bibfield  {author} {\bibinfo {author} {\bibfnamefont {M.~A.}\ \bibnamefont
  {Arroyo-Ure\~na}}, \bibinfo {author} {\bibfnamefont {T.~A.}\ \bibnamefont
  {Valencia-P\'erez}}, \bibinfo {author} {\bibfnamefont {R.}~\bibnamefont
  {Gait\'an}}, \bibinfo {author} {\bibfnamefont {J.~H.}\ \bibnamefont {Montes
  De~Oca}},\ and\ \bibinfo {author} {\bibfnamefont {A.}~\bibnamefont
  {Fern\'andez-T\'ellez}},\ }\bibfield  {title} {\bibinfo {title}
  {Flavor-changing decay $h\to \tau\mu$ at super hadron colliders},\ }\href
  {https://doi.org/10.1007/JHEP08(2020)170} {\bibfield  {journal} {\bibinfo
  {journal} {JHEP}\ }\textbf {\bibinfo {volume} {08}},\ \bibinfo {pages}
  {170}},\ \Eprint {https://arxiv.org/abs/2002.04120} {arXiv:2002.04120
  [hep-ph]} \BibitemShut {NoStop}%
\bibitem [{\citenamefont {Arroyo-Ure\~na}\ \emph {et~al.}(2019)\citenamefont
  {Arroyo-Ure\~na}, \citenamefont {Gait\'an}, \citenamefont {Herrera-Chac\'on},
  \citenamefont {Montes~de Oca~Y.},\ and\ \citenamefont
  {Valencia-P\'erez}}]{Arroyo-Urena:2019qhl}%
  \BibitemOpen
  \bibfield  {author} {\bibinfo {author} {\bibfnamefont {M.~A.}\ \bibnamefont
  {Arroyo-Ure\~na}}, \bibinfo {author} {\bibfnamefont {R.}~\bibnamefont
  {Gait\'an}}, \bibinfo {author} {\bibfnamefont {E.~A.}\ \bibnamefont
  {Herrera-Chac\'on}}, \bibinfo {author} {\bibfnamefont {J.~H.}\ \bibnamefont
  {Montes~de Oca~Y.}},\ and\ \bibinfo {author} {\bibfnamefont {T.~A.}\
  \bibnamefont {Valencia-P\'erez}},\ }\bibfield  {title} {\bibinfo {title}
  {Search for the $t\to ch$ decay at hadron colliders},\ }\href
  {https://doi.org/10.1007/JHEP07(2019)041} {\bibfield  {journal} {\bibinfo
  {journal} {JHEP}\ }\textbf {\bibinfo {volume} {07}},\ \bibinfo {pages}
  {041}},\ \Eprint {https://arxiv.org/abs/1903.02718} {arXiv:1903.02718
  [hep-ph]} \BibitemShut {NoStop}%
\bibitem [{\citenamefont {Arroyo-Ure\~na}\ \emph {et~al.}(2016)\citenamefont
  {Arroyo-Ure\~na}, \citenamefont {Diaz-Cruz}, \citenamefont {D\'\i{}az},\ and\
  \citenamefont {Orduz-Ducuara}}]{Arroyo-Urena:2013cyf}%
  \BibitemOpen
  \bibfield  {author} {\bibinfo {author} {\bibfnamefont {M.~A.}\ \bibnamefont
  {Arroyo-Ure\~na}}, \bibinfo {author} {\bibfnamefont {J.~L.}\ \bibnamefont
  {Diaz-Cruz}}, \bibinfo {author} {\bibfnamefont {E.}~\bibnamefont
  {D\'\i{}az}},\ and\ \bibinfo {author} {\bibfnamefont {J.~A.}\ \bibnamefont
  {Orduz-Ducuara}},\ }\bibfield  {title} {\bibinfo {title} {Flavor violating
  higgs signals in the texturized two-higgs doublet model (thdm-tx)},\ }\href
  {https://doi.org/10.1088/1674-1137/40/12/123103} {\bibfield  {journal}
  {\bibinfo  {journal} {Chin. Phys. C}\ }\textbf {\bibinfo {volume} {40}},\
  \bibinfo {pages} {123103} (\bibinfo {year} {2016})},\ \Eprint
  {https://arxiv.org/abs/1306.2343} {arXiv:1306.2343 [hep-ph]} \BibitemShut
  {NoStop}%
\bibitem [{\citenamefont {Sirunyan}\ \emph
  {et~al.}(2018{\natexlab{a}})\citenamefont {Sirunyan} \emph
  {et~al.}}]{CMS:2017con}%
  \BibitemOpen
  \bibfield  {author} {\bibinfo {author} {\bibfnamefont {A.~M.}\ \bibnamefont
  {Sirunyan}} \emph {et~al.} (\bibinfo {collaboration} {CMS}),\ }\bibfield
  {title} {\bibinfo {title} {Search for lepton flavour violating decays of the
  higgs boson to $\mu\tau$ and e$\tau$ in proton-proton collisions at
  $\sqrt{s}=13$ tev},\ }\href {https://doi.org/10.1007/JHEP06(2018)001}
  {\bibfield  {journal} {\bibinfo  {journal} {JHEP}\ }\textbf {\bibinfo
  {volume} {06}},\ \bibinfo {pages} {001}},\ \Eprint
  {https://arxiv.org/abs/1712.07173} {arXiv:1712.07173 [hep-ex]} \BibitemShut
  {NoStop}%
\bibitem [{\citenamefont {Aad}\ \emph {et~al.}(2020)\citenamefont {Aad} \emph
  {et~al.}}]{ATLAS:2019pmk}%
  \BibitemOpen
  \bibfield  {author} {\bibinfo {author} {\bibfnamefont {G.}~\bibnamefont
  {Aad}} \emph {et~al.} (\bibinfo {collaboration} {ATLAS}),\ }\bibfield
  {title} {\bibinfo {title} {Searches for lepton-flavour-violating decays of
  the higgs boson in $\sqrt{s}=13$ tev pp collisions with the atlas detector},\
  }\href {https://doi.org/10.1016/j.physletb.2019.135069} {\bibfield  {journal}
  {\bibinfo  {journal} {Phys. Lett. B}\ }\textbf {\bibinfo {volume} {800}},\
  \bibinfo {pages} {135069} (\bibinfo {year} {2020})},\ \Eprint
  {https://arxiv.org/abs/1907.06131} {arXiv:1907.06131 [hep-ex]} \BibitemShut
  {NoStop}%
\bibitem [{\citenamefont {Tumasyan}\ \emph {et~al.}(2023)\citenamefont
  {Tumasyan} \emph {et~al.}}]{CMS:2022mgd}%
  \BibitemOpen
  \bibfield  {author} {\bibinfo {author} {\bibfnamefont {A.}~\bibnamefont
  {Tumasyan}} \emph {et~al.} (\bibinfo {collaboration} {CMS}),\ }\bibfield
  {title} {\bibinfo {title} {Measurement of the
  b$^0_\mathrm{S}$$\to$$\mu^+\mu^-$ decay properties and search for the
  b$^0$$\to$$\mu^+\mu^-$ decay in proton-proton collisions at $\sqrt{s}$ = 13
  tev},\ }\href {https://doi.org/10.1016/j.physletb.2023.137955} {\bibfield
  {journal} {\bibinfo  {journal} {Phys. Lett. B}\ }\textbf {\bibinfo {volume}
  {842}},\ \bibinfo {pages} {137955} (\bibinfo {year} {2023})},\ \Eprint
  {https://arxiv.org/abs/2212.10311} {arXiv:2212.10311 [hep-ex]} \BibitemShut
  {NoStop}%
\bibitem [{\citenamefont {Workman}\ \emph {et~al.}(2022)\citenamefont {Workman}
  \emph {et~al.}}]{Workman:2022ynf}%
  \BibitemOpen
  \bibfield  {author} {\bibinfo {author} {\bibfnamefont {R.~L.}\ \bibnamefont
  {Workman}} \emph {et~al.} (\bibinfo {collaboration} {Particle Data Group}),\
  }\bibfield  {title} {\bibinfo {title} {Review of particle physics},\ }\href
  {https://doi.org/10.1093/ptep/ptac097} {\bibfield  {journal} {\bibinfo
  {journal} {PTEP}\ }\textbf {\bibinfo {volume} {2022}},\ \bibinfo {pages}
  {083C01} (\bibinfo {year} {2022})}\BibitemShut {NoStop}%
\bibitem [{Note1()}]{Note1}%
  \BibitemOpen
  \bibinfo {note} {$a_\mu $ also receives contributions coming from a charged
  scalar boson, however its contribution is subdominant.}\BibitemShut {Stop}%
\bibitem [{\citenamefont {Abi}\ \emph {et~al.}(2021)\citenamefont {Abi} \emph
  {et~al.}}]{Muong-2:2021ojo}%
  \BibitemOpen
  \bibfield  {author} {\bibinfo {author} {\bibfnamefont {B.}~\bibnamefont
  {Abi}} \emph {et~al.} (\bibinfo {collaboration} {Muon g-2}),\ }\bibfield
  {title} {\bibinfo {title} {Measurement of the positive muon anomalous
  magnetic moment to 0.46 ppm},\ }\href
  {https://doi.org/10.1103/PhysRevLett.126.141801} {\bibfield  {journal}
  {\bibinfo  {journal} {Phys. Rev. Lett.}\ }\textbf {\bibinfo {volume} {126}},\
  \bibinfo {pages} {141801} (\bibinfo {year} {2021})},\ \Eprint
  {https://arxiv.org/abs/2104.03281} {arXiv:2104.03281 [hep-ex]} \BibitemShut
  {NoStop}%
\bibitem [{\citenamefont {Aad}\ \emph {et~al.}(2022)\citenamefont {Aad} \emph
  {et~al.}}]{ATLAS:2021ifb}%
  \BibitemOpen
  \bibfield  {author} {\bibinfo {author} {\bibfnamefont {G.}~\bibnamefont
  {Aad}} \emph {et~al.} (\bibinfo {collaboration} {ATLAS}),\ }\bibfield
  {title} {\bibinfo {title} {Search for higgs boson pair production in the two
  bottom quarks plus two photons final state in $pp$ collisions at
  $\sqrt{s}=13$ tev with the atlas detector},\ }\href
  {https://doi.org/10.1103/PhysRevD.106.052001} {\bibfield  {journal} {\bibinfo
   {journal} {Phys. Rev. D}\ }\textbf {\bibinfo {volume} {106}},\ \bibinfo
  {pages} {052001} (\bibinfo {year} {2022})},\ \Eprint
  {https://arxiv.org/abs/2112.11876} {arXiv:2112.11876 [hep-ex]} \BibitemShut
  {NoStop}%
\bibitem [{\citenamefont {Collaboration}(2017)}]{ATLAS-CONF}%
  \BibitemOpen
  \bibfield  {author} {\bibinfo {author} {\bibfnamefont {A.}~\bibnamefont
  {Collaboration}},\ }\href@noop {} {\emph {\bibinfo {title} {CERN Rpt.
  ATLAS-CONF-2017-055}}},\ \bibinfo {type} {Tech. Rep.}\ (\bibinfo
  {institution} {CERN},\ \bibinfo {year} {2017})\BibitemShut {NoStop}%
\bibitem [{\citenamefont {Sirunyan}\ \emph
  {et~al.}(2018{\natexlab{b}})\citenamefont {Sirunyan} \emph
  {et~al.}}]{Sirunyan:2018zut}%
  \BibitemOpen
  \bibfield  {author} {\bibinfo {author} {\bibfnamefont {A.~M.}\ \bibnamefont
  {Sirunyan}} \emph {et~al.} (\bibinfo {collaboration} {CMS}),\ }\bibfield
  {title} {\bibinfo {title} {Search for additional neutral mssm higgs bosons in
  the $\tau\tau$ final state in proton-proton collisions at $\sqrt{s}=$ 13
  tev},\ }\href {https://doi.org/10.1007/JHEP09(2018)007} {\bibfield  {journal}
  {\bibinfo  {journal} {JHEP}\ }\textbf {\bibinfo {volume} {1809}},\ \bibinfo
  {pages} {007}},\ \Eprint {https://arxiv.org/abs/1803.06553} {arXiv:1803.06553
  [hep-ex]} \BibitemShut {NoStop}%
\bibitem [{\citenamefont {Aaboud}\ \emph {et~al.}(2018)\citenamefont {Aaboud}
  \emph {et~al.}}]{ATLAS:2018gfm}%
  \BibitemOpen
  \bibfield  {author} {\bibinfo {author} {\bibfnamefont {M.}~\bibnamefont
  {Aaboud}} \emph {et~al.} (\bibinfo {collaboration} {ATLAS}),\ }\bibfield
  {title} {\bibinfo {title} {Search for charged higgs bosons decaying via
  $h^{\pm} \to \tau^{\pm}\nu_{\tau}$ in the $\tau$+jets and $\tau$+lepton final
  states with 36 fb$^{-1}$ of $pp$ collision data recorded at $\sqrt{s} = 13$
  tev with the atlas experiment},\ }\href
  {https://doi.org/10.1007/JHEP09(2018)139} {\bibfield  {journal} {\bibinfo
  {journal} {JHEP}\ }\textbf {\bibinfo {volume} {09}},\ \bibinfo {pages}
  {139}},\ \Eprint {https://arxiv.org/abs/1807.07915} {arXiv:1807.07915
  [hep-ex]} \BibitemShut {NoStop}%
\bibitem [{\citenamefont {Ciuchini}\ \emph {et~al.}(1998)\citenamefont
  {Ciuchini}, \citenamefont {Degrassi}, \citenamefont {Gambino},\ and\
  \citenamefont {Giudice}}]{Ciuchini:1997xe}%
  \BibitemOpen
  \bibfield  {author} {\bibinfo {author} {\bibfnamefont {M.}~\bibnamefont
  {Ciuchini}}, \bibinfo {author} {\bibfnamefont {G.}~\bibnamefont {Degrassi}},
  \bibinfo {author} {\bibfnamefont {P.}~\bibnamefont {Gambino}},\ and\ \bibinfo
  {author} {\bibfnamefont {G.~F.}\ \bibnamefont {Giudice}},\ }\bibfield
  {title} {\bibinfo {title} {Next-to-leading qcd corrections to $b \to x_s
  \gamma$: Standard model and two higgs doublet model},\ }\href
  {https://doi.org/10.1016/S0550-3213(98)00244-2} {\bibfield  {journal}
  {\bibinfo  {journal} {Nucl. Phys. B}\ }\textbf {\bibinfo {volume} {527}},\
  \bibinfo {pages} {21} (\bibinfo {year} {1998})},\ \Eprint
  {https://arxiv.org/abs/hep-ph/9710335} {arXiv:hep-ph/9710335} \BibitemShut
  {NoStop}%
\bibitem [{\citenamefont {Misiak}\ and\ \citenamefont
  {Steinhauser}(2017)}]{Misiak:2017bgg}%
  \BibitemOpen
  \bibfield  {author} {\bibinfo {author} {\bibfnamefont {M.}~\bibnamefont
  {Misiak}}\ and\ \bibinfo {author} {\bibfnamefont {M.}~\bibnamefont
  {Steinhauser}},\ }\bibfield  {title} {\bibinfo {title} {Weak radiative decays
  of the b meson and bounds on $m_{H^\pm }$ in the two-higgs-doublet model},\
  }\href {https://doi.org/10.1140/epjc/s10052-017-4776-y} {\bibfield  {journal}
  {\bibinfo  {journal} {Eur. Phys. J. C}\ }\textbf {\bibinfo {volume} {77}},\
  \bibinfo {pages} {201} (\bibinfo {year} {2017})},\ \Eprint
  {https://arxiv.org/abs/1702.04571} {arXiv:1702.04571 [hep-ph]} \BibitemShut
  {NoStop}%
\bibitem [{Note2()}]{Note2}%
  \BibitemOpen
  \bibinfo {note} {In general, all of them can induce FCNC at
  tree-level.}\BibitemShut {Stop}%
\bibitem [{\citenamefont {Beneke}\ \emph {et~al.}(2019)\citenamefont {Beneke},
  \citenamefont {Bobeth},\ and\ \citenamefont {Szafron}}]{Beneke:2019slt}%
  \BibitemOpen
  \bibfield  {author} {\bibinfo {author} {\bibfnamefont {M.}~\bibnamefont
  {Beneke}}, \bibinfo {author} {\bibfnamefont {C.}~\bibnamefont {Bobeth}},\
  and\ \bibinfo {author} {\bibfnamefont {R.}~\bibnamefont {Szafron}},\
  }\bibfield  {title} {\bibinfo {title} {Power-enhanced leading-logarithmic qed
  corrections to $b_q \to \mu^+\mu^-$},\ }\href
  {https://doi.org/10.1007/JHEP10(2019)232} {\bibfield  {journal} {\bibinfo
  {journal} {JHEP}\ }\textbf {\bibinfo {volume} {10}},\ \bibinfo {pages}
  {232}},\ \Eprint {https://arxiv.org/abs/1908.07011} {arXiv:1908.07011
  [hep-ph]} \BibitemShut {NoStop}%
\bibitem [{\citenamefont {Buras}\ \emph {et~al.}(2012)\citenamefont {Buras},
  \citenamefont {Girrbach}, \citenamefont {Guadagnoli},\ and\ \citenamefont
  {Isidori}}]{Buras:2012ru}%
  \BibitemOpen
  \bibfield  {author} {\bibinfo {author} {\bibfnamefont {A.~J.}\ \bibnamefont
  {Buras}}, \bibinfo {author} {\bibfnamefont {J.}~\bibnamefont {Girrbach}},
  \bibinfo {author} {\bibfnamefont {D.}~\bibnamefont {Guadagnoli}},\ and\
  \bibinfo {author} {\bibfnamefont {G.}~\bibnamefont {Isidori}},\ }\bibfield
  {title} {\bibinfo {title} {On the standard model prediction for br(b{s,d} to
  mu+ mu-)},\ }\href {https://doi.org/10.1140/epjc/s10052-012-2172-1}
  {\bibfield  {journal} {\bibinfo  {journal} {Eur. Phys. J. C}\ }\textbf
  {\bibinfo {volume} {72}},\ \bibinfo {pages} {2172} (\bibinfo {year}
  {2012})},\ \Eprint {https://arxiv.org/abs/1208.0934} {arXiv:1208.0934
  [hep-ph]} \BibitemShut {NoStop}%
\bibitem [{\citenamefont {Inami}\ and\ \citenamefont
  {Lim}(1981)}]{Inami:1980fz}%
  \BibitemOpen
  \bibfield  {author} {\bibinfo {author} {\bibfnamefont {T.}~\bibnamefont
  {Inami}}\ and\ \bibinfo {author} {\bibfnamefont {C.~S.}\ \bibnamefont
  {Lim}},\ }\bibfield  {title} {\bibinfo {title} {Effects of superheavy quarks
  and leptons in low-energy weak processes k(l) ---\ensuremath{>} mu anti-mu,
  k+ ---\ensuremath{>} pi+ neutrino anti-neutrino and k0
  \ensuremath{<}---\ensuremath{>} anti-k0},\ }\href
  {https://doi.org/10.1143/PTP.65.297} {\bibfield  {journal} {\bibinfo
  {journal} {Prog. Theor. Phys.}\ }\textbf {\bibinfo {volume} {65}},\ \bibinfo
  {pages} {297} (\bibinfo {year} {1981})},\ \bibinfo {note} {[erratum: Prog.
  Theor. Phys. 65 (1981) 1772]}\BibitemShut {NoStop}%
\bibitem [{\citenamefont {Harnik}\ \emph {et~al.}(2013)\citenamefont {Harnik},
  \citenamefont {Kopp},\ and\ \citenamefont {Zupan}}]{Harnik:2012pb}%
  \BibitemOpen
  \bibfield  {author} {\bibinfo {author} {\bibfnamefont {R.}~\bibnamefont
  {Harnik}}, \bibinfo {author} {\bibfnamefont {J.}~\bibnamefont {Kopp}},\ and\
  \bibinfo {author} {\bibfnamefont {J.}~\bibnamefont {Zupan}},\ }\bibfield
  {title} {\bibinfo {title} {Flavor violating higgs decays},\ }\href
  {https://doi.org/10.1007/JHEP03(2013)026} {\bibfield  {journal} {\bibinfo
  {journal} {JHEP}\ }\textbf {\bibinfo {volume} {03}},\ \bibinfo {pages}
  {026}},\ \Eprint {https://arxiv.org/abs/1209.1397} {arXiv:1209.1397 [hep-ph]}
  \BibitemShut {NoStop}%
\bibitem [{\citenamefont {Blankenburg}\ \emph {et~al.}(2012)\citenamefont
  {Blankenburg}, \citenamefont {Ellis},\ and\ \citenamefont
  {Isidori}}]{Blankenburg:2012ex}%
  \BibitemOpen
  \bibfield  {author} {\bibinfo {author} {\bibfnamefont {G.}~\bibnamefont
  {Blankenburg}}, \bibinfo {author} {\bibfnamefont {J.}~\bibnamefont {Ellis}},\
  and\ \bibinfo {author} {\bibfnamefont {G.}~\bibnamefont {Isidori}},\
  }\bibfield  {title} {\bibinfo {title} {Flavour-changing decays of a 125 gev
  higgs-like particle},\ }\href
  {https://doi.org/10.1016/j.physletb.2012.05.007} {\bibfield  {journal}
  {\bibinfo  {journal} {Phys. Lett. B}\ }\textbf {\bibinfo {volume} {712}},\
  \bibinfo {pages} {386} (\bibinfo {year} {2012})},\ \Eprint
  {https://arxiv.org/abs/1202.5704} {arXiv:1202.5704 [hep-ph]} \BibitemShut
  {NoStop}%
\bibitem [{Note3()}]{Note3}%
  \BibitemOpen
  \bibinfo {note} {This software has implemented all the experimental
  constraints considered in our project.}\BibitemShut {Stop}%
\bibitem [{\citenamefont {Arroyo-Ure\~na}\ \emph {et~al.}(2022)\citenamefont
  {Arroyo-Ure\~na}, \citenamefont {Gait\'an},\ and\ \citenamefont
  {Valencia-P\'erez}}]{Arroyo-Urena:2020qup}%
  \BibitemOpen
  \bibfield  {author} {\bibinfo {author} {\bibfnamefont {M.~A.}\ \bibnamefont
  {Arroyo-Ure\~na}}, \bibinfo {author} {\bibfnamefont {R.}~\bibnamefont
  {Gait\'an}},\ and\ \bibinfo {author} {\bibfnamefont {T.~A.}\ \bibnamefont
  {Valencia-P\'erez}},\ }\bibfield  {title} {\bibinfo {title} {Spacemath
  version 1.0 a mathematica package for beyond the standard model parameter
  space searches},\ }\href {https://doi.org/10.31349/RevMexFisE.19.020206}
  {\bibfield  {journal} {\bibinfo  {journal} {Rev. Mex. Fis. E}\ }\textbf
  {\bibinfo {volume} {19}},\ \bibinfo {pages} {020206} (\bibinfo {year}
  {2022})},\ \Eprint {https://arxiv.org/abs/2008.00564} {arXiv:2008.00564
  [hep-ph]} \BibitemShut {NoStop}%
\bibitem [{\citenamefont {Aad}\ \emph {et~al.}()\citenamefont {Aad} \emph
  {et~al.}}]{ATLAS:2020tlo}%
  \BibitemOpen
  \bibfield  {author} {\bibinfo {author} {\bibfnamefont {G.}~\bibnamefont
  {Aad}} \emph {et~al.} (\bibinfo {collaboration} {ATLAS}),\ }\bibfield
  {title} {\bibinfo {title} {{Search for heavy resonances decaying into a pair
  of Z bosons in the $\ell ^+\ell ^-\ell '^+\ell '^-$ and $\ell ^+\ell ^-\nu
  {{\bar{\nu }}}$ final states using 139 $\mathrm {fb}^{-1}$ of
  proton\textendash{}proton collisions at $\sqrt{s} = 13\,$TeV with the ATLAS
  detector}},\ }\href@noop {} {\ }\Eprint {https://arxiv.org/abs/2009.14791}
  {arXiv:2009.14791 [hep-ex]} \BibitemShut {NoStop}%
\bibitem [{\citenamefont {Sirunyan}\ \emph {et~al.}(2020)\citenamefont
  {Sirunyan} \emph {et~al.}}]{CMS:2019bnu}%
  \BibitemOpen
  \bibfield  {author} {\bibinfo {author} {\bibfnamefont {A.~M.}\ \bibnamefont
  {Sirunyan}} \emph {et~al.} (\bibinfo {collaboration} {CMS}),\ }\bibfield
  {title} {\bibinfo {title} {{Search for a heavy Higgs boson decaying to a pair
  of W bosons in proton-proton collisions at $\sqrt{s} =$ 13 TeV}},\ }\href
  {https://doi.org/10.1007/JHEP03(2020)034} {\bibfield  {journal} {\bibinfo
  {journal} {JHEP}\ }\textbf {\bibinfo {volume} {03}},\ \bibinfo {pages}
  {034}},\ \Eprint {https://arxiv.org/abs/1912.01594} {arXiv:1912.01594
  [hep-ex]} \BibitemShut {NoStop}%
\bibitem [{\citenamefont {Aad}\ \emph {et~al.}(2024)\citenamefont {Aad} \emph
  {et~al.}}]{ATLAS:2024jja}%
  \BibitemOpen
  \bibfield  {author} {\bibinfo {author} {\bibfnamefont {G.}~\bibnamefont
  {Aad}} \emph {et~al.} (\bibinfo {collaboration} {ATLAS}),\ }\bibfield
  {title} {\bibinfo {title} {{Search for $t\bar{t}H/A \rightarrow
  t\bar{t}t\bar{t}$ production in proton-proton collisions at $\sqrt{s}=13$ TeV
  with the ATLAS detector}},\ }\href@noop {} {\  (\bibinfo {year} {2024})},\
  \Eprint {https://arxiv.org/abs/2408.17164} {arXiv:2408.17164 [hep-ex]}
  \BibitemShut {NoStop}%
\bibitem [{\citenamefont {Navas}\ \emph {et~al.}(2024)\citenamefont {Navas}
  \emph {et~al.}}]{ParticleDataGroup:2024cfk}%
  \BibitemOpen
  \bibfield  {author} {\bibinfo {author} {\bibfnamefont {S.}~\bibnamefont
  {Navas}} \emph {et~al.} (\bibinfo {collaboration} {Particle Data Group}),\
  }\bibfield  {title} {\bibinfo {title} {{Review of particle physics}},\ }\href
  {https://doi.org/10.1103/PhysRevD.110.030001} {\bibfield  {journal} {\bibinfo
   {journal} {Phys. Rev. D}\ }\textbf {\bibinfo {volume} {110}},\ \bibinfo
  {pages} {030001} (\bibinfo {year} {2024})}\BibitemShut {NoStop}%
\bibitem [{\citenamefont {Alloul}\ \emph {et~al.}(2014)\citenamefont {Alloul},
  \citenamefont {Christensen}, \citenamefont {Degrande}, \citenamefont {Duhr},\
  and\ \citenamefont {Fuks}}]{Alloul:2013bka}%
  \BibitemOpen
  \bibfield  {author} {\bibinfo {author} {\bibfnamefont {A.}~\bibnamefont
  {Alloul}}, \bibinfo {author} {\bibfnamefont {N.~D.}\ \bibnamefont
  {Christensen}}, \bibinfo {author} {\bibfnamefont {C.}~\bibnamefont
  {Degrande}}, \bibinfo {author} {\bibfnamefont {C.}~\bibnamefont {Duhr}},\
  and\ \bibinfo {author} {\bibfnamefont {B.}~\bibnamefont {Fuks}},\ }\bibfield
  {title} {\bibinfo {title} {Feynrules 2.0 - a complete toolbox for tree-level
  phenomenology},\ }\href {https://doi.org/10.1016/j.cpc.2014.04.012}
  {\bibfield  {journal} {\bibinfo  {journal} {Comput. Phys. Commun.}\ }\textbf
  {\bibinfo {volume} {185}},\ \bibinfo {pages} {2250} (\bibinfo {year}
  {2014})},\ \Eprint {https://arxiv.org/abs/1310.1921} {arXiv:1310.1921
  [hep-ph]} \BibitemShut {NoStop}%
\bibitem [{\citenamefont {Alwall}\ \emph {et~al.}(2011)\citenamefont {Alwall},
  \citenamefont {Herquet}, \citenamefont {Maltoni}, \citenamefont {Mattelaer},\
  and\ \citenamefont {Stelzer}}]{MadGraphNLO}%
  \BibitemOpen
  \bibfield  {author} {\bibinfo {author} {\bibfnamefont {J.}~\bibnamefont
  {Alwall}}, \bibinfo {author} {\bibfnamefont {M.}~\bibnamefont {Herquet}},
  \bibinfo {author} {\bibfnamefont {F.}~\bibnamefont {Maltoni}}, \bibinfo
  {author} {\bibfnamefont {O.}~\bibnamefont {Mattelaer}},\ and\ \bibinfo
  {author} {\bibfnamefont {T.}~\bibnamefont {Stelzer}},\ }\bibfield  {title}
  {\bibinfo {title} {Madgraph 5: going beyond},\ }\href
  {https://doi.org/10.1007/JHEP06(2011)128} {\bibfield  {journal} {\bibinfo
  {journal} {JHEP}\ }\textbf {\bibinfo {volume} {06}},\ \bibinfo {pages}
  {128}}\BibitemShut {NoStop}%
\bibitem [{\citenamefont {Sjostrand}(2009)}]{Sjostrand:2008vc}%
  \BibitemOpen
  \bibfield  {author} {\bibinfo {author} {\bibfnamefont {T.}~\bibnamefont
  {Sjostrand}},\ }\href {https://doi.org/10.3204/DESY-PROC-2009-02/41} {\emph
  {\bibinfo {title} {PYTHIA 8 Status Report}}},\ \bibinfo {type} {Tech. Rep.}\
  (\bibinfo  {institution} {DESY},\ \bibinfo {year} {2009})\ \Eprint
  {https://arxiv.org/abs/0809.0303} {arXiv:0809.0303 [hep-ph]} \BibitemShut
  {NoStop}%
\bibitem [{\citenamefont {de~Favereau}\ \emph {et~al.}(2014)\citenamefont
  {de~Favereau}, \citenamefont {Delaere}, \citenamefont {Demin}, \citenamefont
  {Giammanco}, \citenamefont {Lemaitre}, \citenamefont {Mertens},\ and\
  \citenamefont {Selvaggi}}]{delphes}%
  \BibitemOpen
  \bibfield  {author} {\bibinfo {author} {\bibfnamefont {J.}~\bibnamefont
  {de~Favereau}}, \bibinfo {author} {\bibfnamefont {C.}~\bibnamefont
  {Delaere}}, \bibinfo {author} {\bibfnamefont {P.}~\bibnamefont {Demin}},
  \bibinfo {author} {\bibfnamefont {A.}~\bibnamefont {Giammanco}}, \bibinfo
  {author} {\bibfnamefont {V.}~\bibnamefont {Lemaitre}}, \bibinfo {author}
  {\bibfnamefont {A.}~\bibnamefont {Mertens}},\ and\ \bibinfo {author}
  {\bibfnamefont {M.}~\bibnamefont {Selvaggi}},\ }\bibfield  {title} {\bibinfo
  {title} {Delphes 3, a modular framework for fast simulation of a generic
  collider experiment},\ }\href {https://doi.org/10.1007/JHEP02(2014)057}
  {\bibfield  {journal} {\bibinfo  {journal} {JHEP}\ }\textbf {\bibinfo
  {volume} {02}},\ \bibinfo {pages} {057}}\BibitemShut {NoStop}%
\bibitem [{\citenamefont {Cacciari}\ \emph {et~al.}(2012)\citenamefont
  {Cacciari}, \citenamefont {Salam},\ and\ \citenamefont
  {Soyez}}]{Cacciari:2011ma}%
  \BibitemOpen
  \bibfield  {author} {\bibinfo {author} {\bibfnamefont {M.}~\bibnamefont
  {Cacciari}}, \bibinfo {author} {\bibfnamefont {G.~P.}\ \bibnamefont
  {Salam}},\ and\ \bibinfo {author} {\bibfnamefont {G.}~\bibnamefont {Soyez}},\
  }\bibfield  {title} {\bibinfo {title} {Fastjet user manual},\ }\href
  {https://doi.org/10.1140/epjc/s10052-012-1896-2} {\bibfield  {journal}
  {\bibinfo  {journal} {Eur. Phys. J. C}\ }\textbf {\bibinfo {volume} {72}},\
  \bibinfo {pages} {1896} (\bibinfo {year} {2012})},\ \Eprint
  {https://arxiv.org/abs/1111.6097} {arXiv:1111.6097 [hep-ph]} \BibitemShut
  {NoStop}%
\bibitem [{\citenamefont {Cacciari}\ \emph {et~al.}(2008)\citenamefont
  {Cacciari}, \citenamefont {Salam},\ and\ \citenamefont
  {Soyez}}]{Cacciari:2008gp}%
  \BibitemOpen
  \bibfield  {author} {\bibinfo {author} {\bibfnamefont {M.}~\bibnamefont
  {Cacciari}}, \bibinfo {author} {\bibfnamefont {G.~P.}\ \bibnamefont
  {Salam}},\ and\ \bibinfo {author} {\bibfnamefont {G.}~\bibnamefont {Soyez}},\
  }\bibfield  {title} {\bibinfo {title} {The anti-$k_t$ jet clustering
  algorithm},\ }\href {https://doi.org/10.1088/1126-6708/2008/04/063}
  {\bibfield  {journal} {\bibinfo  {journal} {JHEP}\ }\textbf {\bibinfo
  {volume} {04}},\ \bibinfo {pages} {063}},\ \Eprint
  {https://arxiv.org/abs/0802.1189} {arXiv:0802.1189 [hep-ph]} \BibitemShut
  {NoStop}%
\bibitem [{\citenamefont {Woodruff}(2018)}]{Woodruff2018}%
  \BibitemOpen
  \bibfield  {author} {\bibinfo {author} {\bibfnamefont {K.}~\bibnamefont
  {Woodruff}},\ }\href {https://github.com/k-woodruff/bdt-tutorial} {\emph
  {\bibinfo {title} {BDT Tutorial}}},\ \bibinfo {type} {Tech. Rep.}\ (\bibinfo
  {institution} {GitHub repository},\ \bibinfo {year} {2018})\BibitemShut
  {NoStop}%
\bibitem [{\citenamefont {Conte}\ \emph {et~al.}(2013)\citenamefont {Conte},
  \citenamefont {Fuks},\ and\ \citenamefont {Serret}}]{Conte:2012fm}%
  \BibitemOpen
  \bibfield  {author} {\bibinfo {author} {\bibfnamefont {E.}~\bibnamefont
  {Conte}}, \bibinfo {author} {\bibfnamefont {B.}~\bibnamefont {Fuks}},\ and\
  \bibinfo {author} {\bibfnamefont {G.}~\bibnamefont {Serret}},\ }\bibfield
  {title} {\bibinfo {title} {Madanalysis 5, a user-friendly framework for
  collider phenomenology},\ }\href {https://doi.org/10.1016/j.cpc.2012.09.009}
  {\bibfield  {journal} {\bibinfo  {journal} {Comput. Phys. Commun.}\ }\textbf
  {\bibinfo {volume} {184}},\ \bibinfo {pages} {222} (\bibinfo {year}
  {2013})},\ \Eprint {https://arxiv.org/abs/1206.1599} {arXiv:1206.1599
  [hep-ph]} \BibitemShut {NoStop}%
\bibitem [{Note4()}]{Note4}%
  \BibitemOpen
  \bibinfo {note} {We explore up to $M_{H^{\pm }}=200$ GeV, however we found a
  tiny signal significance.}\BibitemShut {Stop}%
\end{thebibliography}%

\end{document}